\documentclass[useAMS,usenatbib,referee]{mn2e}
\usepackage{graphicx}
\usepackage{natbib}
\usepackage{times}
\usepackage{float}
\usepackage{amsmath}
\usepackage{epsfig}
\usepackage{xcolor}
\definecolor{RED}{cmyk}{0,0,0,1}
\definecolor{ROUGE}{cmyk}{0,0,0,1}
\definecolor{red}{cmyk}{0,0,0,1}
\definecolor{blue}{cmyk}{0,0,0,1}
\definecolor{green}{cmyk}{0,0,0,1}
\definecolor{white}{cmyk}{0,0,0,0}
\newcommand{\bmth}[1]{\mbox{\boldmath${#1}$}}
\newcommand{\pdrv}[2]{\frac{\partial #1}{\partial #2}}
\newcommand{\drv}[2]{{{\mathrm{d} #1}\over {\mathrm{d} #2}}}
\newcommand{\re}{\mathrm{Re}}
\newcommand{\im}{\mathrm{Im}}
\newcommand{\bsf}[1]{\mbox{\boldmath$\mathsf{#1}$}}

\begin{document}

\title[The tidal excitation of $r$ modes % in a binary system 
%with arbitrarily misaligned  orbital and stellar angular momenta
%with solar type primary and giant planet secondary
 and  orbital evolution]
 % I: The aligned case]
{The tidal excitation of $r$ modes  in a 
\textcolor{RED}
%binary system 
%with arbitrarily misaligned  orbital and stellar angular momenta
%with 
solar type \textcolor{RED}{star orbited by a}
% primary and 
 giant planet \textcolor{RED}{companion}
 % secondary 
 and the effect on orbital evolution I: The aligned case}
%{On the evolution of a binary due to quasi-stationary tides }

\author[J. C. B. Papaloizou and  G.J. Savonije ]
{ J. C. B. Papaloizou $^{1}$\thanks{E-mail:
		J.C.B.Papaloizou@damtp.cam.ac.uk (JCBP)}, 
{G.J. Savonije$^{2}$\thanks{E-mail: g.j.savonije@uva.nl (GJS)},}
	    \\
	$^{1}$ DAMTP, Centre for Mathematical Sciences, University of
           Cambridge, Wilberforce Road, Cambridge CB3 0WA 	 \\    
	$^{2}$Anton Pannekoek Institute of Astronomy, University of Amsterdam, Science Park 904,
	NL-1098 XH, Amsterdam }

%opening
\maketitle
\maketitle

\date{Accepted. Received; in original form}

\pagerange{\pageref{firstpage}--\pageref{lastpage}} \pubyear{2010}

%\maketitle

\label{firstpage}

\textcolor{blue}{\begin{abstract}
It has been suggested  that tidal interaction is important
for shaping the orbital configurations of close orbiting giant planets.
The excitation of propagating waves and normal modes (dynamical tide)
will be important
for estimating time scales for orbital evolution. %39
We consider the tidal interaction of a Jupiter mass planet
orbiting a solar type primary. Tidal and rotational frequencies are
assumed comparable making the effect of  rotation  important.
Although centrifugal distortion is neglected,  Coriolis forces are fully taken into account. %40
We focus in detail on the potentially resonant  excitation of $r$ modes
associated with spherical harmonics of degrees three and five. These are mostly sited in the radiative core
but with a significant response in the convective envelope where dissipation occurs.
Away from resonance significant orbital evolution over the system lifetime is unlikely. %53
 However, tidal interaction is enhanced near resonances and the orbital evolution
 accelerated as they are passed through. This 
 speed up may be sustained if near  resonance can be maintained.
 For close orbits with primaries  rotating sufficiently rapidly,  this could arise from  angular momentum loss and stellar spin down  
 through a stellar wind bringing about significant orbital evolution
 over the system lifetime.%61
\end{abstract}}

\begin{keywords}
%hydrodynamics - celestial mechanics - planetary systems:
%formation, planet -star interactions, stars: binaries: close,
%rotation, oscillations, solar-type
hydrodynamics - celestial mechanics  - planet-star interactions
 -stars: rotation - stars: oscillations (including pulsations) - stars: solar-type

\end{keywords}

\section{Introduction}

\textcolor{blue}{
Tidal interactions are important in close binary systems.
They lead to orbital circularisation and 
 alignment of the component spin and angular momentum vectors.
When these have  been completed, tidal interactions
 can result in  orbital  evolution
leading to synchronisation of the orbital and component spins.
\citep[see][for a review]{O2014}.
% Recent attention has focused on stars with
%planetary companions such as hot Jupiters where tides have been postulated to play an important
%role in shaping the system. Tidal interactions have been considered in a low tidal forcing equilib-
%rium tide regime or alternatively in a regime where so called dynamical tides and the excitation of
%normal modes in one or both of the components is important.
%In this paper we consider tidal interactions in the equilibrium tide regime in a binary system
%consisting of a primary component with spin angular momentum that is arbitrarily misaligned with
%that of the binary orbit. 
In this paper we study the tidal interaction of a close binary containing both orbital and  spin angular momenta with a view to application to exoplanets where it has been suggested that tides have played an important role
in determining currently observed orbital parameters.}

 \textcolor{blue} {For example it has been suggested that tides cause orbital and spin angular momentum alignment  
for cool stars with convective envelopes \citep[eg.][]{Winn2010}.
This alignment should occur more rapidly than tidal evolution can operate for aligned circularised orbits.
In this context
\citet{Albrecht2012}, noting that  synchronisation and alignment times  should be similar  for equilibrium tides,
imply that the latter will not be effective  \citep[see also][]{O2014}.
However this limitation does not apply to dynamical tides \citep[eg.][]{Terquem} as different normal modes may be excited 
 or wave propagation occur in these cases.}

 \textcolor{blue}{We focus on the case of a  solar mass primary and Jupiter
 mass secondary in a close circular orbit
 \textcolor{RED}{, though application to lower mass seconaries is briefly discussed.} The spin and orbital angular momenta are taken to be aligned. 
 Subsequently confirming that in the absence of the excitation of normal modes, tidal interaction is unlikely to lead to significant orbital evolution,
 we aim to  perform in depth  numerical and semi-analytic studies  of the spectrum of $r$ modes  associated with spherical harmonics of order 
 $3$ and $5$ that are mostly sited in the radiative core.   Rotation is dominant in determining their
 properties and they  have eigenfrequencies that lead to \textcolor{RED}{resonant}  excitation as the secondary approaches synchronisation.
 \textcolor{RED}{ Such an approach may occur through an initially more rapidly rotating star spinning down through a process such the magnetic braking
 undergone by cool stars near the main sequence,}
 }
 
 \textcolor{blue}{We also obtain the response of the convective envelope which is associated with inertial modes \citep[eg.][]{PP81, IP2007,Rieutord2010, LO2021}.
  Notably we do not employ the traditional approximation as this is inapplicable there.
 Under the assumption that turbulent viscosity operates \citep[eg.][]{Zahn1977,Duguid2020} this region provides most of the energy dissipation
 associated with the tides that results in orbital evolution.
 We  determine the effect of resonant mode excitation on orbital evolution and investigate the possibility
 of this occurring at a significant rate while resonance is maintained \citep{SP83,WS02, Zanazzi2021}.}

\textcolor{blue}{ The plan of this paper is as follows.
 In Section \ref{BasicM} we describe the basic model for the 
 % binary%
  system with
 the  coordinates
  %system 
  adopted given  in Section   \ref{Geometry}.
 %the perturbing tidal potential due to the secondary  \ref{pertpoten}
% induced perturbation to the  external gravitational potential due to the primary   \ref{pertpot}
} \textcolor{blue}{ \textcolor{RED}{The perturbing tidal potential due to the secondary,
 is  given  in Section \ref{pertpoten}.
 In Section \ref{pertpot} we derive the perturbed external
gravitational potential of the primary at the position of the planetary
companion and use that to calculate the tidal torque on the planet.
 The formulation of the calculation of the primary's response
  is described in Section \ref {Respd}.}
%In section {Boundary Conditions} \ref {Bdrycons}
The numerical solution procedure is then outlined in Section \ref {Numproc}.
Quantities derived from this include the rate of viscous dissipation 
assumed to be produced by turbulence in the convective envelope (Section   \ref {viscstress})
and the imaginary part of the overlap integral determining the tidal torque (Section \ref{Imagovlap}).
%In section {Interpretation of the terms in equation} (\ref{lequmot2})
}

\textcolor{blue}{Numerical results  are then given in Section \ref{Results}.
These include the determination of the $r$ mode resonances associated with spherical harmonic degrees $l'=3$ and $l'=5$  in Sections \ref{Findr}
% - \ref{l'5res}. 
For each of these values there is a spectrum of modes with increasing number of radial nodes 
 that is closely separated in eigenfrequency
(Section \ref{rmodespec}).}
\textcolor{RED}{For a specified, $l',$ in a frame corotating with the star these eigenfrequencies are very near to the value specified by
\begin{equation}
			\omega_f  = 
			\frac{ 2 \, m \, \Omega_s}{l'\, (l'+1)},
			\label{rmodef}
		\end{equation}
where $\Omega_s$ is the stellar angular velocity and $m$ is the azimuthal mode number \citep[see][]{PP78}. In a non rotating frame
this frequency is shifted to $\omega_f-m\Omega_s.$}	
 \textcolor{blue}{A semi-analytic treatment of the origin of these spectra 
is given in appendix \ref {rmodeapp} and successfully compared with our results. These modes are for the most part \textcolor{red}{ex}cited in the
radiative core. The response they produce in the
convective envelope, where most of the dissipation occurs,  is described in Section \ref{convResp}.
This response is compared to a semi-analytic discussion applicable in the low tidal forcing frequency
limit in appendix  \ref{ConvResp} which is able to explain some of the calculated features.}

\textcolor{blue}{We then go on to formulate the effects of the tidal response on the orbital and spin evolution
of the system in Sections \ref{orbspinevol} - \ref{orbevol}.
%In section {Torque and dissipation resulting from tidal forcing} \ref{TorandD}
%In section {Orbital evolution} \ref{orbevol}
The effect of r mode resonances  that greatly speed up the orbital evolution  
over  a narrow frequency range in their vicinity is described in \ref{rmodeeffect},  with
the  results of numerical calculations of spin and orbit evolution given in Section \ref{Numevol}.
Off-resonant tidal forcing inside and outside the inertial regime is discussed in Sections  \ref{offresinert}
%In \textcolor{red}{ section{Viscous dissipation rate for tidal forcing outside inertial range}
and \ref{offresnoninert}.
The possibility of evolution with resonant interaction maintained, a situation that is required to obtain
significant orbital evolution over a realistic lifetime of the system, is discussed in  \ref{Resmaintained}.
Finally in  Section \ref{Discussion} we discuss our results, \textcolor{RED}{ considering their potential application to the Kepler 1643 and COROT-4 systems and} outlining a direction for future investigations.}

%Appendices
%In appendix {The components of the viscous stress tensor}  \ref{stresscomp}
%In appendix {The   {\MakeLowercase {r}}  mode  spectrum in radiative regions in  the low  frequency adiabatic limit}  \ref {rmodeapp}
%In appendix { The tidal response of the convective envelope in the limit of low forcing frequency as viewed in a frame co-rotating with the star:
 %a critical latitude singularity}   \ref{ConvResp}

\section{Basic model}\label{BasicM}
%\textcolor{red}{ For simplicity,} 

We consider a binary system in which one
component, described as the primary,  possesses an internal density structure  and a spin angular momentum which 
is distinct from the orbital angular momentum. 
Both these angular momenta
 evolve with time with the resultant total angular momentum being conserved.
The other component, described as the companion, is approximated as a point mass.
\textcolor{RED}{Although we focus on a planetary mass companion,} this could model a compact object such as a planet, white dwarf or neutron star. 
In this paper we focus on the tidal response of the primary allowing for the possibility of
resonances associated with, $r,$ modes based,  \textcolor{blue}{for the most part,} in the radiative interior and the response of the convective envelope
without making the simplifying traditional approximation \citep[eg.][]{Alberts}.

As this work involves isolating these resonances and characterising their effect on the tidal response, requiring an involved analysis, 
we simplify matters by  restricting  further discussion 
in this paper to the consideration of circular orbits and aligned spin and orbital angular 
momenta.
The more general case  where the orbital and spin angular momenta are misaligned 
and the orbit eccentric will be considered in a future publication.
We further remark that as  we are interested in the situation where the  normal mode response is significant  and  potentially resonant, a treatment based on
 quasi-static or equilibrium tides such as that considered recently in \citep[eg.][]{IP2021} is inappropriate.
 \textcolor{RED}{The reason for this is that although Coriolis and inertial forces are considered, they are assumed to lead to small
 corrections to the quasi-static response, on account of the tidal forcing period being long compared to the star's dynamical time scale,  and dealt with using a perturbation theory that takes  account only the spheroidal component of the generated response.
 This approach cannot lead to a complete description of the excitation of normal modes for which inertial and Coriolis forces are not perturbations,
 such as $r$ modes, especially when they are resonant.}   
After determining the complete tidal response \textcolor{RED}{, without making such approximations,}  and characterising it,  employing 
 both numerical and semi-analytic methods, \textcolor{blue}{we apply it
 to determine}  the induced tidal evolution of the orbit and primary's spin 
  for  cases of interest.

\subsection{Coordinate system and notation}\label{Geometry}
As we are concerned with a circular orbit and aligned spin and orbital angular momenta
we adopt  a non rotating  Cartesian coordinate system $(X,Y,Z)$  with origin $O$ at the centre of mass of the primary of mass $M_*,$
and  such that  the total angular momentum of the system, ${\bf J}$, 
defines the direction of the $Z$ axis
\footnote{As the spin and orbital angular momenta are aligned
they  can also be used to define the direction of the $Z$ axis.}. 
The  $X$ and $Y$ axes lie in the orthogonal plane passing through $O.$.

%The coordinate systems are illustrated in Fig. \ref{coordinates}.
%\begin{figure}
%\begin{center}
%\vspace{1cm}
%\includegraphics[width=14.0cm,height= 18.0cm,angle=0]{newercoordinates1.pdf}
%\includegraphics[width=8.0cm,angle=270]{RadiusMass.ps}
%\vspace{-7cm}
%\caption{Illustration of the $ (X,Y,Z)$  and   $(X',Y',Z')$ coordinate systems
%together with the direction of the total angular momentum, which coincides with the $Z''$ axis
%of  a coordinate system that is fixed in { the primary centred}  frame .
%Note that the $ X', Y', $ and $Y$ axes are coplanar as are the  $Z, Z' $ and $Z''$ axes .
%The $X'$ and $X''$ axes are co-linear.
 %The angle between the angular momentum vectors ${\bf L},$ directed along the $Z'$ axis 
%and ${\bf S}$ directed along the $Z$ axis  is $\beta.$ { The angle between  ${\bf L}$ and the $Z''$ axis directed along ${\bf J}$ is $i$
%and $\delta=\beta - i.$.   The angle between the $Y''$ axis and the $Y$ axis $2\pi -\alpha_r.$
%The apsidal line, the location of pericentre and  an orbital arc in its neighbourhood 
%are shown.}
%}
%\label{coordinates}
%\end{center}
%\end{figure}

\section{The perturbing tidal potential}\label{pertpoten}
The perturbing tidal potential due to the companion , $U$, can be readily found in the   $(X,Y,Z)$ frame. 
We adopt spherical polar coordinates $(r,\theta, \phi)$ related to $(X,Y,Z)$ in the usual way. 
In the quadrupole approximation we have 
\begin{equation}
 U=-\frac{GM_pr^2}{a^3}P_2(\cos\psi),
\label{jpe0}
\end{equation}
where $M_p$ is the mass of perturbing companion, $a$ is distance between the 
%binary
\textcolor{RED}{orbiting} components,  $P_2$ is the usual Legendre polynomial
and $\cos\psi =\sin\theta\cos(\Phi-\phi)$.
Here the orbit is taken to be in the $\theta = \pi/2$  plane 
 with $\Phi$ being the  azimuthal angle of the line joining the 
 %binary
 \textcolor{RED}{orbiting}  components.
For convenience we shall measure both $\Phi$ and $\phi$ from the $X$ axis without loss of generality.
In addition for a circular orbit we have, $\Phi = n_0 t,$ where $n_0$ is he mean motion and without loss of generality
we choose the origin of time such that $\Phi$ coincides with the $X$ axis at $t=0.$

Equation (\ref{jpe0}) may also written as
\begin{equation}
 U=-\frac{GM_p r^2}{a^3}
 \left(\frac{4\pi}{5}\right) 
\sum^{m=2}_{m=-2}Y_{2,m}(\theta, \phi) Y_{2,m}(\pi/2, 0)\exp(-{\rm i}m \Phi)
 %\equiv r^2\sum_n A_n Y_{2,n}(\theta', \phi'),
\label{jpe1}
\end{equation}  
where $ Y_{l,m}(\theta, \phi),$  is the usual spherical harmonic, \textcolor{blue}{here}  $l=2$.
On account of the primary being axisymmetric we can consider the response to each 
value of $m$ separately and then linearly superpose.
In this case only when  $|m|=2$ does the potential  vary in time and produce a response of interest.
Accordingly we restrict attention to that case and note that

\begin{equation}
 U=-\frac{2GM_p r^2}{a^3}
 \left(\frac{4\pi}{5}\right) Re[
Y_{2,m}(\theta, \phi) Y_{2,m}(\pi/2, 0)\exp(-{\rm i}m\Phi)],
 %\equiv r^2\sum_n A_n Y_{2,n}(\theta', \phi'),
\label{jpe1a}
\end{equation}  
where $Re$ denotes that the real part is to be taken and $m$ may be taken to be either $2$ or $-2.$ Thus only one value of $m$ has to be considered in practice.
We define the tidal factor
\begin{equation}
	c_{tid}\, = -\frac{8 \pi G M_p }{5 \,a^3}\, N_{sph} \, P^{2}_2(\pi/2) 
\end{equation}	
where $N_{sph}=(-1)^m \, \sqrt{5/(4 \pi) (2-|m|)!/(2+|m|)!)}$ is the spherical harmonics normalisation constant. To construct the primary's  full linear response in accordance with  equation (\ref{jpe1}) we \textcolor{blue}{perform}  numerical calculations (see Section \ref{Respd}) to determine the primary´s response to  harmonically varying tidal potentials  of the form 
\begin{equation}
	U_{m,\sigma} \, \exp({\rm i}\sigma t ) = c_{tid} \,\, r^2Y_{2,m}(\theta, \phi) \, \exp({\rm i}\sigma t )\label{potpert}
\end{equation}
 where the forcing frequency $\sigma$ is chosen  to be such that 
$\sigma = - m n_o $
%\textcolor{red}{footnote can be deleted? (U is now defined including $c_{tid}$)}
%\footnote[11]{Here $m$ may be $2$ or $-2$ and $U_{m,\sigma}$ may be multiplied by an appropriate amplitude factor}.
We denote the Lagrangian displacement associated with the response to the perturbing potential $U_{m,\sigma}$ by $\mbox{{\boldmath$\xi$}}_{m,\sigma}  \exp({\rm i}\sigma t)$. The associated Eulerian perturbation of the density is $\rho^\prime_{m,\sigma} \exp({\rm i} \sigma t)$ with similar expressions for the other perturbed state variables.

\section{The perturbation to the  external gravitational potential due to the primary  }\label{pertpot}
After separating out the time dependent factor $\exp({\rm i}\sigma t),$
the perturbation to the external gravitational potential  produced by the tidal  potential $U_{m,\sigma}$  at position vector ${\bf R}$ 
is
\begin{equation}
\psi_{m,\sigma}' = -G\int_V \frac{\rho'_{m,\sigma}({\bf r})}{| {\bf R} -{\bf r}|}dV,       
\end{equation} 
where the integral is taken over the volume of the star.

\noindent For $R= |{\bf R}| >> |{\bf r}|$ we perform a multipole expansion in which the successive terms scale as  inverse powers of $ R.$
The dominant term then takes the form of a quadrupole in the form
\begin{equation}
\psi_{m,\sigma}' = -G\int_V{\rho'_{m,\sigma}({\bf r})}\frac{r^2}{2  R^3}\left(3\frac{ ({\bf r}\cdot{\bf R} )^2}{r^2   R^2} -1\right)  dV  .    
\end{equation} 
%where $R = |{\bf R}|.$
%Evaluating  this at the location of the companion by taking ${\bf R}$ to be the position vector of the orbit
% and 
 Making use of the spherical harmonic addition theorem, 
this can be written in the form
\begin{equation}
\psi_{m,\sigma}' = -\frac{4\pi G}{5R^3} \sum^{m'=2}_{m'=-2}Y_{2,m'}(\theta,\phi) \int_V {\rho'_{m,\sigma}({\bf r})}{r^2}Y_{2,m'}^*(\theta,\phi) dV      
\end{equation} 
Note that in the above where they appear outside the integral,
 $\theta$ and $\phi$ are the spherical polar angles associated with ${\bf R}$ which will be used to define the location of the companion.
Doing this while noting that here we are dealing with circular orbits,  we replace, $R,$ by $a.$
 To avoid additional notation these angles are also used as dummy variables in the integrand.
 We further note that on account of separability in, $\phi,$ in fact only the term with $m'=m$ survives in the summation
 over $m'.$ After making use of this simplification, we may write the gravitational potential 
  produced as a  response to the tidal potential given by
 equation (\ref{potpert}), at the location of the companion, in the form
 
 \begin{align}
&%\psi'=\sum_{n,m,m} 
%\sum^{n=2}_{n=-2}\sum^{k=\infty}_{k=-\infty}
%{\cal A}_{n,k,m}
\psi_{m,\sigma}'  \exp({\rm i}\sigma t )\equiv
%\nonumber\\
%\frac{4\pi G}{5R^3}
%- \frac{a^3}{R^3}
%\sum^{n=2}_{n=-2}\sum^{k=\infty}_{k= -\infty} 
  -  \frac{4\pi G}{5a^3} Q_{m,\sigma}
 % \sum^{'}_{m=0, |2|} \frac{4\pi GM_p}{5 a^3} \phi_{k,m}D^{(2)}_{n,m}  Y_{2,m}(\pi/2, 0)\exp(-{\rm i} (m\varpi -kn_ot))
 Y_{2,m}(\theta,\phi) \exp({\rm i}\sigma t ),\hspace{3mm}{\rm where}
 %\nonumber\\&
 %\int_V {\rho'_{n,k}({\bf r})}{r^2}Y_{2,n}^*(\theta,\phi) dV
 \label{extpotstar}      
\end{align}
%\begin{align}
%\hspace{-8.3cm} {\rm with}\hspace{5mm}  {\cal B}_{n,k,m} = \frac{4\pi G}{5a^3}{\cal A}_{n,k,m}
% Q_{n,\sigma}\hspace{5mm} {\rm and}
 %\label{calBdef}
%\end{align}
\begin{align}
\hspace{-5.2cm} Q_{m,\sigma}= \int_V {\rho'_{m,\sigma}({\bf r})}{r^2}Y_{2,m}^*(\theta,\phi) dV\hspace{5mm} {\rm defines\hspace{1mm}  the \hspace{1mm} overlap \hspace{1mm} integral.}\label{overlap}
\end{align}

\noindent The component of the of the  specific torque in the $Z$ direction this produces is 
  \begin{align}
&T_Z\equiv -\frac{\partial \psi'_{m,\sigma}}{\partial \phi}=
% \sum^{n=2}_{n=-2}\sum^{k=\infty}_{k=-\infty}{\cal A}_{n,k}\psi_{n,k}'  \exp({\rm i}k n_o t )=\nonumber\\
%\frac{4\pi G}{5R^3}
%\frac{a^3}{R^3} 
%\sum^{n=2}_{n=-2}\sum^{k=\infty}_{k= -\infty} 
%\sum_{n,k,m}
\frac{4\pi G}{5a^3}{\rm i} m Q_{m,\sigma}
 % \sum^{'}_{m=0, |2|} \frac{4\pi GM_p}{5 a^3} \phi_{k,m}D^{(2)}_{n,m}  Y_{2,m}(\pi/2, 0)\exp(-{\rm i} (m\varpi -kn_ot))
 Y_{2,m}(\theta,\phi) \exp({\rm i}\sigma t ),
 %{\cal B}_{n,k.m}
 % \sum^{'}_{m=0, |2|} \frac{4\pi GM_p}{5 a^3} \phi_{k,m}D^{(2)}_{n,m}  Y_{2,m}(\pi/2, 0)\exp(-{\rm i} (m\varpi -kn_ot))
% \exp({\rm i} \sigma t )
% Y_{2,n}(\theta,\phi)
  %\nonumber\\ &
  %\int_V {\rho'_{n,k}({\bf r})}{r^2}Y_{2,n}^*(\theta,\phi) dV  
    \label{torqstar}  
\end{align} 
Thus  $T_Z$ is obtained  from 
for $\psi'_{m,\sigma}$ by multiplying it  by $-{\rm i}m.$
 
 We remark that from the properties of spherical harmonics
 we have $\rho'_{-m,-\sigma}=(-1)^m\rho^{'*}_{m,\sigma},\\
  \psi'_{-m,-\sigma}= 
(-1)^m \psi^{'*}_{m,\sigma},
  $ 
 and $Q_{-m,-\sigma}=(-1)^m Q_{m,\sigma}^*.$ These results taken together with the forms of 
 (\ref{jpe0}) and (\ref{potpert}) imply that at the location of the companion where
$\theta=\pi/2$ and  $\phi=n_0t,$   
the specific torque $T_Z$ produced in response to the forcing potential
(\ref{potpert})  is found to be
\begin{align}
T_Z = - \left(\frac{4\pi G}{5a^3}\right) (Y_{2,m}(\pi/2,0))m Im ( Q_{m,\sigma}), \label{torque0}
 % \sum^{'}_{m=0, |2|} \frac{4\pi GM_p}{5 a^3} \phi_{k,m}D^{(2)}_{n,m}  Y_{2,m}(\pi/2, 0)\exp(-{\rm i} (m\varpi -kn_ot))
\end{align}
where $Im$ denotes that the imaginary part is to be taken and $m$ may be taken to be either $2$ or $-2.$

%\textcolor{blue}{However, we find it convenient numerically to remove the factor $c_{tid}$ from (\ref{potpert}) 
%such that the numerical response does not depend on $a$ or $M_p.$ Incorporating this factor into $T_Z,$
%the expression for it given by (\ref{torque0}) is scaled to become
%\begin{align}
%T_Z = 2\left(\frac{4\pi G}{5a^3}\right)^2M_p (Y_{2,m}(\pi/2,0))^2m Im ( Q_{m,\sigma}). \label{torque}
 % \sum^{'}_{m=0, |2|} \frac{4\pi GM_p}{5 a^3} \phi_{k,m}D^{(2)}_{n,m}  Y_{2,m}(\pi/2, 0)\exp(-{\rm i} (m\varpi -kn_ot))
%\end{align}
%Thus from now on we proceed with  $c_{tid}$ being replaced by unity in equation (\ref{potpert}).
\textcolor{blue} {Note that the imaginary part of the overlap
 integral,  $Q_{m,\sigma}$ defined here is $\propto c_{tid}\,$ through its occurrence  as a factor of  the forcing potential (\ref{potpert}). 
  The overlap integral}  plays a key role in determining the tidal evolution.

\section{Calculation of the primary's Response }\label{Respd}

%\section{Numerical calculation of the primary's Response }\label{Respd}
%Effectively summing over contributions from normal modes}

%\textcolor{white}{ 
We now formulate the  calculation of the tidal response of the primary to  the forcing  tidal potential 
\textcolor{RED}{of the planet}
in the linear approximation. This will be obtained numerically.
We focus on the density perturbation that is produced as this will  be used to 
determine the tidal  evolution of
%find tidal torques and the rate of energy transfer from 
the orbit.  As in previous work \citep[eg.][]{PS97} we adopt a stellar model for which Coriolis forces are included but centrifugal distortion, being second order in the angular velocity, is neglected. The model is accordingly spherically symmetric. However, we do not make the traditional approximation and thus all components
of the Coriolis force are taken into account. Nonlinear effects due to wave breaking near the stellar centre are likely to be significant  only for secondary masses 
\textcolor{red}{$>3$~Jupiter masses}, which exceeds that adopted here,  and  very short orbital periods \citep{Barker2010}. However, other nonlinear effects may occur close to the centre of a resonance
( see Section \ref{l'3res} below).

As noted above the linear response problem is separable in $\phi$ such that the tidal perturbations are readily represented as a linear combination of responses with $\phi$ dependence through a factor,  $\exp({\rm i}m\phi)$, where  \textcolor{red}{$m$} is the azimuthal mode number.    
We shall consider the stellar response  to the perturbing potential given by  (\ref{potpert}), taking the associated 
 Lagrangian displacement to be $\mbox{{\boldmath$\xi$}}_{m,\sigma}\exp({\rm i}\sigma t).$
%\begin{equation}
%	\mbox{{\boldmath$\xi$}} =\sum_{n=-2}^{n=2} \mbox{{\boldmath$\xi$}}_{n},
	% \quad U=r^2\sum_n A_n Y_{2,n},
%	\label{e1X}
%\end{equation}  
%\section {Equations governing the stellar response}\label{Respeq}

It is convenient to work in a frame corotating with the unperturbed primary.
In this frame, \textcolor{blue}{where}  the forcing frequency  is no longer $\sigma$ but 
 the Doppler shifted forcing frequency 
\begin{equation}
	\omega_f =\sigma+m\Omega_s,
	\label{omf}
\end{equation}
we can write the linearised equation of motion \textcolor{blue}{in the form}
\begin{equation}
	-\omega^2_f \mbox{{\boldmath$\xi$}}_{m,\sigma} +2{\rm i} \omega_f\Omega_s {\bf \hat{k}}\times\mbox {\boldmath$\xi$}_{m,\sigma}
	= -\frac{1}{\rho}\nabla P'_{m,\sigma}  +\frac{\rho'_{m,\sigma}}{\rho^2}\nabla P + \frac{1}{\rho} \nabla \cdot  \mbox{\boldmath{ $\mathsf{\Sigma}$}}_{m,\sigma}-\nabla U_{m,\sigma}.
	\label{lequmot}
\end{equation}
\textcolor{blue}{This}  includes the contribution from the force per unit mass due to viscosity through the
 Navier-Stokes term  $(1/\rho) \nabla \cdot\mbox{\boldmath{$\mathsf{\Sigma}$}}_{m,\sigma},$ 
where $\mbox{\boldmath{$\mathsf{\Sigma}$}}_{m,\sigma}$
%.THE \bsf command  and $\bsf{\alpha}$
is the  spatial contribution of the viscous stress tensor for compressible flow which is used to model the action of turbulent viscosity in the convective layers of the primary star ( see Section \ref{viscstress}). We neglect the small contribution of  perturbations to the  gravitational potential due to the  the primary induced by the tidal perturbation
(The Cowling approximation).  

The linearised continuity equations is given by
\begin{equation}
	\rho' _{m,\sigma}= -\nabla\cdot ( \rho \mbox {\boldmath$\xi$}_{m,\sigma}). \label{conte}
\end{equation}

\noindent In addition we have the linearised 
energy equation in the form
\begin{equation}
	P'_{m,\sigma}+\Gamma_1 P \nabla\cdot  \mbox {\boldmath$\xi$}_{m,\sigma} +  \mbox {\boldmath$\xi$}_{m,\sigma}\cdot\nabla P 
	= -\frac{(\Gamma_3-1)\nabla{\pmb{ \cdot \cal{F}}'_{m,\sigma}}}{{\rm i}\omega_f}.\label{Nonad}
\end{equation}
Here $\Gamma_1$ and $\Gamma_3$ are the standard adiabatic exponents and ${\pmb {\cal F}}'_{m,\sigma} \exp({\rm i}\sigma t)$ is the
perturbation to the
\textcolor{blue} {energy flux, ${\pmb {\cal F}}$, which may contain contributions from both radiative and convective transport.
 However, we remark that the effect of the  perturbed flux on the tidal dissipation is found to be  very small compared to that arising from turbulent viscosity 
 (see below). Furthermore
 the perturbed convective flux is expected to  become significant  only in a
  low mass non adiabatic region near  the surface which is not expected to be important for the $r$ modes
  we focus on. As there is no rigorous theoretical approach we adopt the simplifying assumption of neglecting this \citep[i.e. frozen convection as in ][]{Bunting2019}.} 
Perturbations of the energy generation rate are neglected. 
Being second order in the perturbations there is no contribution from viscous dissipation.

The  radiative flux perturbation ${\pmb {\cal F}^\prime}_{rad, m,\sigma}$ is calculated by making use of  the $r$, $\theta$ and $\phi$ derivatives of the temperature perturbation   $T^\prime_{m,\sigma}(r,\theta) \exp{(\rm{i}\sigma t)}$ in conjunction with the radiative diffusion approximation. 
	This involves the local opacity derivatives taken from the  output of  the MESA code.  Linearisation of the  radial component of the diffusion equation 
	leads to the equation	
%	$ are \textcolor{blue}{respectively}  added as a separate equation and variable.  \textcolor{blue}{In particular we have}
\begin{equation} 
	\frac{\mathcal{F}^{\prime}_{rad, r, m,\sigma}}{\mathcal{F}}  =\left( {{{\rm d} \ln{T}}\over {{\rm d} r}}\right)^{-1}  \frac{\partial }{\partial r  }\left( \frac{T'_{m,\sigma}}{T}\right)  - \left( \kappa_T -4 \right) \left( 
	\frac{T'_{m,\sigma}}{T}\right)  - \left( \kappa_{\rho} + 1 \right) \left( \frac{\rho^{\prime}_{m,\sigma}}{\rho} \right) 
	\label{EQ6}
\end{equation}
where  $\mathcal{F}^\prime_{rad, r,m,\sigma}$ is the radial component of the radiative flux 
 perturbation and $\kappa_{\rho}$ and $\kappa_T$ are the logarithmic derivatives of the opacity with respect to density and temperature. 
Finally, we use the linearisation of the equation of state in the  form
\begin{equation}
	\frac{P^\prime_{m,\sigma}}{P}= \chi_\rho \, \frac{\rho^\prime_{m,\sigma}}{\rho} + \chi_{T} \,  \frac{T^\prime_{m,\sigma}}{T} -\chi_\mu \drv{\mu}{r} \,\xi_{r,m,\sigma}
	\label{EOS}
\end{equation}
with $\chi_T=(T/P) {dP}/{dT}$, \,\, $\chi_\rho=(\rho/ P) dP/d\rho$ and  $\chi_\mu={dP}/{d\mu}$, where $\mu$ is here the  the local  mean molecular weight of the stellar material.

\subsection{Boundary Conditions}\label{Bdrycons}
At the stellar centre we set both $\xi_{r,m,\sigma}$ and ${\cal{F}^\prime}_{r,m,\sigma}$ equal to zero while at the stellar surface we apply the Stefan-Boltzmann  law and assume the  pressure drops sufficiently rapidly to zero at the moving  surface  \textcolor{blue}{where the}  optical depth $\sim 2/3$. \textcolor{blue}{Thus we have}
\begin{eqnarray}
	\frac{\delta {\cal{F}}_{r,m,\sigma}}{{\cal{F}}_r} = 4 \, \frac{\delta T_{m,\sigma}}{T}  \, ;  &&{\rm and}  \,  \hspace{5mm}   \frac{\delta P_{m,\sigma}}{P} = 0,
\end{eqnarray}
where $\delta$ denotes the Lagrangian  perturbation.

\noindent At the stellar rotation axis we use the \textcolor{blue}{corresponding symmetry}  of the response to the  (anti)symmetry in $\theta$ of the tidal forcing to extrapolate the value of each perturbation to $\theta=0$, while at the stellar equator  for all perturbations we \textcolor{blue}{similarly} apply the expected (anti)symmetry of (odd) even responses as implied by the symmetry of the forcing potential.

\subsection{Numerical solution procedure}\label{Numproc}
For the numerical calculation of the tidal response we adopt equation (\ref{potpert}) for the tidal potential.
\textcolor{red}{The  three level (in both radial and theta direction) difference form of the above set of partial differential equations (\ref{lequmot}-\ref{EQ6}) applied here is based on the non-equidistant radial  grid with about 2400 mesh points constructed with the MESA  \citep{Paxton2015}  stellar evolution code (version 12778). We let the MESA code define the radial coordinates of cell boundaries and  of the intermediate cell centres where the unperturbed thermodynamic variables are defined.  The perturbed thermodynamic variables are also defined at cell centres, while the three components expressing the spatial dependence of the displacement vector $\mbox{{\boldmath$\xi$}}_{m,\sigma}$ together with the perturbed stellar radiative energy flux ${\pmb {\cal F}^\prime}_{rad,m,\sigma}$ are defined at cell boundaries.
	To evaluate the required perturbations of the  thermodynamic variables at cell boundaries  linear interpolation in $r$ was employed.
	We adopt a $\theta$-grid (where all perturbed variables are defined ) covering the domain  $[ 0, \pi/4]$ that is  equidistant in $\sin{\theta}$ together with another covering the domain  $[\pi/4, \pi/2]$ that is equidistant in $\cos{\theta}$.  Thus  $\theta$  derivatives
 in the difference  equations are dealt with by,  respectively,  expressing them in terms of  ${d}/{d\sin \theta}$, and ${d}/{d\cos \theta}$.  
 The solution in the lower hemisphere,  $[\pi/2, \pi]$,  follows from the  (anti)symmetry of the  forcing potential. With 128 grid points in $\theta$ the resolution is usually adequate  near both the rotation axis  and  the equator. }

%\textcolor{red}{***COMMENT**I don't think we need to repeat the stuff about separation of variables as it is already mentioned above***}

%The unperturbed stellar model being axi-symmetric, we solve the set of difference equations for a given forcing frequency $\sigma$ and  azimuthal index $n$ by first separating the $\phi$ and time part of the perturbations. Thus we may express the displacement vector ${\mbox{\boldmath$\xi$}}$ (and similarly the Eulerian perturbed quantities $\Phi^\prime$, $\rho^\prime$, $T^\prime$, $P^\prime$ and $\mathcal{F}^\prime_r$) in the  frame corotating with the primary in complex form  as
%\begin{equation}
%	\hat{{\mbox{\boldmath$\xi$}}}  (r,\theta,\phi,t)\,=\,  {\mbox{\boldmath$\xi$}}(r,\theta)\,\, \exp{\left(\rm{i}(\omega_f t+ n \, \phi)\right) }
%\end{equation}	
Note that the
% separation of the $\phi$ and time factor of all perturbations renders the solution of the difference equations a 2-dimensional problem. The
  quantities such as  ${\mbox{\boldmath$\xi$}}_{m,\sigma}$ attain complex values through the non-adiabatic $\nabla \cdot  {\cal{F}}^\prime_{m, \sigma}$ term in the energy equation (\ref{Nonad}) and by the effect of the viscous terms in the equation of motion (\ref{lequmot}). Ultimately the physical value of e.g. the density perturbation for each forcing component \textcolor{blue}{with a specified} $m$  in the non-rotating frame  is obtained \textcolor{blue}{from}
   $\re{[\rho^\prime_{m,\sigma}(r,\theta) \, \exp{(\rm{i} (\sigma t + m\phi})]}$, \textcolor{blue}{where $\re$ denotes  the real part is to be taken,}
      with $ \sigma t= - m n_ot $.

We solve the difference form of equations (\ref{lequmot}) -(\ref{EOS}) numerically following the procedure outlined in Appendix B of \citet{SP97}.
According to this, a representation of these equations in finite difference form on a $(r,\theta)$  grid leads to the determination of the state variables following the inversion of a  large matrix using a parallelised version of the solution scheme.  
Having found the response to separate Fourier components of the perturbing potential,  we  may use linear superposition  to construct the complete response.
The tidal forces  acting to cause evolution of the orbit and the angular velocity of the primary star may then be determined.

%\textcolor{red}{*****More detail needed here???*******}

\subsection{The viscous force per unit mass and the rate of viscous dissipation}  \label{viscstress}
The viscous force and viscous dissipation induced by the turbulent convection in the envelope of the star are calculated from the viscous stress tensor for compressible flow $\bsf{\Sigma}$ given in appendix A, whereby the kinematic viscosity $\nu(r)$  is taken from \citep{Duguid2020}
\begin{equation}
	\nu(r)= 
	\frac{ {\frac{1}{3}}
	{ \mathcal{L}_{mx} v_{c} }}
	{(1+({\tau_{c}}/{P_{osc}})^{s})}
	\label{turbvisc}
	% \nonumber
\end{equation}
The convective mixing length $ \mathcal{L}_{mx}= \alpha |H_P|$ is scaled by the parameter $\alpha=1$ . The local pressure scale height $|H_P(r)|$  and the local convective velocity $v_{c}(r)$ are taken from the  MESA input stellar  model.  The possible mismatch of the timescale of the forced oscillations ($P_{osc}=2 \pi/\omega_f$) and that of the turbulent convection ($\tau_c=1/\sqrt{|N^2|}$), 
where $N^2=|(1/\rho)(dP/dr) ((1/(\Gamma_1 P)dP/dr)-(1/\rho)d\rho/dr)|,$ 
is taken into account by the term raised to the power $s=2$.\textcolor{blue}{But it should be noted that whether to  implement  such a reduction factor is a matter of some  controversy \citep[see][]{Terquem2021}. However, for most calculations we discuss here, including those of the $r$ modes we focus 
on, it does not play a significant role.}

The components of the viscous stress tensor $\bsf{\Sigma}$ in spherical polar coordinates 
expressed in terms of the components of a general associated  displacement vector $\bmth{\xi}$  are given in  appendix \ref{stresscomp}.
Making use of these, and the expressions for the
$r$, $\theta$ and $\phi$ components of $\nabla \cdot\bsf{\Sigma}$  also given in appendix \ref{stresscomp}, the components of the  viscous force per unit mass $f_{\nu,m,\sigma}$  
given by $(1/\rho) \nabla \cdot\bsf{\Sigma}_{m,\sigma}$ in the equation of motion (\ref{lequmot}) can also be found in terms of $\bmth{\xi}_{m,\sigma}.$

 %\begin{align}
%\hspace{-6cm}  \frac{1}{\rho} \nabla \cdot\bsf{\Sigma} \equiv (f_{r}, f_{\theta}, f_{\phi})\hspace{3mm}  \rm{ in\hspace{2mm}  the\hspace{2mm}  form
%\hspace{2mm}(see\hspace{2mm} equation \hspace{2mm}(\ref{lequmot}))}
%\nonumber
 %\end{align}
%\begin{align}
%&\hspace{-3mm} f_{r} =\rm{i}\omega_f  \left[ \frac{\nabla \cdot (\rho \nu \nabla \xi_{r})}{\rho} -\frac{2 \nu}{r^2 \, \sin{\theta}} \pdrv{(\sin{\theta} \xi_{\theta})} {\theta} - \frac{2 \nu}{r^2 \sin{\theta}} \pdrv{\xi_{\phi}}{\phi} -\frac{2 \nu \, \xi_{r}}	{r^2}  +\frac{1}{\rho} \pdrv{}{r} (\frac{\rho \nu} {3} \nabla \cdot \mbox{{\boldmath$\xi$}} ) \right] \label{viscf1}
%\\
%	& \hspace{-3mm}f_{\theta} =\rm{i} \omega_f \left[\frac{\nabla \cdot (\rho \nu \nabla \xi_{\theta})}{\rho} -\frac{2 \nu \cos{\theta}}{r^2 \, \sin^2{\theta}} \pdrv{\xi_{\phi}}  {\phi} + \frac{2\nu}{r^2 } \pdrv{\xi_{r}} {\theta} -\frac{\nu\xi_{\theta}}{r^2 sin^2{\theta}}  +\frac{1}{\rho r} \pdrv{}{\theta} (\frac{\rho \nu} {3} \nabla \cdot \mbox{{\boldmath$\xi$}} ) \right]  \label{viscf2}
%\\
%&\hspace{-3mm} f_{\phi} =	\rm{i}\omega_f  \left[ \frac{\nabla \cdot (\rho \nu \nabla \xi_{\phi})}{\rho} +\frac{2\nu}{r^2 \sin{\theta}}  \pdrv{\xi_{r} } {\phi} + \frac{2 \nu \cos{\theta}}{r^2 \sin^2{\theta}} \pdrv{\xi_{\phi} }{\phi} -\frac{\nu\xi_{\phi}}{r^2\sin^2{\theta}}   +\frac{1}{\rho \, r \sin{\theta}} \frac{\partial}{\partial \phi} (\frac{\rho \nu} {3}  \nabla \cdot \mbox{{\boldmath$\xi$}} )\right].  \label{viscf3}
%\end{align}
We remark that where this procedure requires the  calculation of the derivatives of $\nabla \cdot \mbox{{\boldmath$\xi$}}_{m,\sigma}$ 
% in equations (\ref{viscf1}-\ref{viscf3}) 
we find it helpful to replace
 $\nabla \cdot \mbox{{\boldmath$\xi$}}_{m,\sigma}$
  by using the equation of continuity (\ref{conte}) in the form 
\begin{equation}
	\nabla \cdot \mbox{{\boldmath$\xi$}}_{m,\sigma}= -\left(\frac{\rho^\prime_{m,\sigma}}{\rho} + \drv{\ln{\rho}} {r} \, \xi_{r,m,\sigma}\right)	
\end{equation}	
The viscous dissipation rate can be expressed in terms of the viscous stress tensor by  the negative definite expression
\begin{equation}
	\drv{E_{kin}}{t} = - \int_V \left( \frac{\frac{1}{2}\bsf{ \Sigma^*}_{r r} \bsf{\Sigma}_{r r} + \frac{1}{2} \bsf{\Sigma}^*_{\theta \theta} \bsf{\Sigma}_{\theta \theta} + \frac{1}{2} \bsf{\Sigma}^*_{\phi \phi} \bsf{\Sigma}_{\phi \phi} + \bsf{\Sigma}^*_{r \theta} \bsf{\Sigma}_{r \theta} +\bsf{\Sigma}^*_{\theta \phi} \bsf{\Sigma}_{\theta \phi} + \bsf{\Sigma}^*_{\phi r} \bsf{\Sigma}_{\phi r}} {\rho \nu} \right) d\tau \label{vdissip}
\end{equation}
where  the integral is taken over the volume, $V,$  of the primary star. 

\subsection{Expression for the imaginary part of the overlap integral}\label{Imagovlap}
The quantity determining the induced secular orbital evolution resulting from 
a component of the forcing potential that is proportional to a spherical harmonic is the 
imaginary part of the overlap integral, $\im(Q_{m,\sigma}),$ specified by equation (\ref{overlap}).
We can find an expression for this by making use of equations (\ref{lequmot}), (\ref{conte})  and (\ref{Nonad}).
\textcolor{red} {We start by multiplying equation (\ref{lequmot}) by  $\rho\mbox{{\boldmath$\xi$}}_{m,\sigma}^{*}$ and integrate over the volume of the star after making use of the boundary conditions  to eliminate surface terms. After integrating by parts we obtain}
\begin{align}
&\int_V\rho\mbox{{\boldmath$\xi$}}_{m,\sigma}^{*}\left(-\omega^2_f \mbox{{\boldmath$\xi$}}_{m,\sigma} 
+2{\rm i} \omega_f\Omega_s {\bf \hat{k}}\times\mbox {\boldmath$\xi$}_{m,\sigma}\right)d\tau+c_{tid}\, Q_{m,\sigma}^*
	=\nonumber\\
&	\int_V\left(\rho\mbox{{\boldmath$\xi$}}_{m,\sigma}^{*}\cdot{\bf f}_{\nu, m,\sigma}+\nabla\cdot \mbox{{\boldmath$\xi$}}_{m,\sigma}^{*} \left(P'_{m,\sigma}
	   +  \mbox{{\boldmath$\xi$}}_{m,\sigma}\cdot\nabla P\right)\right)d\tau
	%+\nabla\psi' _{n,k}
	\nonumber\\
&% \nabla\cdot \mbox{{\boldmath$\xi$}}_{n,\sigma}^{*} \left(P'_{n,\sigma}  +  \mbox{{\boldmath$\xi$}}_{n,\sigma}\cdot\nabla P\right)
-\int_V\left( (\nabla\cdot \mbox{{\boldmath$\xi$}}_{m,\sigma}^{*} ) \mbox{{\boldmath$\xi$}}_{m,\sigma}\cdot\nabla P
+ (\nabla\cdot \mbox{{\boldmath$\xi$}}_{m,\sigma} ) \mbox{{\boldmath$\xi$}}_{m,\sigma}^*\cdot\nabla P
	+\mbox{{\boldmath$\xi$}}_{m,\sigma}\cdot\frac{\nabla \rho}{\rho}\mbox{{\boldmath$\xi$}}_{m,\sigma}^{*}\cdot\nabla P 
	%+ \rho'^*_{n,\sigma} \Phi^\prime_{n,\sigma}
	 \right)d\tau
	\label{lequmot1}
\end{align}
Taking the imaginary part of the above expression, we note that the integral on the left hand side and the second integral on the right hand side are purely real and so do not contribute. We thus obtain
\begin{align}
&c_{tid} \,\im(Q_{m,\sigma}^*)=-c_{tid}\,\im(Q_{m,\sigma})=
%	=\nonumber\\&
Im \left(	\int_V\left(\rho\mbox{{\boldmath$\xi$}}_{m,\sigma}^{*}\cdot{\bf f}_{\nu, m,\sigma}+\nabla\cdot \mbox{{\boldmath$\xi$}}_{m,\sigma}^{*} \left(P'_{m,\sigma}
	   +  \mbox{{\boldmath$\xi$}}_{m,\sigma}\cdot\nabla P\right)\right)d\tau\right)
	%+\nabla\psi' _{n,k}
%	\nonumber\\&
% \nabla\cdot \mbox{{\boldmath$\xi$}}_{n,\sigma}^{*} \left(P'_{n,\sigma}  +  \mbox{{\boldmath$\xi$}}_{n,\sigma}\cdot\nabla P\right)
%-\int_V\left( \nabla\cdot \mbox{{\boldmath$\xi$}}_{n,\sigma}^{*}  \mbox{{\boldmath$\xi$}}_{n,\sigma}\cdot\nabla P
%+ \nabla\cdot \mbox{{\boldmath$\xi$}}_{n,\sigma}  \mbox{{\boldmath$\xi$}}_{n,\sigma}^*\cdot\nabla P
	%+\mbox{{\boldmath$\xi$}}_{n,\sigma}\cdot\frac{\nabla \rho}{\rho}\mbox{{\boldmath$\xi$}}_{n,\sigma}^{*}\cdot\nabla P + \rho'^*_{n,\sigma} \Phi^\prime_{n,\sigma} \right)d\tau
	\label{lequmot2}
	\end{align}
%	\textcolor{red}{***COMMENT***The self gravity term is easily shown not to contribute and it can be removed later as now not there******}

	\subsubsection{Interpretation of the terms in equation (\ref{lequmot2})} 
	\noindent \textcolor{red}{Assuming that no viscous stresses act at the boundaries the first term in}
	\textcolor{blue}{the integral on the right hand side of} \textcolor{red}{
	 (\ref{lequmot2}) can be seen to be  the product of $\omega_f^{-1}$ and  twice  the mean rate of  kinetic energy change arising from viscosity that results from forcing due to 
	the {\it real part}  of the potential (\ref{potpert}).} This is expected to be negative definite 
	corresponding to a positive definite rate of  increase of thermal energy. 
	 The second term similarly corresponds to the product of $\omega_f^{-1}$
	and twice the rate of doing $PdV$ work. From (\ref{Nonad}) we have
\begin{align}
	P'_{m,\sigma}+\Gamma_1 P \nabla\cdot  \mbox {\boldmath$\xi$}_{m,\sigma} 
	= -  \mbox {\boldmath$\xi$}_{m,\sigma}\cdot\nabla P -\frac{(\Gamma_3-1)\nabla{\pmb{ \cdot \cal{F}}'_{m,\sigma}}}{{\rm i}\omega_f}.\label{Nonad1}
\end{align}
Using this to eliminate, $P'_{m,\sigma},$ in (\ref{lequmot2}) we obtain
\begin{align}
&-c_{tid}\, \im(Q_{m,\sigma})= 
%	=\nonumber\\&
\im \left(	\int_V\left(\rho\mbox{{\boldmath$\xi$}}_{m,\sigma}^{*}\cdot{\bf f}_{\nu, m,\sigma}
%\left(P'_{n,\sigma +  \mbox{{\boldmath$\xi$}}_{n,\sigma}\cdot\nabla P\right)
-\frac{(\Gamma_3-1)\nabla\cdot \mbox{{\boldmath$\xi$}}_{m,\sigma}^{*} \nabla{\pmb{ \cdot \cal{F}}'_{m,\sigma}}}{{\rm i}\omega_f}\right)d\tau\right)
	%+\nabla\psi' _{n,k}
%	\nonumber\\&
% \nabla\cdot \mbox{{\boldmath$\xi$}}_{n,\sigma}^{*} \left(P'_{n,\sigma}  +  \mbox{{\boldmath$\xi$}}_{n,\sigma}\cdot\nabla P\right)
%-\int_V\left( \nabla\cdot \mbox{{\boldmath$\xi$}}_{n,\sigma}^{*}  \mbox{{\boldmath$\xi$}}_{n,\sigma}\cdot\nabla P
%+ \nabla\cdot \mbox{{\boldmath$\xi$}}_{n,\sigma}  \mbox{{\boldmath$\xi$}}_{n,\sigma}^*\cdot\nabla P
	%+\mbox{{\boldmath$\xi$}}_{n,\sigma}\cdot\frac{\nabla \rho}{\rho}\mbox{{\boldmath$\xi$}}_{n,\sigma}^{*}\cdot\nabla P + \rho'^*_{n,\sigma} \Phi^\prime_{n,\sigma} \right)d\tau
	\label{lequmot3}
	\end{align}
In this form the role of heat transport can be clearly seen. \textcolor{red}{
The rate of radiative damping due to non-adiabatic effects in the primary follows as
\begin{equation}
	{ \cal D}_r = \im \left(	\int_V	\rm{i} (\Gamma_3-1)\nabla\cdot \mbox{{\boldmath$\xi$}}_{m,\sigma}^{*} \nabla{\pmb{ \cdot \cal{F}}'_{m,\sigma}} d\tau \right) \label{Disp_r}
\end{equation}  
We also note in passing that
\begin{align}
& \im \left(	\int_V\left(\rho\mbox{{\boldmath$\xi$}}_{m,\sigma}^{*}\cdot{\bf f}_{\nu, m,\sigma}d\tau\right)\right)
=  \frac{1}{\omega_f}{\cal D}_v
%\left(P'_{n,\sigma +  \mbox{{\boldmath$\xi$}}_{n,\sigma}\cdot\nabla P\right)
\end{align} 
where   ${\cal D}_v$ is twice the  rate of kinetic energy change due to viscosity in the primary given by  (\ref{vdissip}) but with  $\bsf{\Sigma}$ replaced by $\bsf{\Sigma}_{m,\sigma}.$ }

	In summary we can write for the total rate of change
	of the kinetic energy  that results from forcing due to 
	the {\it real part}  of the potential (\ref{potpert}) as
	\begin{align}
	\frac {dE_{kin}}{dt}= -\frac{c_{tid}\,\omega_f}{2} \im(Q_{m,\sigma})
	\end{align}
		 We remark that  exactly the same expression as (\ref{lequmot2})  occurs when calculating
	the damping or excitation rate of a normal mode. In that case the growth rate follows from
	equation (\ref{lequmot}) as being given by
	\begin{align}
	\gamma =\frac{ {\cal H}}{{\cal N}} \hspace{3mm} {\rm where}
         \end {align}

\begin{align}
{\cal{H}}=
%& Im(Q_{m,\sigma})= -
%	=\nonumber\\&
\im \left(	\int_V\left(\rho\mbox{{\boldmath$\xi$}}_{m,\sigma}^{*}\cdot{\bf f}_{\nu, m,\sigma}
%\left(P'_{n,\sigma +  \mbox{{\boldmath$\xi$}}_{n,\sigma}\cdot\nabla P\right)
-\frac{(\Gamma_3-1)\nabla\cdot \mbox{{\boldmath$\xi$}}_{m,\sigma}^{*} \nabla{\pmb{ \cdot \cal{F}}'_{m,\sigma}}}{{\rm i}\omega_f}\right)d\tau\right)\hspace{3mm}{ and}
	%+\nabla\psi' _{n,k}
%	\nonumber\\&
% \nabla\cdot \mbox{{\boldmath$\xi$}}_{n,\sigma}^{*} \left(P'_{n,\sigma}  +  \mbox{{\boldmath$\xi$}}_{n,\sigma}\cdot\nabla P\right)
%-\int_V\left( \nabla\cdot \mbox{{\boldmath$\xi$}}_{n,\sigma}^{*}  \mbox{{\boldmath$\xi$}}_{n,\sigma}\cdot\nabla P
%+ \nabla\cdot \mbox{{\boldmath$\xi$}}_{n,\sigma}  \mbox{{\boldmath$\xi$}}_{n,\sigma}^*\cdot\nabla P
	%+\mbox{{\boldmath$\xi$}}_{n,\sigma}\cdot\frac{\nabla \rho}{\rho}\mbox{{\boldmath$\xi$}}_{n,\sigma}^{*}\cdot\nabla P + \rho'^*_{n,\sigma} \Phi^\prime_{n,\sigma} \right)d\tau
	\label{shoo1}
	\end{align}
		
		 \begin{align}
&{\cal N}= \int_V\rho\mbox{{\boldmath$\xi$}}_{m,\sigma}^{*}\left(2\omega_f \mbox{{\boldmath$\xi$}}_{m,\sigma} 
-2{\rm i}\Omega_s {\bf \hat{k}}\times\mbox {\boldmath$\xi$}_{m,\sigma}\right)d\tau.\label{shoo}
%+Q_{m,\sigma}^*
	\end{align}
Here we note that at resonance $\bmth{\xi}_{m,\sigma}$ will correspond to the displacement vector for the normal mode.

	 This means that if the tidal response is dominated by a normal mode as at a resonance
	the sign of, $c_{tid}\, \im(Q_{m,\sigma})\equiv {\cal H},$ leads to the conventionally expected direction of  tidal evolution only when the mode is damped
	\footnote{As it is easy to show that for a stable model, $\omega_f{\cal N}$ is non negative, it follows that when $\omega_f{\cal H}<0,$ corresponding to kinetic energy dissipation, the. growth rate $\gamma <0.$
	corresponding to mode damping.  }  
	For an unstable normal mode this would be
	 reversed.

\section{Results}\label{Results}	 
\textcolor{red}{We adopt the following system parameters for the numerical calculations: we assume a circular orbit whereby  the mass of the perturbing planet is $M_p=9.543 \times 10^{-4} M_\odot$ (Jupiter's mass) and the primary star has mass $1.0 M_\odot$ with radius $R_*=0.951 R_\odot$ and  spin rate $\Omega_s =10^{-2}\Omega_c$. The corresponding spin period of the star is $P_s=10.746$ d. The radius of the radiative core $R_c=0.731 \, R_*$. Note that for the rest of this section all frequencies are normalised by the star's critical rotation rate $\Omega_c = 6.7674 \times 10^{-4}$ $ s^{-1}$. }

\subsection{Finding r mode resonances}\label{Findr}
We consider the r modes that can be excited in the radiative core of the star by the dominant tidal component with $l$=2, $|m|$=2. When the forcing frequency in the rotating frame $\omega_f$ approaches a resonant condition the amplitudes of the tidal perturbations in the core become large with a corresponding increase of the tidal amplitude in the convective envelope with an enhanced viscous dissipation there. This effect can cause a speed up of the tidal evolution of the system.

\textcolor{red}{For a given azimuthal mode number, $m,$
	the resonant forcing frequencies seen in a frame  co-rotating with the star  are close to \citep[see][and appendix B]{PP78} 
		\begin{equation}
			\omega_f = \sigma +m \Omega_s = \frac{ 2 \, m \, \Omega_s}{l'\, (l'+1)}
			\label{rmodef}
		\end{equation}
%        \begin{align}
%       \hspace{-2mm} \omega_f = \sigma +m\Omega_s = 2 m\Omega_s/(l'('l+1))  \hspace{3mm}{\rm with}\hspace{1mm}{\rm pattern }  $\omega_p =- \omega_f/m = -2\Omega_s/(l'(l'+1))$.
	The pattern speed of the $r$ modes is thus $\omega_p =- \omega_f/m = -2\Omega_s/(l'(l'+1))$.}
%       \hspace{1mm} {\rm speed} \hspace{2mm} \omega_p =- \omega_f/m = -2\Omega_s/(l'(l'+1))
%        \end{align}
For a forcing potential with $l=2, |m|=2$, as considered here,  the values of, $l' \ge |m|,$ to be used in (\ref{rmodef}) 	are the odd values  $l'=3,5,...$.  However, we note that  if forcing with $ l=2, |m|=1,$  as 	occurs for misaligned orbits is considered we also obtain the forcing frequency with $l'=1$ in (\ref{rmodef}). 
	In particular we note that when $m= -1,l'=1,$ $\omega_p= -\Omega_s$ corresponding to a rigid tilt mode. 
	
	The reason the resonances are close to the frequencies given by (\ref{rmodef}) is that
	the latter  correspond to  the actual resonances in the limit $\Omega_s\rightarrow 0$
	( see appendix \ref{rmodeapp}).  For small but finite $\Omega_s$ these resonances split into a discrete
	set of close frequencies corresponding to modes of increasing radial order.
	An important aspect of the $r$ mode resonances is that the associated pattern speed,
	$\omega_p,$ which is the difference between the orbital and rotation frequencies
	 is negative corresponding to retrograde forcing as seen in the rotating frame.
	This occurs when the \textcolor{red}{ the secondary rotates more slowly in its orbit} than the primary
	and so the tidal interaction will cause angular momentum to be transferred from the primary's
	spin to the orbit. We remark that for the aligned case with, $|m| \ge 2,$ the locations of the resonances are such that
	$0< -\omega_p = \Omega_s - n_0 \le \Omega_s/6$.

In order to find the r mode resonances we start with $\omega_f$ near the r-mode limit given by equation (\ref{rmodef}). For general $l$ we expect primarily  $l'=l-1$ for odd forcing symmetry at the stellar equator  and $l'=l+1$ for even forcing symmetry
\footnote{But note that larger values of $l'$ including $l'=5$ occur for forcing with $l=|m|=2$ at higher order (see appendix B)}. To search for a resonance we apply a numerical algorithm to search the forcing frequency $\omega_f$ for which the kinetic energy of the tidal oscillations is maximal, while keeping the 
% orbital separation 
\textcolor{ROUGE}{semi-major axis} fixed at an arbitrary value $\tilde{a}.$ \textcolor{blue}{Without loss of generality we adopt  $m=-2.$}
 %(for the numerical calculation we replace the  orbital separation $R$ by the parameter $\tilde{ a}$ ). 
 The actual tidal response at resonance is then found by calculating the orbital angular speed $n_o=\Omega_s-\omega_f/m$ corresponding to the resonance,  where  $\omega_f= \omega_0$, the resonance frequency in the corotating frame (note that for $m<0,$ $\omega_0 <0$). With the value of the corresponding %orbital separation 
 \textcolor{ROUGE}{semi-major axis}
 $a$  we can find the actual tidal response  $\mbox{{\boldmath$\xi$}}_{m,\sigma}$ and  $\rho^\prime_{m,\sigma}$ etc. by scaling by a factor $(\tilde{a}/a)^3$, while the viscous dissipation rate, being second order, scales as $(\tilde{a}/a)^6$.
After the primary r mode resonance is found we shift the new starting value of $\omega_f$ to search for the next r mode resonance, etc. We define the number of radial nodes $n_r$ of the resonant r mode  by the number of radial nodes seen in $\rm{Re} (\xi_\phi)$ in the radiative core.

\subsection{ \textcolor{red}{$l^\prime$ = 3 and  $l^\prime$ = 5  r mode resonances}} \label{l'3res}
\begin{figure}
	\includegraphics[width=\columnwidth, height=9cm]{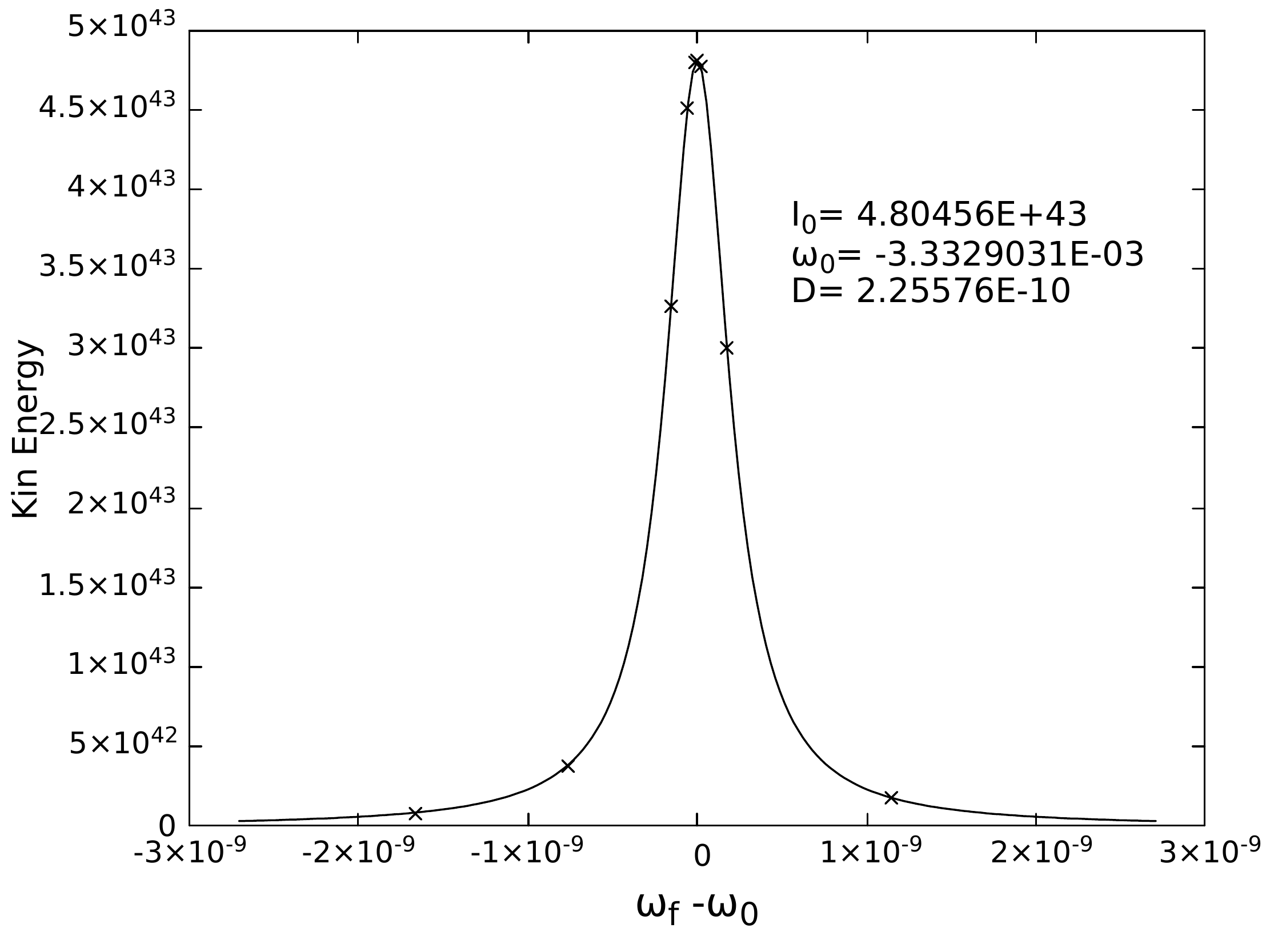}
	%\caption{Resonance curve \textcolor{red}{showing} the Kinetic Energy (erg) \textcolor{red}{associated with}  the $l'$=3 r mode with $n$=0 fitted to the calculated points.  \label{Tkin-0508L1}}
	\caption{This illustrates the resonance curve associated with the Kinetic Energy (erg) for the the $l'$=3 r mode with $n_r$=0 fitted to the calculated points.  The determined parameters $l_o, D,$ and $\omega_0,$ (the latter two in units of $\Omega_c$)  are indicated.} \label{Tkin-0508L1}
\end{figure}

\begin{figure}
\hspace{0cm}\includegraphics[width =\columnwidth, height=9cm]{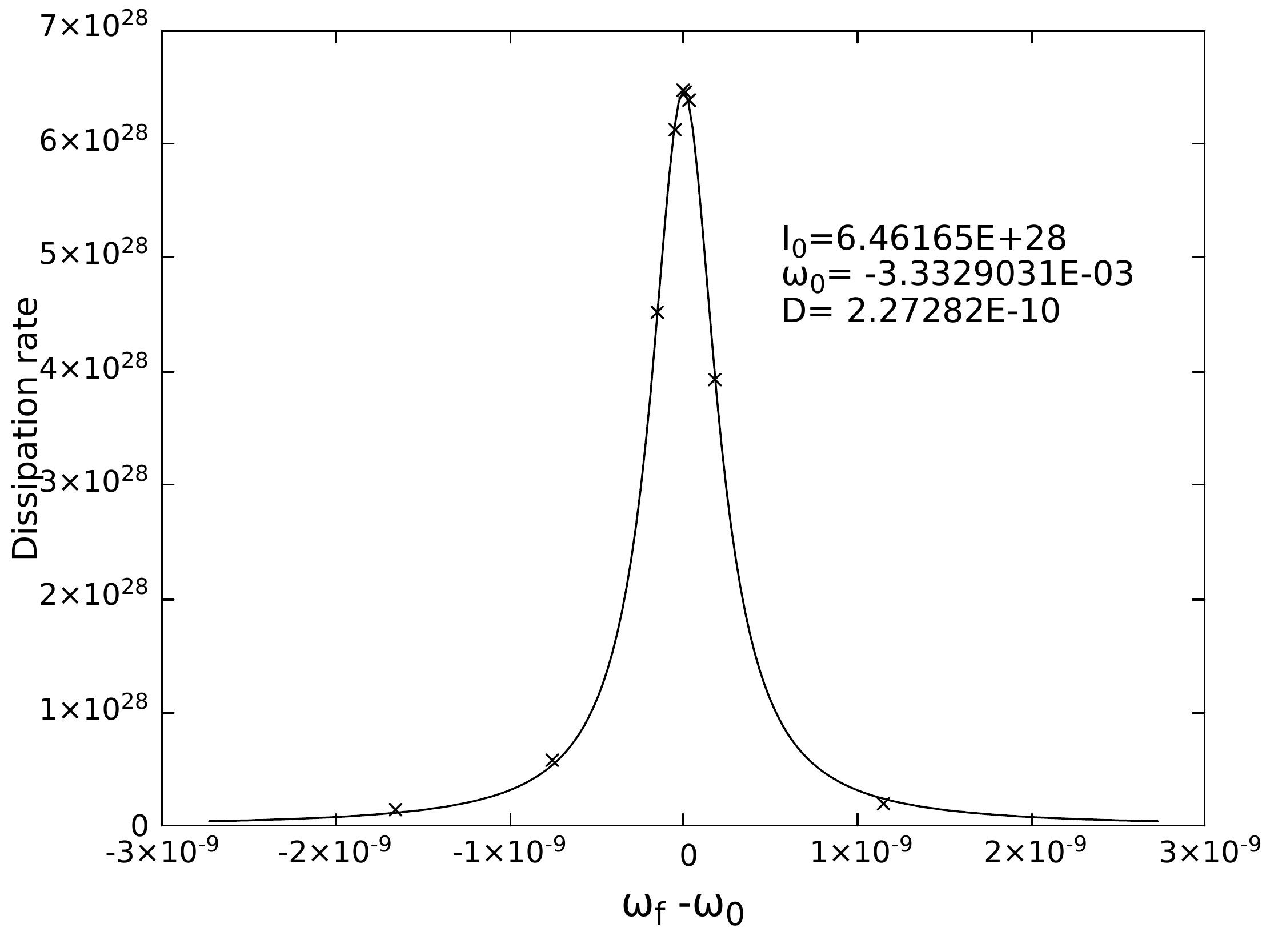}
	\includegraphics[width=\columnwidth, height=9cm]{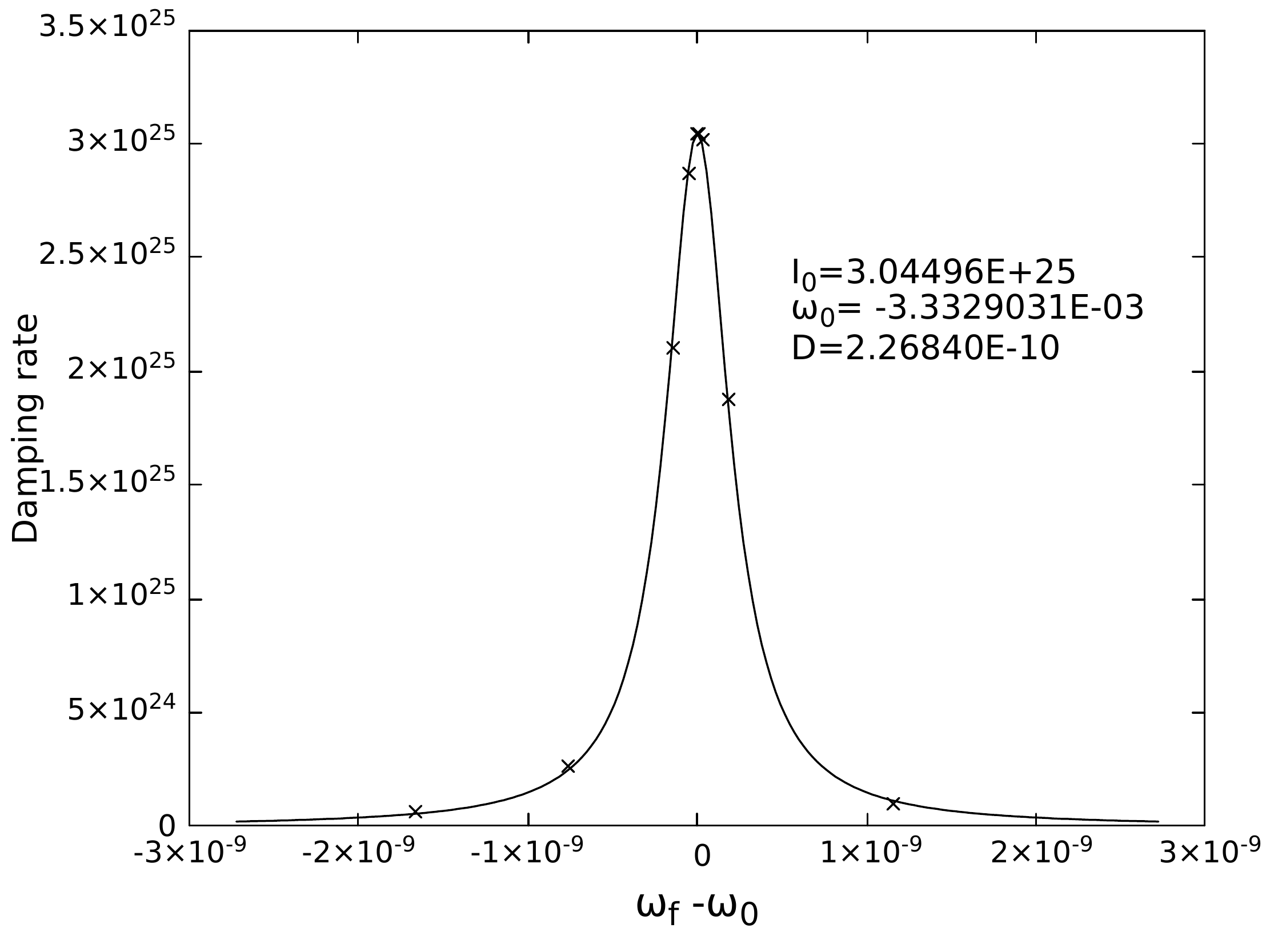}
	\caption{The 
		%	left panel \textcolor{red}{illustrates the} resonance curve \textcolor{red}{showing} the Kinetic Energy (in cgs units) \textcolor{red}{associated with}  the $l'$=3 r mode with $n$=0 fitted to the calculated points. 
		\textcolor{ROUGE}{upper} panel shows the  resonance curve associated with the viscous dissipation rate (erg/s) 
		%of the $l'$=3 r mode with $n$=0 %
		produced in the convective envelope fitted to the data points (crosses) for the mode illustrated in Fig. \ref{Tkin-0508L1}.
		The \textcolor{ROUGE}{lower}  panel shows the corresponding resonance curve obtained for the radiative \textcolor{red}{damping} rate (erg/s) (see text).}  \label{vDisp-0508L1}
\end{figure}

In the left hand panel of Figure \ref{Tkin-0508L1} the crosses show the calculated kinetic energy of the fundamental $l'=3$ r mode as a function of $\omega_f-\omega_0$ during an iterative search  for resonance (centred at a kinetic energy maximum) occurring at the resonant frequency, 
 $\omega_0,$ in the corotating frame. The calculated data points are fitted by a resonance curve \textcolor{blue}{ specified by equation (\ref{ResCurve}) below,} using a least squares method that gives the best values for  the defining parameters $\omega_0$ and  $D$ for the given maximum $I_0$ at resonance. This is given by 
\begin{equation}
	I_{res} \left(\omega_f- \omega_0 \right) =\frac{I_0} {1 + \left(\frac{\omega_f-\omega_0}{D}\right)^2} .\label{ResCurve}
\end{equation}
We remark that $D$ can be interpreted as a damping rate associated with the resonance curve.
Note too that the parameters determined from the fits may vary slightly according to which resonantly excited quantity is considered.
However, this resonance is very narrow, as are all others we find, occupying a frequency range, $\sim 10^{-5}\omega_0 \sim 10^{-8}\Omega_c.$  This is a consequence of the mode being global
and centred in the radiative core. 
\textcolor{red}{Figure  \ref{vDisp-0508L1} %and \ref{rDisp-0508L1}
 shows the resonance curves for the viscous dissipation rate and radiative damping rate for the fundamental r mode resonance  illustrated in Fig. \ref{Tkin-0508L1}.
In all cases the radiative damping (\ref{Disp_r}) is much smaller than the viscous dissipation and accordingly radiative damping will be ignored in the rest of the paper. }
As expected the radial displacement is characteristically less than the horizontal displacement by more than one order of magnitude. But $\xi_{\phi}/R_*$
attains a maximum of $\sim 1.5$ for the mode with $n_r=0$ at the centre of the resonance indicating significant nonlinearity. However, the corresponding velocity $\omega_f\xi_{\phi} \sim 2$ km s$^{-1}$
is very subsonic where the mode is sited. Nevertheless the associated  perturbed vorticity $\sim 2\Omega_s $ is significant.
However, this is reduced in the wings of the resonance. For example  at a separation of $7\times 10^{-9}\omega_0$ from the centre the amplitude is reduced by a factor $\sim 30$ and the energy dissipation rate by a factor $10^{3}$ ( see Fig. \ref{vDisp-0508L1}).
 \textcolor{blue}{ Even so} this may lead to significant effects on  tidal evolution
( see Sections \ref{orbevol}, and \ref{rmodeeffect} below).

\textcolor{RED}{\subsubsection{Reducing the mass of the perturbing planet}\label{reducingmp}
We remark that as our tidal response calculation is linear it can be simply scaled
to apply to different planet masses, taking the amplitude to be proportional to the
planet mass with the energy dissipation rate proportional to its square.
Thus reducing the planet mass by a factor of $30$ placing it in the mini- Neptune
regime would have the same effect on that as moving from the resonance centre to the  wings as
described above. Accordingly we shall assume that the linear results may be used
even at the centre of the resonances for this and smaller masses.}

%Table \ref{tabl3} \textcolor{red}{gives} the properties of calculated $l'=3$ r mode resonances. 
%\textcolor{red}{The quantities $D$ and $D(Disp)$ are respectively  determined from the resonance curves corresponding 
%to $E_{kin}$ and the rate of viscous dissipation in the convective envelope ( see Fig. \ref{vDisp-0508L1}).}
%The orbital separation and the kinetic energy and dissipation rates are scaled at the resonances \textcolor{red}{as indicated above to correspond to an}  orbital period $P_{orb} \simeq$ 12.895 days. 

%$\begin{array} 	{cccccc} 
%	\hline 
%	n_r  & \omega_0 \times 10^{3}  & I_0=E_{kin} (cgs) & D(E_{kin})  & I_0=-\drv{E_{kin}}{t} (cgs) &  D(Disp) \\ 
%	\hline
%	0  & -3.332903054   &  4.80456 \times  10^{43} & 2.25576 \times 10^{-10} & 6.46165 \times 10^{28} &  2.27282 \times 10^{-10} \\ 
%	1  & -3.332861397  & 2.69173 \times 10^{41} & 8.59906 \times 10^{-10} &  1.37056 \times 10^{27} & 8.23732 \times 10^{-10} \\
%	2 & -3.332806443   &  9.35504\times 10^{40} & 1.84798 \times 10^{-9} &  1.02787 \times 10^{27} & 1.85566 \times 10^{-9} \\
%	3  & -3.332738070 & 8.66023 \times 10^{39} & 3.28931 \times 10^{-9} &  1.67101 \times 10^{26} & 4.00689 \times 10^{-9} \\
%	4  & -3.332655910 &  4.42358 \times 10^{39} & 4.89002 \times 10^{-9} & 1.29819 \times 10^{26} & 5.52811 \times 10^{-9}  \\
%	5 & -3.332560283 & 1.04789\times 10^{39} &  7.50894 \times 10^{-9} & 4.61155 \times 10^{25} & 8.98549 \times 10^{-9}  \\
%	\hline
%	\label{tabl3}
%\end{array}	$

\begin{figure}
\hspace{-2.5cm}	\includegraphics[width=1.28\columnwidth]{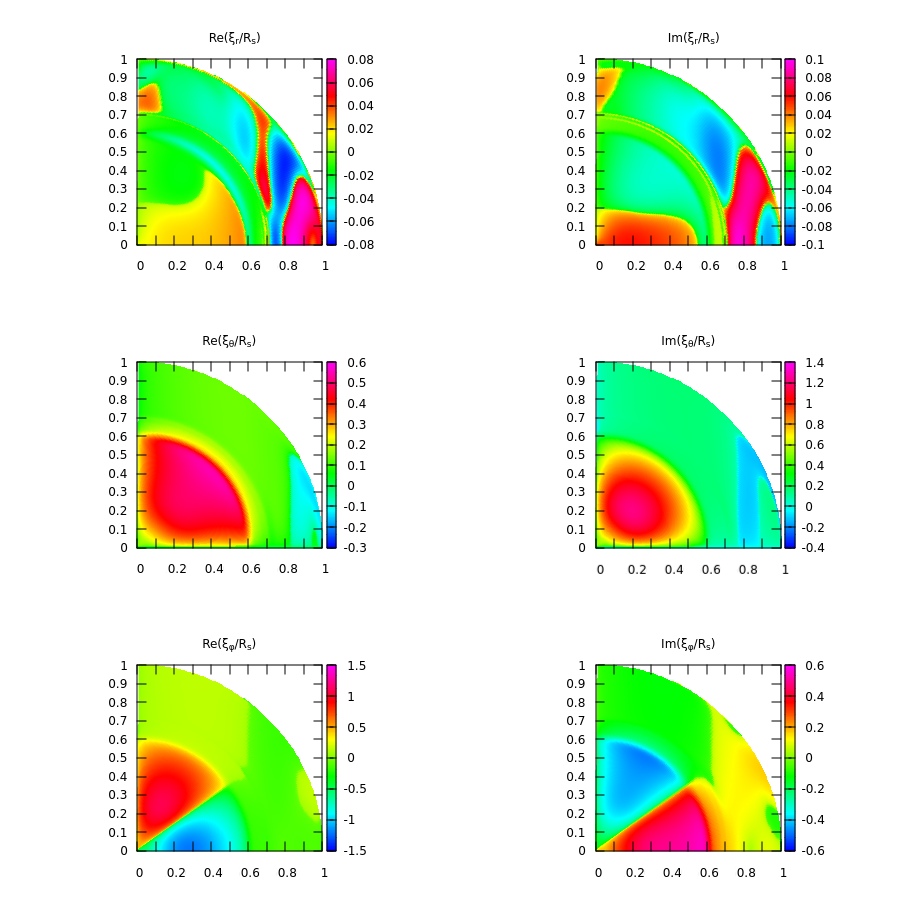}
	\caption{Contour plots  in the primary's meridional plane $\phi=0$ for the resonant $l'$=3 r mode with $n_r$ = 0 at resonance frequency $\omega_0=-3.332903054\times 10^{-3}$. The Cartesian coordinates along the two axes indicate the relative radius $r/R_*$. 
The vertical colour bars on the right indicate the local value of  sign($|\xi_x|^{\frac{1}{4}},\xi_x)$, where  $\xi_x$ is  the  component of the displacement vector illustrated. The base of the convective envelope is at $r_c=0.7313R_*$. \label{c0-0508L1}}
\end{figure}

\begin{figure} 
\hspace{-2.6cm}\includegraphics[width=1.29\columnwidth]{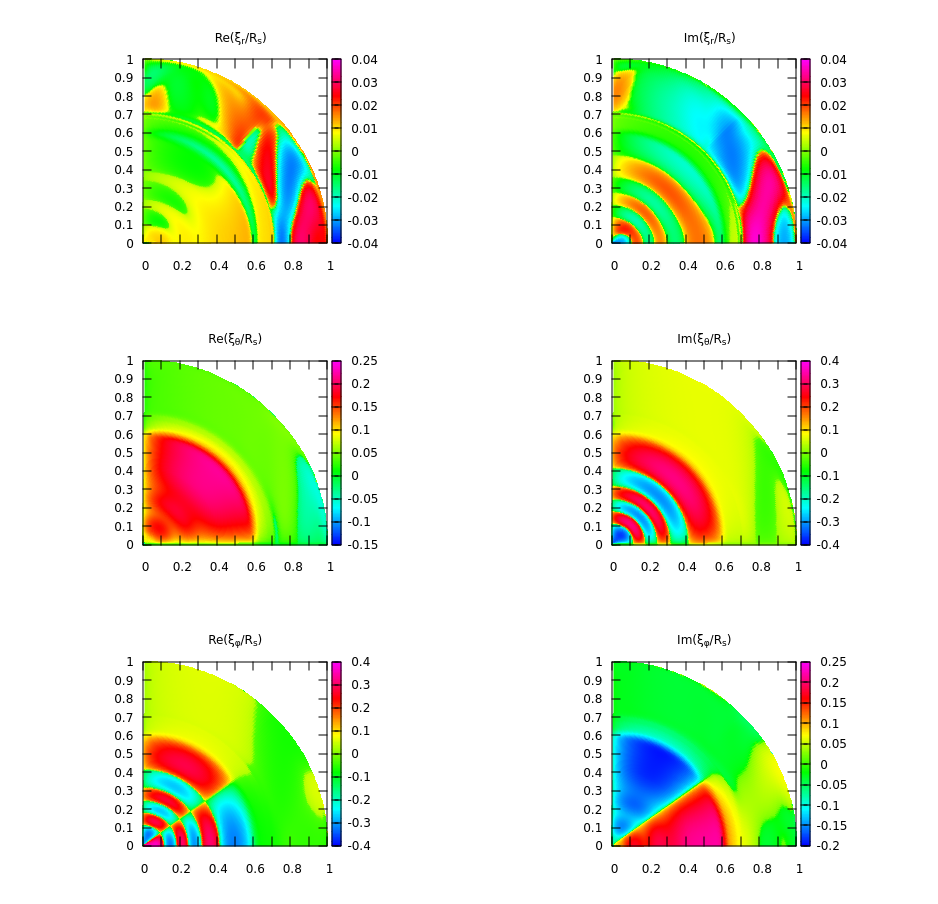}
	\caption{ As in Fig. \ref{c0-0508L1} but showing  contour plots  in the primary's meridional plane $\phi=0$ for the resonant $l'$=3 r mode with $n_r$ = 5 at resonance frequency $\omega_0=-3.332560283 \times 10^{-3}$.} %The Cartesian coordinates along the two axes are given by the relative radius $r/R_*$. The vertical colour bars on the right indicate the local value of  sign($|\xi_x|^{\frac{1}{4}},\xi_x)$, whereby $\xi_x$ is a component of the displacement vector. The base of the convective envelope is at $R_c/R_*=0.7313$. 
	\label{c5-0608L5}
\end{figure}

\begin{table} 	
\begin{center}
\begin{tabular}{cccccc} 
	\hline 
	$n_r$  &$ \omega_0 \times 10^{3}$  &$ I_0=E_{kin} (cgs)$ &$ D(E_{kin})$  &$ I_0=-\drv{E_{kin}}{t} (cgs)$ & $ D(Disp) $\\ 
	\hline
	$0$  &$ -3.332903054   $&$  4.80456 \times  10^{43} $& $2.25576 \times 10^{-10}$ &$ 6.46165 \times 10^{28}$ & $ 2.27282 \times 10^{-10}$ \\ 
	$1$  & $-3.332861397 $ &$ 2.69173 \times 10^{41} $&$ 8.59906 \times 10^{-10} $&$  1.37056 \times 10^{27} $&$ 8.23732 \times 10^{-10}$ \\
	$2$ & $-3.332806443 $  & $ 9.35504\times 10^{40} $&$ 1.84798 \times 10^{-9} $&$  1.02787 \times 10^{27} $&$ 1.85566 \times 10^{-9} $\\
	$3$  & $-3.332738070 $&$  8.66023 \times 10^{39} $&$ 3.28931 \times 10^{-9} $&$  1.67101 \times 10^{26} $&$ 4.00689 \times 10^{-9}$ \\
	$4 $ & $-3.332655910 $&  $4.42358 \times 10^{39} $&$ 4.89002 \times 10^{-9} $&$ 1.29819 \times 10^{26} $&$ 5.52811 \times 10^{-9} $ \\
	$5 $& $-3.332560283$ & $1.04789\times 10^{39}$ & $ 7.50894 \times 10^{-9} $& $4.61155 \times 10^{25} $&$ 8.98549 \times 10^{-9} $ \\
	\hline
	\end{tabular}
	\end{center}
\caption{The properties of calculated $l'=3$ r mode resonances. 
The quantities $D(E_{kin})$ and $D(Disp)$ are, respectively,  determined from the resonance curves corresponding to $E_{kin}$ and the rate of viscous dissipation in the convective envelope ( see Fig. \ref{vDisp-0508L1}).
The %orbital separation
\textcolor{ROUGE}{semi-major axis} and the kinetic energy and dissipation rates are scaled at the resonances as indicated above to correspond to an orbital period $P_{orb} \simeq$ 12.90 days. }
	\label{tabl3}
\end{table}	
Figures \ref{c0-0508L1} \textcolor{RED}{and} \ref{c5-0608L5} show the contour plots in the primary's meridional plane $\phi=0$ of the three components of the displacement vector for the r modes with,
respectively, $n_r = 0 $ \textcolor{RED}{(the fundamental mode), and $n_r = 5$ } for $l'=3$.  %Here $n_r$ is the number of nodes in radius within the radiative core.
It will be seen that in \textcolor{blue}{ the radiative core}  where the modes are sited, that $\xi_{\theta}$ has no nodes  and $\xi_{\phi}$ one node in $\theta$
in the interval $(0,\pi/2).$ This is consistent with the dominant form of these displacement components given by (see appendix \ref{rmodeapp}).
\begin{align}
 \xi_{\theta}  =\frac{r{\cal D}_3}{\sin\theta}\frac{\partial {Y_{l',m}}}{\partial \phi}, \hspace{3mm} {\rm and}\hspace{3mm}  \xi_{\phi}                      
	=-r{\cal D}_3\frac{\partial {Y_{l',m}}}{\partial \theta}  \label{rmodeform}
\end{align}
for $l'=3, m=-2,$ \textcolor{blue}{with ${\cal D}_3$  depending only on $r.$}

%\subsection{ $l^\prime$ = 5 r mode resonances}\label{l'5res}

%Table \ref{tabl5} shows the properties of calculated $l'=5$ r mode resonances.  The orbital separation is scaled at the resonances with corresponding orbital period %$P_{orb} \simeq$ 11.513 days. 

%$\begin{array} {cccccc} 
%	\hline 
%	n_r  & \omega_0 \times 10^{3}  & E_{kin} (cgs) & D(E_{kin})  & -\drv{E_{kin}}{t} (cgs) &  D(Disp) \\ 
%	\hline
%	0  & -1.332426514 & 5.07221 \times 10^{44} & 4.58021 \times 10^{-11} & 
%	4.42512 \times 10^{28}	& 4.54596 \times 10^{-11} \\
%	1  & -1.332415131 &  1.49945 \times 10^{42} &  1.94710 \times 10^{-10} & 
%	5.50060 \times 10^{26} &  1.81742 \times 10^{-10} \\ 
%	2  &  -1.332400899 & 9.75876 \times 10^{42} & 4.27517 \times 10^{-10} &	
%	8.01107 \times 10^{26}	& 3.92256 \times 10^{-10} \\
%	3  & -1.332383763 & 	4.73842 \times 10^{40} & 8.07189 \times 10^{-10} & 
%	7.47518 \times 10^{25} & 8.90890 \times 10^{-10} \\
%	4 &  -1.332363853 &  4.67462 \times 10^{40} & 1.11238 \times 10^{-9} &
%	1.03937 \times 10^{26} & 1.40793 \times 10^{-9} \\
%	5  & -1.332341064 & 6.58980 \times 10^{39} &  1.83775 \times 10^{-9} &
%	2.62449 \times 10^{25} & 2.13029 \times 10^{-9} \\
%	\hline	
%	\label{tabl5}
%\end{array}$	

\begin{table} 
\begin{center}
\begin{tabular} {cccccc} 
	\hline 
	$n_r  $&$ \omega_0 \times 10^{3}$  & $E_{kin} (cgs) $&$ D(E_{kin}) $ & $-\drv{E_{kin}}{t} (cgs) $&$  D(Disp)$ \\ 
	\hline
	$0 $ & $-1.332426514 $&$ 5.07221 \times 10^{44} $&$ 4.58021 \times 10^{-11} $& 
	$4.42512 \times 10^{28}	$&$ 4.54596 \times 10^{-11} $\\
	$1$  & $-1.332415131 $&$  1.49945 \times 10^{42} $&$  1.94710 \times 10^{-10} $&$ 
	5.50060 \times 10^{26} $&$  1.81742 \times 10^{-10} $\\ 
	$2$  & $ -1.332400899 $&$ 9.75876 \times 10^{42} $&$ 4.27517 \times 10^{-10} $&$	
	8.01107 \times 10^{26}	$&$ 3.92256 \times 10^{-10}$ \\
	$3$  & $-1.332383763 $& $	4.73842 \times 10^{40} $&$ 8.07189 \times 10^{-10} $&$ 
	7.47518 \times 10^{25} $& $8.90890 \times 10^{-10} $\\
	$4$ & $ -1.332363853 $& $ 4.67462 \times 10^{40} $&$ 1.11238 \times 10^{-9} $&$
	1.03937 \times 10^{26} $&$ 1.40793 \times 10^{-9} $\\
	$5$  &$ -1.332341064 $&$ 6.58980 \times 10^{39} $&$  1.83775 \times 10^{-9} $&$
	2.62449 \times 10^{25} $&$ 2.13029 \times 10^{-9} $\\
	\hline	

	\end{tabular}
	\end{center}
	\caption{ The properties of calculated $l'=5$ r mode resonances.  The %orbital separation
	\textcolor{ROUGE}{semi-major axis} is scaled at the resonances with corresponding orbital period $P_{orb} \simeq$ 11.51 days. }
		\label{tabl5}
\end{table}
	
\begin{figure} 
\hspace{-2.73cm}	\includegraphics[width=1.31\columnwidth]{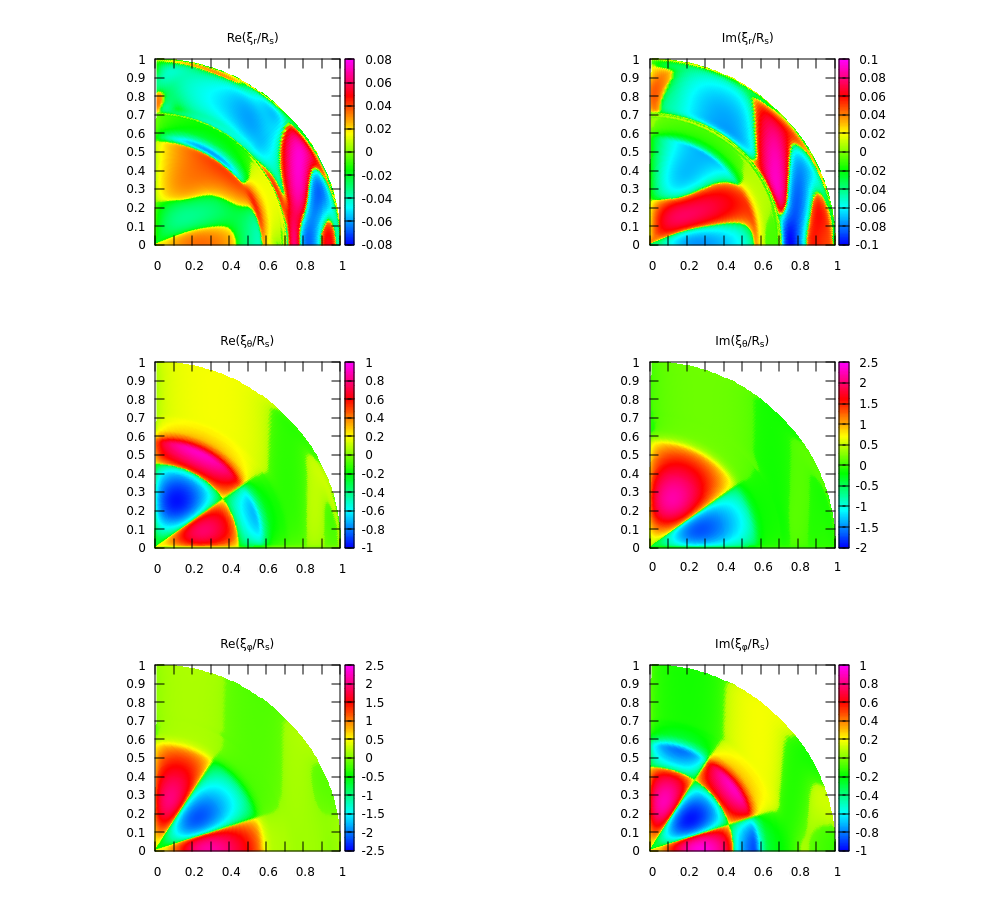}
	\caption{As in Fig. \ref{c0-0508L1} but for  contour plots  in the primary's meridional plane $\phi=0$ of the resonant $l'$=5 r mode with $n_r$ = 0 at resonance frequency $\omega_0=-1.332426514 \times 10^{-3}$. 	
		%	The Cartesian coordinates along the two axes are given by the relative radius $r/R_*$. The vertical colour bars on the right indicate the local value of  sign($|\xi_x|^{\frac{1}{4}},\xi_x)$, whereby $\xi_x$ is a component of the displacement vector. The base of the convective envelope is at $R_c/R_*=0.7313$.
		\label{c0-2807M1}}
\end{figure}

\begin{figure} 
\hspace{-2.73cm}\includegraphics[width=1.31\columnwidth]{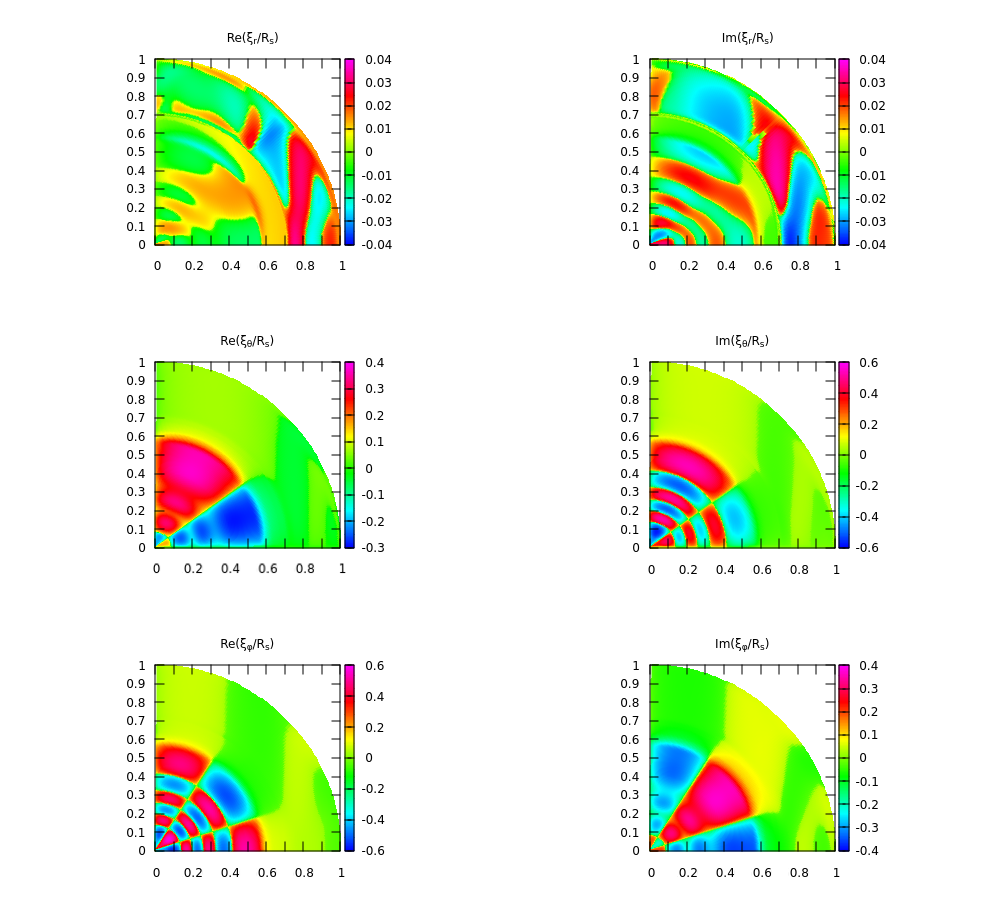}
	\caption{ As in Fig. \ref{c0-0508L1} but for contour plots  in the primary's meridional plane $\phi=0$ of the resonant $l'$=5 r mode with $n_r$ = 5 at resonance frequency $\omega_0=-1.332341064 \times 10^{-3}$. 
		%The Cartesian coordinates along the two axes are given by the relative radius $r/R_*$. The vertical colour bars on the right indicate the local value of  sign($|\xi_x|^{\frac{1}{4}},\xi_x)$, whereby $\xi_x$ is a component of the displacement vector. The base of the convective envelope is at $R_c/R_*=0.7313$. 
		\label{c5-2807L2}}
\end{figure}

Figures \ref{c0-2807M1} \textcolor{RED}{and } \ref{c5-2807L2} show the contour plots in the primary's meridional plane $\phi=0$ of the three components of the displacement  for  the  r modes with $l'=5$ \textcolor{RED} {and $n_r =0,$ and with $l'=5$ and} $n_r =5,$ respectively. The presentation is of the same form as in Figs.  \ref{c0-0508L1} \textcolor{RED}{and} \ref{c5-0608L5}  for the case $l'=3,$
but  in this case there is one node in $\xi_{\theta}$ and  two in $\xi_{\phi}$  as implied by (\ref{rmodeform}) with $l'=5.$

\subsection{$r$ mode spectrum}\label{rmodespec}
From equation (\ref{spectrum}) of appendix \ref{rmodeapp} the spectrum of $r$ mode resonances is  given in a WKBJ approximation  by
\begin{align}
	\omega_f - \frac{2m\Omega_s}{l'(l'+1)} =-\frac{({ n_r}\pi+\psi_{WKBJ})^2\omega_f^3}{C_{l',m} (\int_{0}^{r_c}r^{-1}Ndr)^2}\sim
	-\frac{({ n_r}\pi+\psi_{WKBJ})^2(2m\Omega_s)^3}{C_{l',m}( l'(l'+1))^3(\int_{0}^{r_c}r^{-1}Ndr)^2},\label{spectrum1}
	\end {align}
	where $r_c$ is the radius of the inner convective envelope boundary.
	%From (\ref{spectrum1}) we form the difference between the resonant frequency and its value when $n_r=0$ as the WKBJ approximation
	%is least applicable for estimating the latter, so obtaining
%	\begin{align}
%	\omega_f - \omega_{f}|_{n_r=0} =
%	-\frac{(({ n_r}\pi)^2+2n_r\phi_{WKBJ}\pi)(2m\Omega_s)^3}{C_{l',m}( l'(l'+1))^3\int_{0}^{r_c}r^{-1}N^2dr}.\label{spectrum2}
%	\end {align}
	Although (\ref{spectrum1}) is also not expected to be a good approximation for global modes, with small values of $n_r,$
	 and an atypical boundary condition at the convective envelope boundary,
	 we make a rough comparison with our numerical results.
	 
\textcolor{blue}{	For $l'=3,$ and $m=-2,$  an estimate for the right hand side \textcolor{blue}{of (\ref{spectrum1})   }  (see appendix \ref{rmodeapp}) leads to
%	\begin{align}
%	\omega_f - \omega_{f}|_{n_r=0} \sim 0.37\Omega_s\frac{({ n_r}^2+2n_r\phi_{WKBJ}/\pi)\Omega^2_s}{18C_{3,2}\Omega_c^2}\sim 4\times 10^{-8}\Omega_c
%	({ n_r}^2+2n_r\phi_{WKBJ}/\pi)\hspace{3mm}.% ({\rm see\hspace {2mm} appendix \hspace{2mm} \ref{rmodeapp}}).\nonumber
%	\label{spectrum3}
   %     \end {align}
        	\begin{align}
	\omega_f - \frac{2m\Omega_s}{l'(l'+1)} \sim 0.37\Omega_s\frac{( n_r+\psi_{WKBJ}/\pi)^2\Omega^2_s}{36C_{3,2}\Omega_c^2}\sim 2\times 10^{-8}\Omega_c
	( n_r+\psi_{WKBJ}/\pi)^2\hspace{3mm}.% ({\rm see\hspace {2mm} appendix \hspace{2mm} \ref{rmodeapp}}).\nonumber
	\label{spectrum3}
        \end {align}
	We find that  for $ n_r = 3,4$ and  $5,$  the largest values of $n_r$ available, this gives results within a factor  of $2$ of those in table
	 \ref{tabl3}  for $\psi_{WKBJ}=\pi,$ indicating a reasonable
	order of magnitude estimate for these frequency differences in spite of the obvious limitations of the approach. }
% Similarly, on repeating the discussion for  $l'=5$ we find the analogue of (\ref{spectrum3}) to be 
%\begin{align}
%	\omega_f - \omega_{f}|_{n_r=0}  \sim 2.5\times 10^{-9}\Omega_c
%	({ n_r}^2+2n_r\phi_{WKBJ}/\pi) \nonumber
%	\end{align}
%We find that this gives results within a factor $1.5$ of those in table \ref{tabl5} for $1 \le n_r \le 5,$ and $\Phi_{WKBJ}/\pi=1.5.$ 	

\clearpage
%\begin{figure} 
%	\includegraphics[width=\columnwidth]{Cont5/c22807L3.png}
%	\caption{As in Fig. \ref{c0-0508L1} but for contour plot  in the primary's meridional plane $\phi=0$ of the resonant $l'$=5 r mode with $n_r$ = 2 at resonance frequency $\omega_0=-1.332400899 \times 10^{-3}$.
		% The Cartesian coordinates along the two axes are given by the relative radius $r/R_*$. The vertical colour bars on the right indicate the local value of  sign($|\xi_x|^{\frac{1}{4}},\xi_x)$, whereby $\xi_x$ is a component of the displacement vector. The base of the convective envelope is at $R_c/R_*=0.7313$. 
%		\label{c2-2807L3}}
%\end{figure}

\subsection{Convective envelope response}\label{convResp}
An $r$ mode sited in the radiative core excites a response in the convective envelope
In appendix \ref{ConvResp} we consider the response of the convective envelope 
in the limit of vanishing $|\omega_f|/(2\Omega_s)$ \footnote{The maximum value of this is unity in the inertial range.}.\newline
 This limit strictly applies only for very large $l',$ though we find it of interest to make some comparison
for $l'=3,$ for which $ |\omega_f|/(2\Omega_s)\sim1/6,$ and for $l'=5,$ for which
 $|\omega_f|/(2\Omega_s)\sim 1/15.$  In appendix \ref{ConvResp} we find that in this limit
 $\xi_{\phi}$ should depend only on the cylindrical radius $\bar r.$ This tendency can be seen through the near vertical contour lines 
 in Figs. \ref{c0-0508L1} - \ref{c5-2807L2}
 for all cases with $l'=3$ and $l'=5.$
 
In addition in the limit of vanishing $\omega_f$ we also find a critical latitude type singularity along the line $\bar r=r_c$
commencing at $z=0.$ However, we remark that in our case the Eckman number, $\nu/(2r_c^2\Omega_s),$ in this region
is $\sim 10^{-4} $ which is significantly higher than those typically considered in inertial mode calculations \citep[see eg.][]{Rieutord2010, O2014}.
This would lead to an estimated thickness $\sim  r_c (\nu /(2r_c^2\Omega_s))^{1/4}\sim 0.1r_c.$ Furthermore this scale is less than
 $r_c\sqrt{|\omega_f|/ (2\Omega_s)}\sim  0.2 r_c$ for $l'=5$ indicating that inertial waves play a part in the response (see discussion in Section \ref{Strict})  which will be of quite a large scale.
 Nonetheless there is evidence of an incipient  critical latitude phenomenon in $\xi_r,$  particularly for $l'=5$  ( see Figs. \ref{c0-0508L1} - \ref{c5-2807L2}).

\section{Effects on orbital and spin evolution}\label{orbspinevol}
	 \label{tidalevol}
	 We now consider the effects of the tidal response of the primary on the orbital evolution of the 
	 %binary
	 \textcolor{RED}{system}  and the spin of the primary.
	 As stated previously we here restrict consideration to aligned angular momenta and circular orbits.
	  
	 \subsection{Torque and dissipation resulting from tidal forcing}\label{TorandD}
%	 We begin by  remarking  that by comparing the amplitude factors associated with the potentials (\ref{jpe1a}) and  (\ref{potpert}) 
%	 that the rate of change of kinetic energy resulting from the potential (\ref{jpe1a}) can be seen to be given by 

%\textcolor{green}{	 In this subsection changes were made: possible confusion of $T_*$ as torque primary/planet:  $T_*$ similar to $M_*$!   Eqn 39 seems superfluous/incorrect}

	From (\ref{torque0}) we  see that the torque acting on $M_p$ is given by 
%	 \begin{align}
%       M_pT_Z = 2\left(\frac{4\pi GM_p}{5a^3}\right)^2 
 %    (Y_{2,m}(\pi/2,0))^2 \, m \, \im(Q_{m,\sigma})  =  -\frac{m}{\omega_f} \frac{dE_{kin}}
%	{dt},
%	 \end{align}
	 \begin{align}
          M_pT_Z = - \left(\frac{4\pi GM_p}{5a^3}\right) (Y_{2,m}(\pi/2,0))m Im ( Q_{m,\sigma})=   -\frac{m}{\omega_f} \frac{dE_{kin}} {dt} ,% \label{torque0}
          % \sum^{'}_{m=0, |2|} \frac{4\pi GM_p}{5 a^3} \phi_{k,m}D^{(2)}_{n,m}  Y_{2,m}(\pi/2, 0)\exp(-{\rm i} (m\varpi -kn_ot))
          \end{align}
%\textcolor{blue}{where the rate of viscous dissipation is}
%	 	 \begin{align}
%	\frac{dE_{kin}} 	{dt}=  \omega_f
	%\left(\frac{4\pi GM_p}{5a^3}\right)^2
         %(Y_{2,m}(\pi/2,0))^2 \, \im(Q_{m,\sigma}).
   %       \left(\frac{2\pi GM_p}{5a^3}\right) (Y_{2,m}(\pi/2,0))m Im ( Q_{m,\sigma})
	%\end{align}
	 Note that for a positive dissipation rate, ${dE_{kin}}/{dt} =- dE_{dissip}/dt <0$ and by conservation of angular momentum
	 the torque acting on the primary is, $T_* =-M_pT_Z,$ accordingly we have
	 \begin{align}
	 \hspace{-3.8cm} T_*=\frac{1}{\omega_p}\frac{dE_{dissip}}{dt},\hspace{3mm}{\rm where\hspace{1mm}the\hspace{1mm} pattern
	 \hspace{1																															mm} speed \hspace{1mm}
	 of\hspace{2mm} the\hspace{1mm}forcing \hspace{2mm} }  \omega_p= -\omega_f/m.\nonumber
	 \end{align}
%	where the pattern speed of the forcing  $\omega_p= -\omega_f/n.$
         This leads to a putative rate of evolution of the orbit given by
           \begin{align}
	 \hspace{-3.8cm} \frac{M_pM_*}{2(M_*+M_p)}\sqrt{\frac{G(M_*+M_p)}{a}}\frac{da}{dt} =\frac{ \textcolor{red}{|E_{orb}|}}{n_0} \frac{1}{a}\frac{da}{dt} 
	 =-\frac{1}{\omega_p}\frac{dE_{dissip}}{dt}.
	% \hspace{3mm}{\rm where\hspace{1mm}the\hspace{1mm} pattern
	% \hspace{1mm} speed \hspace{1mm}
	% of\hspace{2mm} the\hspace{1mm}forcing \hspace{2mm} }  \omega_p= -\omega_f/m.\nonumber
	 \end{align}         
         with $E_{orb}$ being the orbital energy.

	\subsection{Orbital evolution}\label{orbevol}
	The torque acting on the primary results in evolution of its spin while its reaction causes
	evolution of the orbit. If only mutual gravitation and tidal forces act, the
	 sum of the spin and orbital angular momentum is conserved. Thus we may write
	\begin{align}
	 J= |{\bf J}| = \mathcal{I} \, \Omega_s+\frac{ M_*M_p}{(M_*+M_p)}\sqrt{G(M_*+ M_p)a},
	\end{align}
          with $\mathcal{I}$  being the moment of inertia of the primary  and, $J,$ the magnitude of
          the total angular momentum being constant (note that changes to, $J,$ induced by non tidal effects will be considered below).
          Assuming the system attains synchronisation, we set $\Omega_s= \sqrt{G(M_*+ M_p)/a^3}.$ Then we find
          \begin{align}
	 J(a) = \mathcal{I} \frac{\sqrt{G(M_*+M_p)}}{a^{3/2}}+\frac{ M_*M_p}{(M_*+M_p)}\sqrt{G(M_*+ M_p)a},\label{Jform}
	\end{align}
	From this one can see that, once, $J,$ is specified there are either two 
	values, or no values, of $a$ for which the system is synchronised.
	For the former situation to apply we require
	   \begin{align}
	 J >\frac{ 4}{3^{3/4}} \left(\frac{ M_*M_p}{(M_*+M_p)}\right)^{3/4}\sqrt{G(M_*+M_p)} \, \mathcal{I}^{1/4}.\label{Jmin}
	\end{align}
	In that case the solution with the smaller value of $a$ is unstable.
	This means that for synchronisation, the \textcolor{blue}{semi-major axis} must exceed the value one obtains when $J$ is
	specified to be the value given by (\ref{Jmin}) when the inequality is replaced by equality.
	This leads to
	\begin{align}
	a \ge \left( \frac{3 \, \mathcal{I} \, (M_*+M_p)}{M_*M_p}\right)^{1/2}.
	\end {align}
	Expressing this in terms of the orbital period, $P_{orb},$  and a solar like primary, this gives
	\begin{align}
	P_{orb} \ge 0.036 \left( \frac{(1+q)^{1/4}}{q^{3/4}}\right) d.\label{synccrit}
	\end {align}
        For smaller periods in-spiral occurs with no approach to synchronisation
        especially when the star is slowly rotating.
        From the above it is clear that for a solar mass primary and $q\sim 1$  inequality (\ref{synccrit}) 
        will in general  be satisfied. On the other hand it is quite reasonable that it is not satisfied for $q\sim 0.001$
        corresponding to secondaries in the giant planet mass regime and (\ref{synccrit}) turns out not to be  satisfied for almost all objects classified as hot Jupiters. Thus these objects will not attain synchronisation under the above assumption of conservation of angular momentum given by Equation (\ref{Jform}). 

In this context we remark that the system for which we described the r mode resonances with $l'=3$ in Section \ref{l'3res} has $a= 1.61 \times 10^{12}$ cm which exceeds the  value given above  \textcolor{red}{($1.04 \times 10^{12}$ cm for a Jupiter mass planet)
so allowing for the possibility of synchronisation. In addition $\Omega_s > n_o$ so that tidal effects will increase $a$, thereby  slowing down the  orbit while spinning down the primary more rapidly. This would move the system towards
synchronisation. Only if the primary  spins down (beyond synchronism) to the point where $\Omega_s < n_0$, for example due to magnetic braking, will  tides lead to smaller $a$ values}.

         \subsection{The effect of r mode resonances}\label{rmodeeffect}
	The discussion in Section (\ref{orbevol}) is important when assessing the effects of normal mode
	resonances on the orbital evolution. The form of the  $r$ modes that can be resonantly excited by tides
	is  discussed in appendix \ref{rmodeapp}. For a given azimuthal mode number, $m,$
	the resonant forcing frequencies seen in a frame  co-rotating with the star  are close to the values given by
	equation (\ref{rmodef}) ( see Section \ref{Findr}).

	 Recalling that when the total angular momentum of the system is conserved, if
	  condition (\ref{synccrit}) is satisfied the spin and orbital periods approach
	each other more closely and the system moves towards synchronisation.
	In that case for a system with conserved total angular momentum the effect of encountering  the resonances
	will be to cause the system to pass through brief periods of rapid tidal evolution moving the 
	system  towards synchronisation. On the other hand if condition (\ref{synccrit})
	is not satisfied the system moves away from synchronisation undergoing periods of brief acceleration
	as the resonances are encountered. These processes occur in addition
	to effects due to the response of the convective envelope which may also be erratic.

%\begin{figure}
%	\includegraphics[width=8cm]{plots/a-l=3-n=0.pdf}
%	\includegraphics[width=8cm]{plots/Omg_s-l=3-n=0.pdf}
%	\caption{The left hand panel shows the tidal evolution of the orbital separation $a$ during resonance passage of the $l'$=3 r mode with $n$=0.  
%	The right hand panel shows the corresponding evolution  of the stellar angular velocity. 
%	\label{eva-l=3-n=0}}
%\end{figure}

%\begin{figure}
%	\includegraphics[width=8cm]{plots/a-MB-l=3-n=0.pdf}
%	\includegraphics[width=8cm]{plots/Omg_s-MB-l=3-n=0.pdf}
%	\caption{Tidal evolution of the orbital separation $a$ during resonance passage of the $l'$=3 r mode with $n$=0. Magnetic braking speeds up the tidal evolution.  
%	The left hand panel shows the corresponding evolution of the primary's angular velocity.
%	\label{eva-MB-l=3-n=0}}
%\end{figure}

	\subsection{Numerical calculation of spin and orbit evolution \textcolor{RED}{in the case of a Jupiter mass companion}}\label{Numevol}
\textcolor{red}{The tidal evolution of the star/planet system caused by the viscous dissipation in the primary's convective envelope is given by the following two equations where $\omega_p$ is the pattern speed of the considered $r$ mode
\begin{equation}
	%\drv{\Omega_s}{t}= -\frac{m}{\omega_f \, \mathcal{I}} \,\drv{E_{diss}}{t}
	 \mathcal{I} \, \drv{\Omega_s}{t}= \frac{1}{\omega_p } \,\drv{E_{\rm{disp}}}{t}
	\label{DE1}
\end{equation}
where $\mathcal{I} = 6.851 \times 10^{53}$ g cm$^2 = 0.0782M_*R_*^2 ,$ is the primary's moment of inertia.  The rate of change of the %orbital separation
\textcolor{ROUGE}{semi-major axis}  is (see section (\ref{TorandD})
\begin{equation}
	%\drv{a}{t} = \frac{2 m n_o}{\omega_f} \, \frac{a^2}{G M_p M_*} \, \drv{E_{diss}}{t}
	\drv{a}{t} =- \frac{2  n_o}{\omega_p} \, \frac{a^2}{G M_p M_*} \, \drv{E_{\rm{disp}}}{t}
	\label{DE2}
\end{equation}   }
where the viscous dissipation rate $dE_{\rm{disp}}(\omega_f-\omega_0)/dt=-dE_{\rm{kin}}/dt$ follows from the fitted resonance curve for the considered r mode resonance, see tables \ref{tabl3} and \ref{tabl5}.
It is important to take  into account that  the resonance frequency $\omega_0$  of the r mode depends on the actual spin rate $\Omega_s$ of the primary \textcolor{red}{during the tidal evolution of the system}. 
To lowest order we simply assume that the resonance frequency $\omega_0$ scales linearly with $\Omega_s.$ Thus
\begin{equation}
	\omega_0(t)= \omega_0(*) \, \frac{\Omega_s(t)}{\Omega_s(*)}
\end{equation}
whereby $\omega_0(*)$  is the value given in table \ref{tabl3} and $\Omega_s(*)$ is the associated stellar angular velocity.
The differential equations (\ref{DE1}) - (\ref{DE2}) are solved over time using a fifth order Runge-Kutta integration algorithm with adaptive step size control. 
As an example we calculated the  passage through the strongest r mode resonance ($l'=3$  with $n_r=0$ )   starting at $\omega_f - \omega_0= -1.0 \times 10^{-8}$
% ( for which the orbital period is slightly  smaller than at resonance) 
up to  $\omega_f - \omega_0 \simeq  +1.0 \times 10^{-8}$.
%\textcolor{red}{For simplicity, to begin with we shall assume that the resonant frequency $\omega_0$ is fixed.}

%\begin{figure}
%	\includegraphics[width=8cm]{plots/a-frqsh-l=3-n=0.pdf}
%	\includegraphics[width=8cm]{plots/Omg_s-frqsh-l=3-n=0.pdf}
%	\caption{The left panel shows the tidal evolution of the orbital 
%separation $a$ during resonance passage of the 
%$l'$=3 r mode with $n_r=0.$ The right hand panel shows the 
%corresponding  evolution of the primary's angular velocity.
%		\label{eva-frqsh-l=3-n=0}}
%\end{figure}

\begin{figure}
  \hspace{0cm}   \includegraphics[width=8cm,height=10cm]{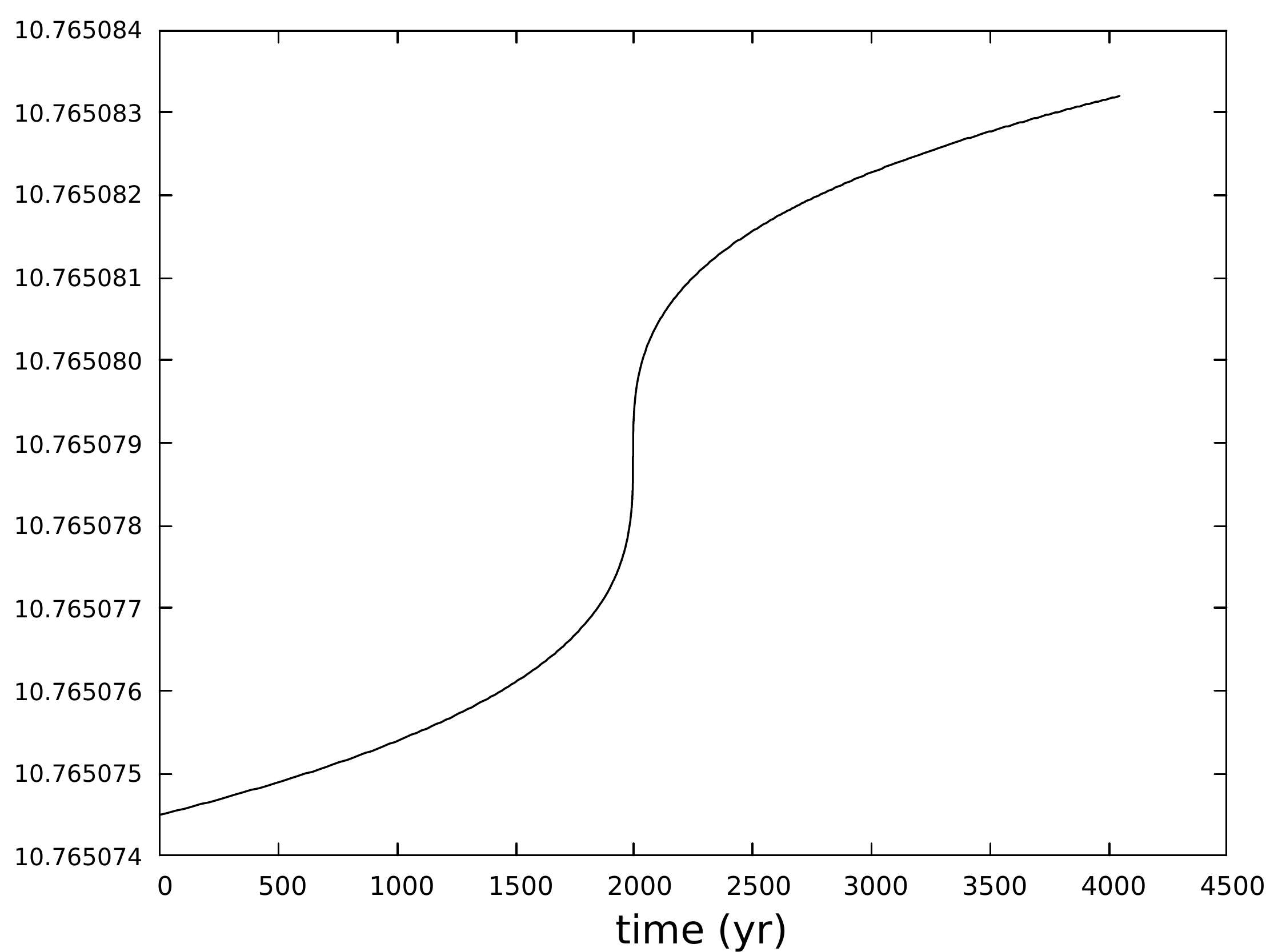}
     \includegraphics[width=8cm, height=10cm]{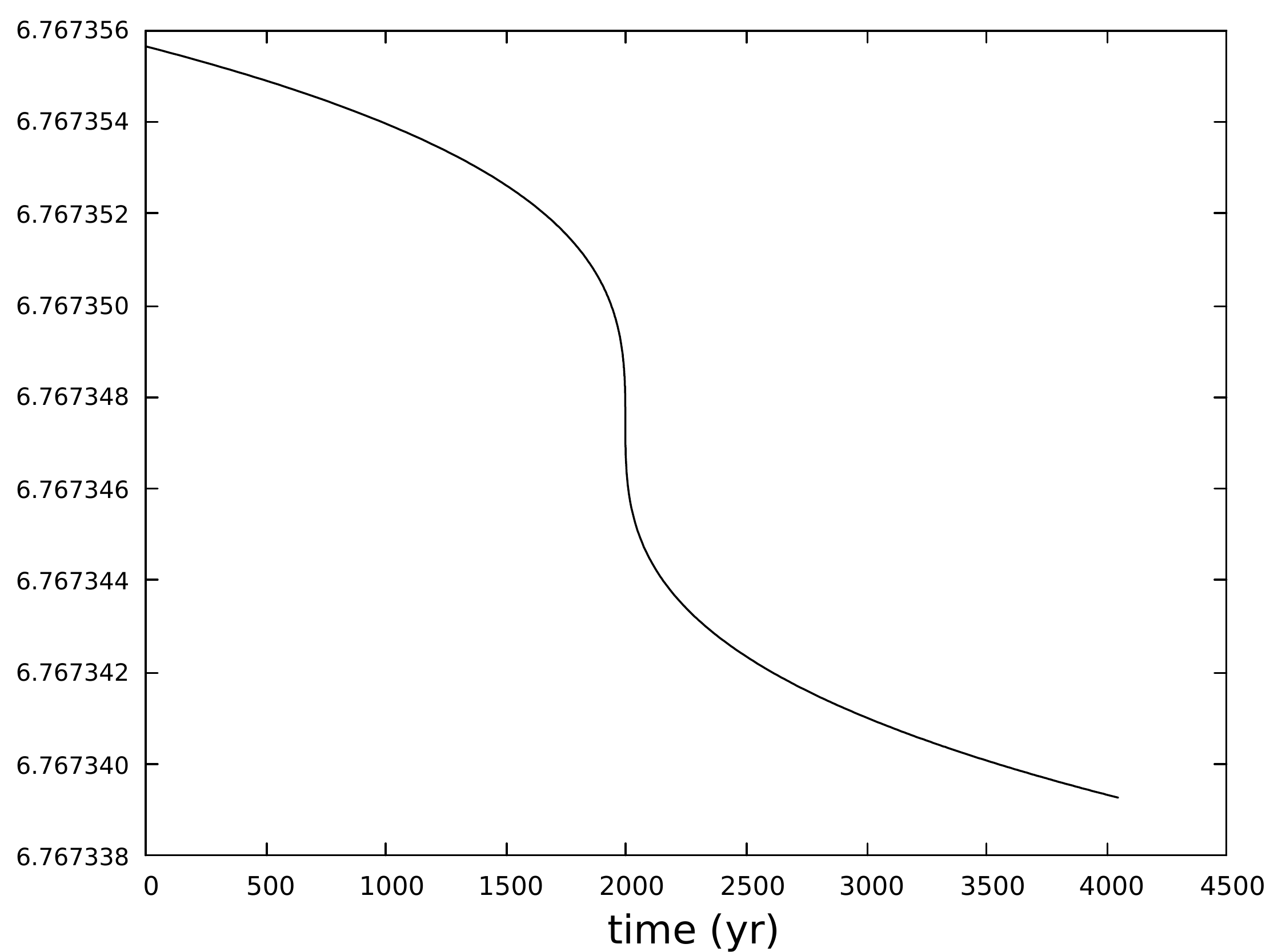}
     \caption{The left panel shows the tidal evolution of the % orbital separation 
 \textcolor{ROUGE}{ semi-major axis} $a$ in units of $0.01 \ au$ during resonance passage of the
$l'$=3 r mode with $n_r=0.$ The right hand panel shows the
corresponding  evolution of the primary's angular velocity \textcolor{RED}{in units of
$10^{-6}$ Hz.}
         \label{eva-frqsh-l=3-n=0}}
\end{figure}

%\begin{figure}
%	\includegraphics[width=8cm]{plots/a-MB-frqsh-l=3-n=0.pdf}
%	\includegraphics[width=8cm]{plots/Omg_s-MB-frqsh-l=3-n=0.pdf}
%	\caption{The left panel shows the tidal evolution of the orbital separation $a$ 
%during resonance passage of the
% $l'$=3 r mode with $n_r=0.$ Magnetic braking speeds up the tidal 
% evolution. The right hand panel shows the corresponding  evolution of 
 %the primary's angular velocity.
%		\label{eva-MB-frqsh-l=3-n=0}}
%\end{figure}

\begin{figure}
    \hspace{0cm} \includegraphics[width=8cm,height=9cm]{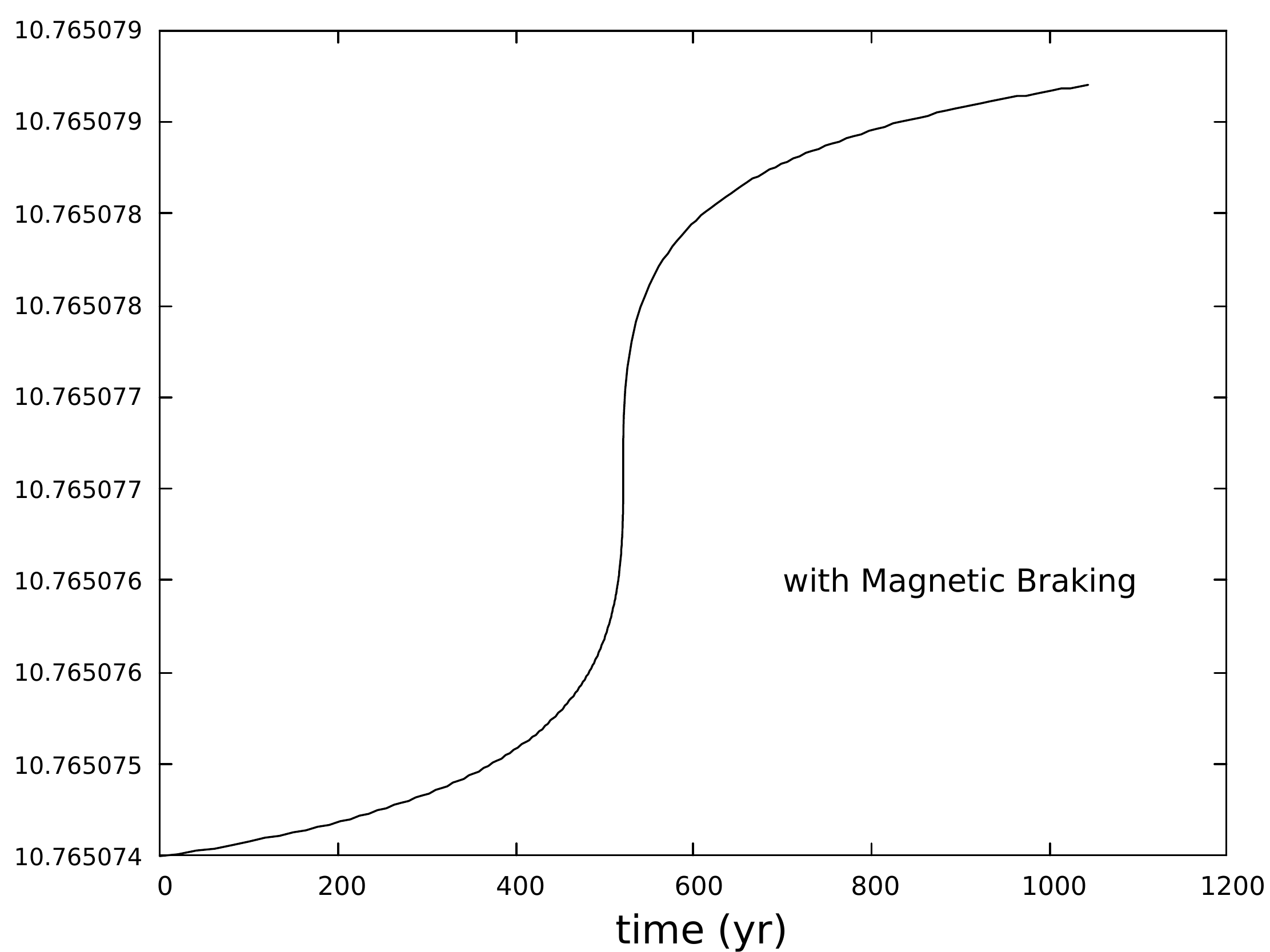}
     \includegraphics[width=8cm,height=9cm]{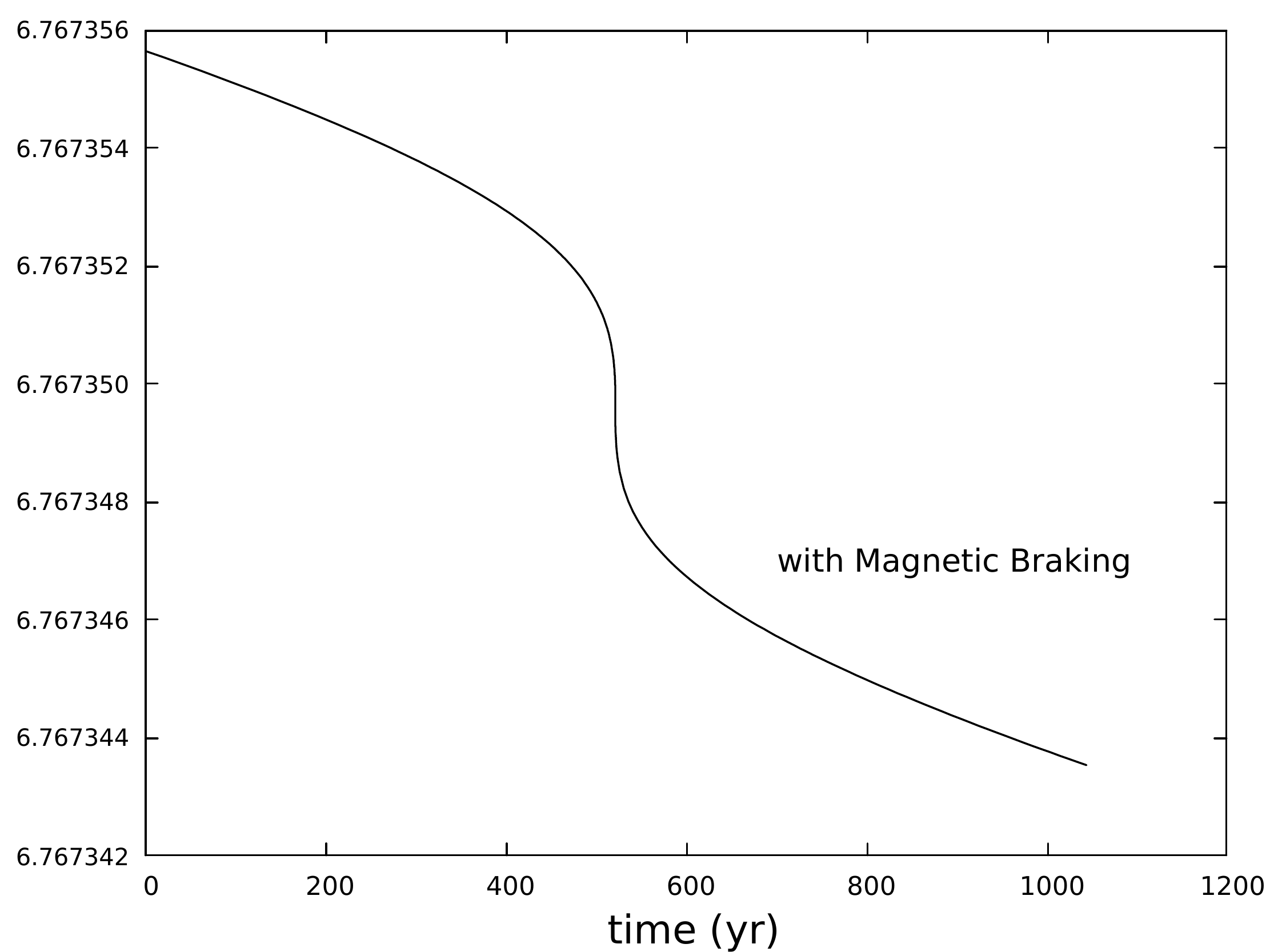}
     \caption{The left panel shows the tidal evolution of the % orbital separation 
 \textcolor{ROUGE}{ semi-major axis} $a$ in units of $0.01  \ au$ %\textcolor{RED}{in units of $10^{12}$ cm}
 during resonance passage of the
$l'$=3 r mode with $n_r=0.$ Magnetic braking speeds up the tidal
evolution. The right hand panel shows the corresponding  evolution of
the primary's angular velocity \textcolor{RED}{in units of $10^{-6}$ Hz}.
         \label{eva-MB-frqsh-l=3-n=0}}
\end{figure}

%\begin{figure}
%	\includegraphics[width=8cm]{plots/a-l=5-n=0.pdf}
%	\includegraphics[width=8cm]{plots/Omg_s-l=5-n=0.pdf}
%	\caption{The left panel shows the tidal evolution of the orbital 
%	separation $a$ during resonance passage of the
%	 $l'$=5 r mode with $n_r=0.$ The right hand panel shows the 
%	 corresponding  evolution of the primary's angular velocity.
%		\label{eva-l=5-n=0}}
%\end{figure}

\begin{figure}
  \includegraphics[width=8cm,height=9cm]{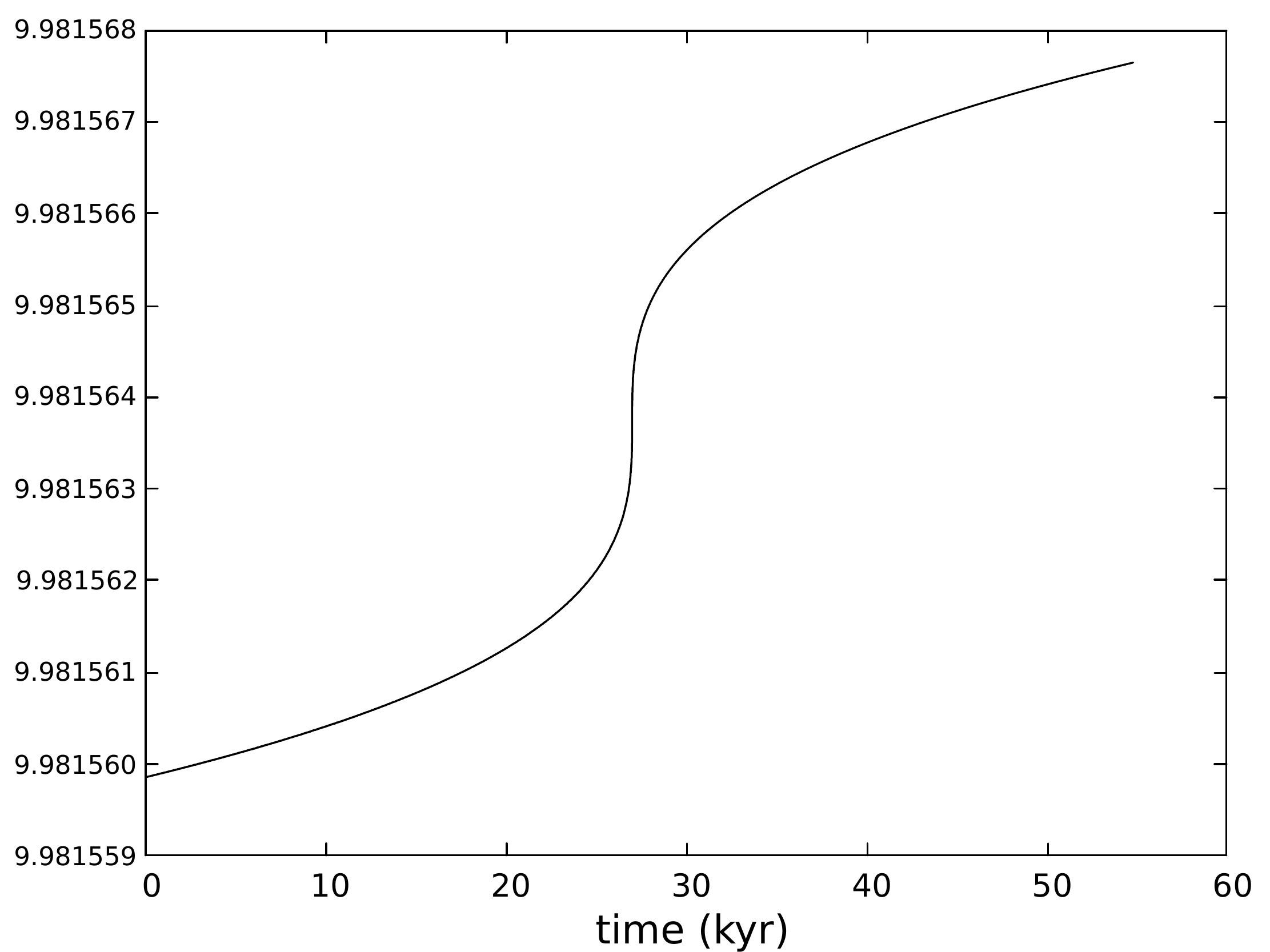}
    \includegraphics[width=8cm,height=9cm]{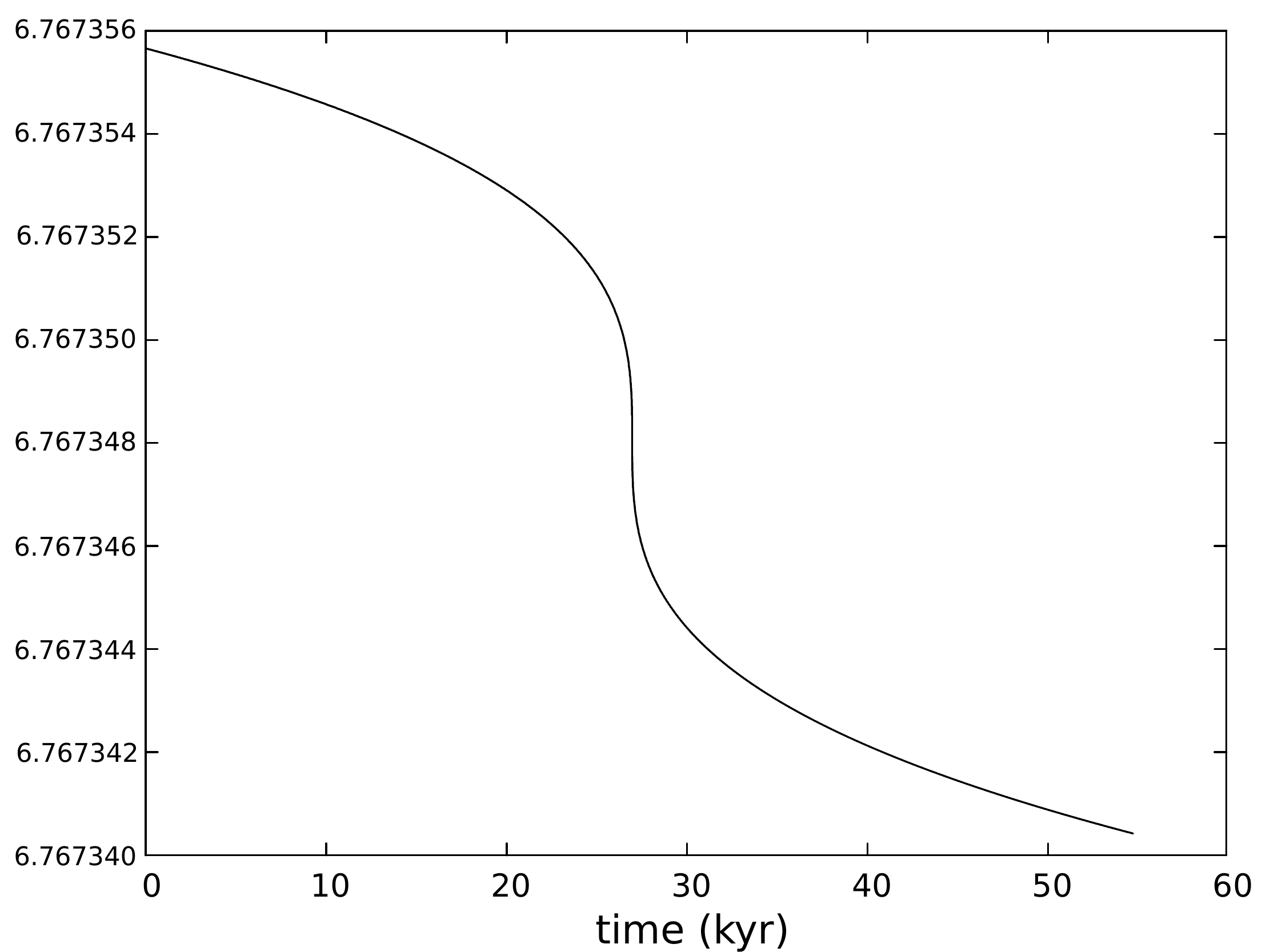}
     \caption{The left panel shows the tidal evolution of the %orbital separation 
 \textcolor{ROUGE}{ semi-major axis} $a$ in units of $0.01 \ au$  during resonance passage of the
$l'$=5 r mode with $n_r=0.$ % The right hand panel shows the
The right hand panel shows corresponding  evolution of the primary's angular velocity \textcolor{RED}{ in units of
$10^{-6}$ Hz. \textcolor{ROUGE}{The unit of time is $10^3 y.$} }
         \label{eva-l=5-n=0}}
\end{figure}

\textcolor{red}{The results are displayed in  Fig. \ref{eva-frqsh-l=3-n=0}. The orbit rapidly accelerates through the resonance  with an average evolution timescale $a/\dot{a} \simeq 3 \times 10^7$ y (from reaching 10 \% of the full resonant dissipation rate  left of the resonance up to the same rate on the right).}
\textcolor{RED}{This can be seen from applying equation (\ref{DE2}) directly making use of the data plotted in Fig. \ref{vDisp-0508L1}, which leads to $a/\dot{a} \simeq 8 \times 10^7$ y at. 10\% of the full dissipation rate. }
 However, as noted in section \ref{l'3res}, nonlinear effects are likely to play a role during resonance passage and effectively cause an extension of this time scale. But we remark that initial and final forcing frequencies  are far into  the wings of the resonance (see Fig.\ref{vDisp-0508L1}), 
where we might expect the linear approach to be valid. 

\textcolor{RED}{Nevertheless, as noted in Section \ref{reducingmp} we can expect the linear results to be valid for lower mass planets
with masses $\sim 30$ times smaller, thus in the mini-Neptune range. Applying equation (\ref{DE2}) at the centre of the resonance in this case yields
$a/\dot{a} \simeq 2.4 \times 10^8$ y indicating the potential significance of resonances in producing strong local torques.
In corroboration of this,  simple scaling of (\ref{DE2}) indicates that $a/\dot{a}$ at the centre of the resonance is $\sim 30$ times smaller
than it is for a Jupiter mass planet at the point where the energy dissipation rate is reduced by a factor $10^3.$ }

% The evolution times in the  wings of the resonance corresponding with  0.02\% and 0.2\% of the maximum resonant dissipation rate are respectively $a/\dot{a} \simeq 1.5 \times 10^{10}$ y and  $\simeq 1.5 \times 10^9$ y. This result indicates the potential significance of resonances for tidal evolution for orbital periods up to $\sim 12$  d 
%for Jupiter mass companions if proximity to them can be maintained. }	

\textcolor{blue}{
	\textcolor{RED}{Returning to consideration of Jupiter mass planets,} observations of G type dwarfs in star clusters \citep{Skum72, Smith79} indicate that their rotational velocity on the Main Sequence decreases in time as $v_{eq}= A/\sqrt{t}$ cm/sec, where the proportionality factor $A$ is not accurately known.
	The stellar spin down is thought to be caused by magnetic braking due to the coupling between the stellar magnetic field and the outflowing stellar wind. 
	To investigate  its possible effect on the tidal evolution  we use Skumanich's result in the form
	\begin{equation}
		\drv{\Omega_s}{t}= -\frac{1}{2} \left(\frac{R_*}{A}\right)^2 \,  \Omega^3_s
	\end{equation} 
	and adopt $A = 7.3 \times 10^{13}$ cgs. The extra rate of magnetic down spinning is added to the right side of equation (\ref{DE1}). 
	Figure \ref{eva-MB-frqsh-l=3-n=0}  shows the same $l' = 3$ with  $n_r=0$ resonance passage as mentioned above but now including magnetic braking of the primary star. The system now moves faster into and out of full resonance due to the stronger spin down of the primary whereby in the outer wings of the resonance the magnetic braking contributes about $3/4$  of the total spin down.
	The time scale for stellar spin down at  viscous dissipation rates of $10^{-3}$ and   $2\times10^{-3}$ of the full resonance value, taking into account both magnetic braking and tidal effects, is   $\Omega_s/\dot{\Omega}_s\sim 10^9$ y and $\sim 7 \times 10^8$ y, respectively,  compared to  $\sim 2.5 \times 10^9$ y and $\sim 1.4 \times 10^9$ y, respectively, without magnetic braking.
	Corresponding results for the orbital expansion timescale are $a/\dot{a} \simeq 7.7 \times 10^9$ y and  $\simeq 4 \times 10^9$ y, respectively,
	 without magnetic braking with only marginally smaller timescales when magnetic braking is included. 
	By comparing these results with those in sections (\ref{offresinert}) and (\ref{offresnoninert})  the potential significance of resonances for tidal evolution for orbital periods up to $\sim 12$ d for Jupiter mass companions is evident under the condition that proximity to them can be maintained. }
%Observations of G type dwarfs in star clusters \citep{Skum72} indicate that their rotational velocity decreases in time as $v_{eq}= A/\sqrt{t}$ cm/sec, with $A \sim 7 \times 10^{13}$
%during their Main Sequence life.
 %This is thought to be caused by magnetic braking due to the coupling between the stellar magnetic field and the outflowing stellar wind. 
  %To investigate  its possible effect on the tidal evolution  we use Skumanich's result in the form
%\begin{equation}
%	\drv{\Omega_s}{t}= -\frac{1}{2} \left(\frac{R_*}{A}\right)^2 \, \,\, \Omega^3_s
%\end{equation} 
%and add the extra rate of down spinning to the right side of equation (\ref{DE1}). Figure \ref{eva-MB-frqsh-l=3-n=0}  shows the same $l=3$ with  $n_r=0$ resonance passage as mentioned above but now including magnetic braking of the primary star. The system now moves faster into and out of full resonance   due to the stronger spin down of the primary. This is on account of the faster evolution of $\Omega_s.$  
%\textcolor{red}{The rate of evolution of the orbital separation $a$  in the resonance wings is somewhat faster than  in the case without magnetic breaking. The time scale for spin down at corresponding points in the wings taking into account both magnetic breaking and tidal effects is   $\Omega_s/\dot{\Omega}_s\sim 8 \times 10^8$ y, compared to  $\sim 2 \times 10^9$ y without magnetic braking.}
As results are sensitive to the specification of the way the resonance frequency  changes and the magnetic braking prescription it may be of interest to consider these processes using a more accurate expression for the  changing resonance frequency obtained numerically during the tidal evolution.

\subsubsection{Resonance passage when $l'=5$}
The strongest resonance with $l'=5$ has $n_r=0$ as is the case for $l' = 3.$ As the $l'=5$ resonance is weaker we we expect the resonance passage to be slower.  
Figure \ref{eva-l=5-n=0} shows the results of the calculation with $l' = 5$ corresponding to that  with $l' = 3$ 
illustrated in in Figure \ref{eva-frqsh-l=3-n=0}. A comparison of these indicates the rate of evolution in the wings is slowed down by approximately one order
of magnitude. This is mainly due to the resonance with $l' = 5$ being significantly narrower.  However, evolution  near the centre of resonance is still rapid, thus such a resonance may be significant for tidal evolution if it can be  maintained.

\subsection{Viscous dissipation rate for off-resonant tidal forcing}\label{offresinert}
Up to now we have considered conditions in the neighbourhood of $r$ mode resonances and noted the possibility of significant tidal evolution.
We now consider the tidal evolution expected out of resonance, in particular between the  $l' = 3$ and  $l' = 5$ r mode resonances considered above.

\begin{figure}	
	\includegraphics[width=\columnwidth,height=9cm]{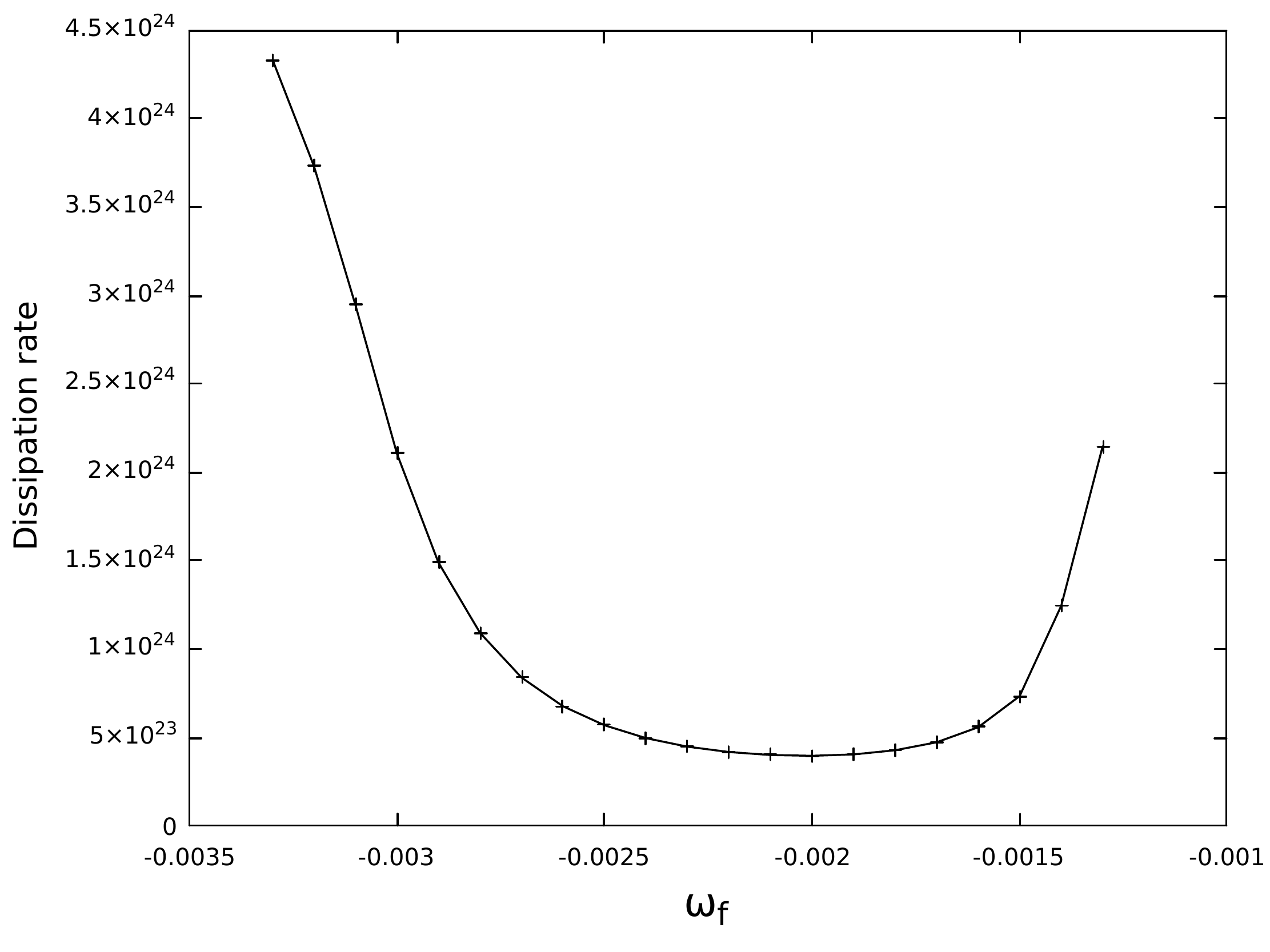}
	\includegraphics[width=\columnwidth,height=9cm]{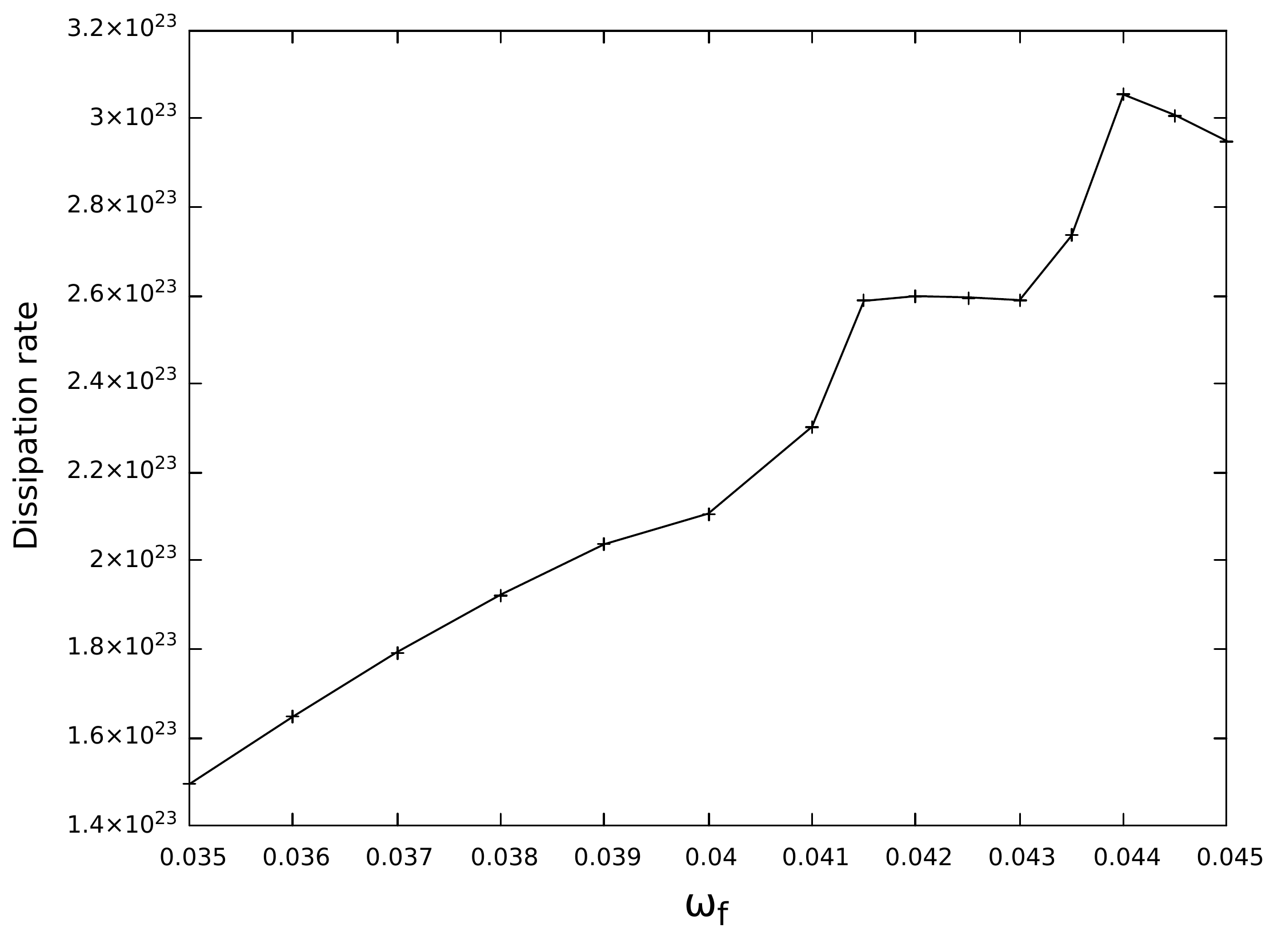}  
	\caption{\textcolor{red}{The \textcolor{ROUGE}{upper}  panel shows the  viscous dissipation rate (erg/s) with $\omega_f$ in the frequency range between the $l'=3$ and $l'=5$ r mode resonances, being inside the inertial range. The \textcolor{ROUGE}{lower}  panel shows the viscous dissipation rate (erg/s) for tidal forcing with frequencies $\omega_f$ outside the inertial range. The forcing frequency, here $>0,$  varies between $3.5$ and $4.5$ times the rotation frequency, the limit of the inertial range being twice the rotation frequency. } }   \label{viscD-off-res} 
\end{figure}

Figure \ref{viscD-off-res} shows the viscous dissipation rate for  forcing frequencies $\omega_f$ in between the $l' = 3$ and  $l' = 5$ r mode resonances. 
The dissipation rate is seen to attain a minimum about half way between the resonances of around $5\times 10^{23}$ erg s$^{-1}.$
The corresponding forcing frequency is \\ $\omega_f=- m\omega_p=-0.002 \, \Omega_c.$
From equation (\ref{DE2}) the time scale for evolution of the % orbital separation 
 \textcolor{ROUGE}{ semi-major axis}  is related to this by
\begin{align}
a\left(\frac{da}{dt}\right)^{-1}  = -\left(\frac{dE_{dissip}}{dt}\right)^{-1}\frac{\omega_p |E_{orb}|}{n_o}\label{DE3}
\end{align}
From (\ref{DE3}) we find that  $a/\dot{a}= 1.25\times 10^{12} \, y$ for an orbital period of $13.4\,d.$
Thus we find, as expected, that the global tidal evolution  significantly exceeds the life time of the system. The time to move between the $l'=3$ and $l'=5$
resonances at $\sim 4\times 10^{10} y$ is still long  compared to the expected age but approaching becoming marginal.

\textcolor{blue}{In contrast to the situation   off resonance,   from }the results given in tables \ref{tabl3} and \ref{tabl5}
\textcolor{red}{we find that   r mode resonances tidally excited in the radiative core of the primary can lead to rapid tidal evolution. However, the large dissipation rate drives the system quickly through the resonance so that the global effects are limited unless mechanisms  act to maintain the system close to resonance (see below).}

\subsubsection{Viscous dissipation rate for tidal forcing outside the inertial range}\label{offresnoninert}
\textcolor{red}{
The off resonant dissipation considered above occurs in the inertial regime such that inertial modes
can be excited in the convective envelope. Here we consider the situation where this does not apply.
Figure \ref{viscD-off-res} shows the viscous dissipation rate for forcing frequencies $|\omega_f| > 2 \,\Omega_s$ outside the inertial range.
The now chosen positive forcing frequency varies from $3.5$ to $4.5$ times the rotation frequency and the viscous dissipation rate varies from $\simeq 1.5 \times 10^{23}$ at orbital period 3.91 d,  to $3 \times 10^{23}$ ergs$^{-1}$  at orbital period 3.31 d. With $m=-2$ the forcing is now prograde. In this frequency range there is some contribution to the tidal response by high radial order ($n \simeq 100 $) $g$ modes which were damped through the  numerical procedure near the stellar centre where their wavelength becomes shorter than the grid spacing.
	From equation (\ref{DE3}) we find that the viscous dissipation in the stellar envelope corresponds to an orbital decay rate $a/\dot{a}= -2.3 \times 10^{13}$ y for $\omega_f =3.5 \times 10^{-2}$ and $-1.4 \times 10^{13}$ y for $\omega_f =4.5 \times 10^{-2}$, again implying that evolution in this regime will be negligible. 
	It is of interest to compare this with the prediction of equation (6.1)
	of \citet{Zahn1977}. According to this for tides exerted on a solar mass star by a Jupiter mass planet with an orbital period of 3.31 d the synchronisation time is $3\times 10^{11}y$ which, although also predicting negligible evolution, is shorter by a factor of \textcolor{green}{$50.$}  However, equation (6.1) of \citet{Zahn1977} does not include the expected but very uncertain reduction in turbulent viscosity resulting from a mismatch between convective and orbital time scales,  see equation (\ref{turbvisc}). This is significant for these short orbital periods and accounts for the discrepancy.}
%	This can account for the discrepancy under an assumption of $\sim 100d$ for the convective time scale, $\tau_c.$ }}

\subsection{The possibility of evolution with resonant interaction maintained}\label{Resmaintained}
For the $r$ mode resonances to be maintained as the system evolves some parameters defining the system 	such as  the total angular momentum, the masses of the components, and/or the moment of inertia of the primary	have to be envisaged to change.
Here for simplicity we shall only consider variation of the total angular momentum.  
Suppose that
resonance is maintained with the mode with $l'=3,|m|=2$ with frequency 
$\omega_p =  -\Omega_s/6$ and hence $ \Omega_s =6n_0/5.$ Using this condition rather
than $\Omega_s = n_o$ the expression for the total angular momentum, $J,$  given by (\ref{Jform})
becomes 	
\begin{align}
	J(a) =\frac{6 \, \mathcal{I}}{5}\frac{\sqrt{G(M_*+M_p)}}{a^{3/2}}+\frac{ M_*M_p}{(M_*+M_p)} \sqrt{G(M_*+ M_p) a}  \label{Jform1} 
\end{align}
This has a minimum when, $a=a_{min},$ given by	
\begin{align}
	a_{min} = \left( \frac{18 \,\mathcal{I} (M_*+M_p)}{5M_*M_p}\right)^{1/2}.
	\end {align}
	For a solar mass primary this corresponds to an  orbital period, $P_{orb,min},$ where  
	%\begin{align}
		%P_{orb,min} = 0.041 \left( \frac{(1+q)^{1/4}}{q^{3/4}}\right) d \label{synccrit1}
	%	\end {align}
\textcolor{RED}{	\begin{align}
P_{orb,min}  = 0.045\left(\frac{\mathcal{I}}{0.0782M_*R_*^2}\right)^{3/4}\frac{(1+q)^{1/4}}{q^{3/4}}
\left (\frac{M_{\odot}}{M_*}\right)^{1/2}\left(\frac{R_*}{R_{\odot}}\right)^{3/2} d
\end{align}}
		
		When, $a>a_{min},$ evolution maintaining resonance requires, $a,$ to increase while the primary spins down.
		Equation (\ref{Jform1}) then indicates that, with other parameters fixed,
		the magnitude of the total angular momentum, $J,$ increases.
		Accordingly this kind of evolution requires angular momentum transfer to the system.
		But note as indicated above that changes of other parameters could have the same effect. 
		
		On the other hand when $a<a_{min},$ evolution maintaining resonance also requires, $a,$ to increase while the primary spins down.
		In this case  Equation (\ref{Jform1}) indicates that, with other parameters fixed,
		the magnitude of the total angular momentum, $J,$ decreases. 
		Some mechanism  for bringing this about such as the effect of a stellar wind needs to be invoked.
		Note that in this case
		$a$ cannot increase beyond $a_{min}.$  As noted above the case $a<a_{min}$ applies to almost all hot Jupiters
		making this form of evolution a possibility at some phase of their lifetime. However, we note that the condition of being close to
		synchronisation requires a rapidly rotating primary, a possibility that may not be realised.  In addition the ability
		to maintain a state of uniform rotation has to be assumed.  
	
\textcolor{blue}{\section{Discussion}\label{Discussion}
		In this paper  we have calculated the tidal response of a rotating solar type primary to 
  the tidal forcing of a
  % binary
  \textcolor{RED}{planetary mass}  companion.
  Although rotation is considered small in the sense that Coriolis forces are retained 
  with centrifugal forces being neglected, we do not make the traditional approximation
  which is not valid in convective regions. Tidal forcing frequencies as seen in the  frame
  corotating with the primary that are smaller in magnitude than $2\Omega_s$, 
  and so are in the inertial regime, are considered \citep{PP81,OL2007}.
  For a first set of detailed calculations,
  we limited consideration to a circular
   %binary
    orbit in the  period range of $10-14d,$  leaving extensions to future work.}

 \textcolor{blue}{ With a view to application to exoplanets we focused on a Jupiter mass companion
  though results can be scaled to apply to  companions of arbitrary mass.
  Resonantly excited modes of oscillation can be important for tidal evolution \citep[see e.g.][]{SP83, WS02, Zanazzi2021}.
  Hence we focused on identifying and calculating the resonant response associated with $r$ modes \citep{PP78} which are
  predominantly sited in the radiative core.  We gave the properties of this spectrum of modes
  for $l'=3$ and $l'=5$ obtained numerically  in Section \ref{l'3res}
  with a semi-analytic treatment in appendix \ref{rmodeapp}.    
 %the  coordinate system adopted given  in Section   \ref{Geometry}.
 %the perturbing tidal potential due to the secondary  \ref{pertpoten}
% induced perturbation to the  external gravitational potential due to the primary  } \ref{pertpot}
%The formulation of calculation of the primary's Response to the perturbing tidal potential 
%due to the secondary given which is given  in Section \ref{pertpoten} 
%is described in Section \ref {Respd}.
%In section {Boundary Conditions} \ref {Bdrycons}
%The numerical solution procedure is then outlined in Section \ref {Numproc}.
%Quantities derived from this include the rate of viscous dissipation 
%assumed to be produced by turbulence in the convective envelope (Section   \ref {viscstress})
%and the imaginary part of the overlap integral determining the tidal torque (Section \ref{Imagovlap}).
%In section {Interpretation of the terms in equation} (\ref{lequmot2})
%Numerical results  are then given in Section \ref{Results}.
%These include the determination of the $r$ mode resonances associated with $l'=3$ and $l'=5$  in Sections \ref{Findr}-
% \ref{l'5res}. For each of these values there is a spectrum of modes with increasing number of radial nodes 
 %that is closely separated in eigenfrequency
%(Section \ref{rmodespec}). A semi-analytic treatment of the origin of these spectra 
%is given in appendix \ref {rmodeapp} and successfully compared with our results. These modes are for the most part cited in the
%radiative core. 
We described the response of the
convective envelope, where most of the dissipation occurs through the action of turbulent viscosity in Section \ref{convResp}.
 A semi-analytic discussion applicable in the low tidal forcing frequency
limit is given in appendix  \ref{ConvResp} which relates to the behaviour of the horizontal displacements
and an incipient critical latitude phenomenon.}

\textcolor{blue}{We formulated the effects of the tidal response on the orbital and spin evolution
of the system in Sections \ref{orbspinevol} - \ref{orbevol}.
%In section {Torque and dissipation resulting from tidal forcing} \ref{TorandD}
%In section {Orbital evolution} \ref{orbevol}
The effect of r mode resonances  is to  greatly speed up the orbital evolution  
over  a narrow frequency range in their vicinity and  their effect on orbital evolution is limited
if the primary's structure is fixed and the total angular momentum is conserved.}

\textcolor{blue}{We described the  results of numerical calculations of spin and orbit evolution in Section \ref{Numevol}.
Non resonant tidal forcing between the $l'=3$ and $l'=5$ resonances which lies within in the inertial regime gives rise to orbital evolution times $\sim  10^{12}y$ which greatly exceeds potential system lifetimes. However, the  evolution time 
between the $l'=3$ and $l'=5$ resonances with are separated by only $ \delta a \sim 0.07 a$ approaches 
this to within an order of magnitude for orbital periods $\sim 13.4 d.$}

\textcolor{blue}{Non-resonant tidal forcing at higher frequencies lying  outside the inertial regime
 is even slower occuring on a time scale \textcolor{red}{ $ \sim 2 \times 10^{13}y$} ( see  Sections  \ref{offresinert}
%In \textcolor{red}{ section{Viscous dissipation rate for tidal forcing outside inertial range}
and \ref{offresnoninert}).  As rotation is expected to have  a small effect in this regime,  with making allowance for reductions
in turbulent viscosity due to frequency mispatch,
the result  was found to be consistent with the results of \citet{Zahn1977} for a non rotating star.}

\textcolor{blue}{Significant orbital evolution over a realistic lifetime of the system can only be found
if  resonant interaction \textcolor{RED}{that provides enhanced tidal interaction } is maintained. 
\textcolor{RED}{ Such resonance locking has been invoked as a method of speeding up tidal evolution
in a number of astrophysical contexts including binary star and exoplanet systems \citep[see][]{ F2021, Zanazzi2021}.   
  For a discussion of the process of resonance locking see e.g. \citet{F2016}.}
%\citep{ WS02, Zanazzi2021}.
Resonance locking   \textcolor{RED}{either requires a substantial orbital
eccentricity \citep{WS99}} or   parameters defining the system which could be  the total angular momentum
or the stellar moment of inertia to change.}

Such changes could be produced by \textcolor{RED}{well established phenomena such as}
 magnetic braking  \textcolor{RED}{ in the former case} or the effects of stellar evolution \textcolor{RED}{in the latter.}
For systems such as hot or warm Jupiters it was found that significant evolution within the system lifetime could occur at a separation from the resonance centre where the dissipation rate was $1000$ times smaller than at the peak and the linear response plausibly applicable. 
For these systems there is the possibility of sustained tidal evolution with increasing orbital period and maintained proximity to resonance driven by angular momentum loss due to a stellar wind at some phase of their lifetime (Section  \ref{Resmaintained}), provided the primary spins rapidly enough initially and the orbital period is. \textcolor{RED}{ such that 
% $<  0.041(1+q)^{1/4}/q^{3/4} d$
\begin{align}
P_{orb}  <  P_{orb,min}=0.045\left(\frac{\mathcal{I}}{0.0782M_*R_*^2}\right)^{3/4}\frac{(1+q)^{1/4}}{q^{3/4}}\left (\frac{M_{\odot}}{M_*}\right)^{1/2}\left(\frac{R_*}{R_{\odot}}\right)^{3/2} d.\label{Porbmin}
\end{align}}
The extent of this is dependent on the angular \textcolor{green}{momentum} loss mechanism (see Section  \ref{Resmaintained}). 
\textcolor{RED}{When the inequality in (\ref{Porbmin}) reversed resonance locked evolution requires the  orbital period and the system angular momentum to increase with time as could possibly be  driven by mass accretion.}

\textcolor{RED}{Processes of the type described above are expected to be effective only when resonance locking is sustained on account of  evolutionary changes
to the system such as angular momentum loss through a wind,  and the central star is rapidly rotating.  Systems in which they
are currently  operating are likely to be young with
the central stars  possessing a significant radiative core.}

\textcolor{RED}{In addition an  expected observational signature should be that the stellar rotation period is shorter than the orbital period
and satisfies $(5P_{rorb})/6 \le P_{rot} \le P_{orb},$ allowing a resonance condition to be satisfied. Then the system
loses/gains angular momentum according to whether the inequality (\ref{Porbmin}) is satisfied/not satisfied.}

%\textcolor{RED}{A tentative candidate system is CI Tauri   \citep[see][]{JK2016} which has $M_*=0.8M_{\odot}, $ $R_*=1.4R_{\odot}$
% with an estimated age
%of $2\times 10^6 y$. The planet mass is $M_p = 11.3M_J$ for which
%our response calculations can be scaled and would indicate comparable torques to the $1M_J$ case
%somewhat further away from resonance centres. The orbital period is close to $9d$ with the stellar rotation period being
%highly uncertain but $<9d$ with one estimate at $~7d.$ On the other hand, assuming the star has a significant
%radiative core,  the resonance condition suggest $7.5d < P_{rot}<  9d.$
%The inequality (\ref{Porbmin}), evaluated assuming the first term in brackets on the right hand side is unity,  indicates increasing angular momentum of the system which could arise from mass accretion onto the central star or planet.}

\textcolor{RED}{A  tentative candidate system is Kepler 1643 \citep[see][]{Bouma2022}.
This has $M_*= 0.92M_{\odot}$,  $R_* = 0.88R_{\odot},$ with a rotation period $P_{rot}=5.106d, $
and an estimated age $\sim 4\times 10^7 y.$ The orbital period is $5.3426d.$
The mass of the planet described as a mini-Neptune is uncertain, accordingly 
for illustrative purposes we adopt the characteristic value $M_p= M_J/30.$
The condition,\\
 $(5P_{rorb})/6 \le P_{rot}\le P_{orb},$ is satisfied with the  expected resonance for $l'=5,$ and $n_r=0$ being close to
$P_{rot}=4.99d.$ Noting theoretical and observational uncertainties arising from effects such as the relation of surface to interior rotation,
this is in reasonable agreement with the quoted value. In this case the inequality (\ref{Porbmin}), 
 evaluated here and below assuming the first term in brackets on the right hand side is unity, indicates the system should be 
losing angular momentum possibly due to a stellar wind.}
 %This is close to the quoted value%Resonances with $l'=7$  have $P_{rot}$ closer to the orbital period with $P_{rot}=5.15d.$

\textcolor{RED}{Another system in which tidal evolution with resonance locking may have occurred is\\  COROT~-~4 \citep[see][]{M2008}.
This has $P_{orb}=9.20d,$  $M_p=0.72M_J$ with  $M_*=1.16M_{\odot},$  $R_*=1,17R_{\odot}.$ 
and  $P_{rot}=8.87\pm 1.12d.$  With an estimated age  $\sim 1Gy,$ angular momentum loss 
through a wind may have slowed significantly.  The inequality (\ref{Porbmin}) is satisfied with $P_{orb} < 13.44d $
so the resonance condition should be maintained. The $l' = 3 $ resonance is near $P_{rot}= 5/6P_{orb}=7.67d,$
and the $l'=5$ resonance is near to $P_{rot}=14P_{orb}/15 = 8.59d$ both of which could be possible in view of the uncertainties.
In particular resonance locking may explain why this system has been able to attain a near synchronous state
at its relatively long orbital period.}
  
In this paper we have concentrated on aligned systems in circular \textcolor{blue}{ orbits over a limited period range.
The effects of  resonant $r$ mode excitation found here should be more pronounced at shorter periods making  a detailed survey is of interest. 
In addition,} the extension to misaligned systems is important in the context of
the possibility of tidal alignment of hot Jupiter orbits, in particular whether this happens
 more rapidly than  tides acting in aligned systems can operate. 
The existence of $r$ modes with $l'= |m|=1$ is promising as these may always be close to resonance.
\textcolor{blue}{These aspects} will be the subject of future work.

\section{Data Availability}
The data underlying this article will be shared on reasonable request to the corresponding author.

%\clearpage

%\subsection{Gauss equation 5}
%\begin{align}
%&\frac{d \Omega}{dt}=
%-F_{\perp}
%\frac{R\cos(\Phi+\gamma)}{\sin(i)\sqrt{G(M_p+M_*)a(1-e^2)}}
%\end{align}

\begin{appendix}

	\section{The components of the viscous stress tensor}\label{stresscomp}
	Recalling that it is symmetric, the components of the viscous stress tensor, $\bsf{\Sigma},$  expressed in spherical coordinates 
	in terms of the components of the  associated displacement vector, $\mbox{{\boldmath$\xi$}} ,$ are given by 
	\begin{align}
		&\bsf{\Sigma}_{r r} = 2 \rm i \, \omega_f \, \rho \nu \left[ \pdrv{\xi_r}{r} - \frac{1}{3} \nabla \cdot \mbox{{\boldmath$\xi$}} \right]\nonumber\\
		&\bsf{\Sigma}_{\theta,\theta} =  2 \rm i \, \omega_f \, \rho \nu \left[ \left( \frac{1}{r} \pdrv{\xi_\theta}{\theta} +\frac{\xi_r}{r} \right)  - \frac{1}{3} \nabla \cdot \mbox{{\boldmath$\xi$}} \right]\nonumber\\
		&\bsf{\Sigma}_{\varphi,\varphi} = 2 \rm i \, \omega_f \, \rho \nu \left[ \frac{1}{r \, \sin \theta} \pdrv{\xi_{\varphi}}{\varphi} + \frac{\xi_r}{r} + \frac{\xi_{\theta}}{r \, \tan \theta} - \frac{1}{3} \nabla \cdot \mbox{{\boldmath$\xi$}} \right]\nonumber\\
		&\bsf{\Sigma}_{r \theta}= \rm i \, \omega_f \, \rho \nu \left[\pdrv{\xi_\theta}{r} +\frac{1}{r} \pdrv{\xi_r}{\theta}-\frac{\xi_\theta}{r} \right]\nonumber\\
		&\bsf	{\Sigma}_{\theta \varphi}=\rm i \, \omega_f \, \rho \nu \left[\frac{1}{r \, \sin \theta} \pdrv{\xi_\theta}{\varphi} + \frac{1}{r} \pdrv{\xi_\varphi}{\theta} -\frac{\xi_{\varphi}}{r \, \tan\theta}\right]\nonumber\\
		&\bsf{\Sigma}_{\varphi r} = \rm i \, \omega_f \, \rho \nu \left[\pdrv{\xi_\varphi}{r} +\frac{1}{r \sin \theta} \pdrv{\xi_r}{\varphi} - \frac{\xi_\varphi}{r} \right]\nonumber
	\end{align}	
	\subsection{The divergence of the tensor, $\bsf{\Sigma} .$ in spherical coordinates}
	The divergence of $\bsf{\Sigma} $ is required to evaluate the viscous force per unit mass associated with the displacement
	$\bmth{\xi}$ indicated above. This is given by
	\begin{align}
		\nabla\cdot \bsf{\Sigma} =
		&\left( \frac{1}{r^2}\frac{\partial}{\partial r} (r^2\bsf{\Sigma} _{r,r} )+\frac{1}{r\sin\theta}\frac{\partial}{\partial \theta} (\sin\theta\bsf{\Sigma} _{r,\theta} )
		+\frac{1}{r\sin\theta}\frac{\partial}{\partial \phi} (\bsf{\Sigma} _{r,\phi} )-\frac{(\bsf{\Sigma} _{\theta,\theta}+\bsf{\Sigma} _{\phi,\phi})}
		{r}\right){\hat {\bf r}}\nonumber\\
		&+\left( \frac{1}{r^2}\frac{\partial}{\partial r} (r^2\bsf{\Sigma} _{r,\theta} )+\frac{1}{r\sin\theta}\frac{\partial}{\partial \theta} (\sin\theta\bsf{\Sigma} _{\theta,\theta} )
		+\frac{1}{r\sin\theta}\frac{\partial}{\partial \phi} (\bsf{\Sigma} _{\theta,\phi} )+\frac{(\bsf{\Sigma} _{r,\theta}-\bsf{\Sigma} _{\phi,\phi}\cot\theta)}{r}\right)
		{\hat {\bf \mbox{\boldmath$\theta$}}}\nonumber\\
		&+ \left(\frac{1}{r^2}\frac{\partial}{\partial r} (r^2\bsf{\Sigma} _{r,\phi} )+\frac{1}{r\sin\theta}\frac{\partial}{\partial \theta} (\sin\theta\bsf{\Sigma} _{\phi,\theta} )
		+\frac{1}{r\sin\theta}\frac{\partial}{\partial \phi} (\bsf{\Sigma} _{\phi,\phi} )+\frac{(\bsf{\Sigma} _{r,\phi}+\bsf{\Sigma} _{\theta,\phi}\cot\theta)}{r}\right)
		{\hat {\bf \mbox{\boldmath$\phi$}}}\nonumber
		\end {align}
		where, ${\bf \mbox{\boldmath$r$}}, {\bf \mbox{\boldmath$\theta$}}, {\bf \mbox{\boldmath$\phi$}}$, respectively denote unit vectors in the, $(r,\theta,\phi),$ directions

		\section{The   {\MakeLowercase {r}}  mode  spectrum in radiative regions in  the low  frequency adiabatic limit}\label{rmodeapp}
		We begin by formulating the basic equations governing the tidal response and then proceed
		to consider their solution in the limit that the forcing frequency as seen in the rotating frame
		is small while being comparable in magnitude to the rotation frequency \textcolor{blue}{ while adopting the Cowling approximation in which
			response perturbations to the gravitational potential are neglected.}
		\subsection{Linearised equation of motion}
		The linearised equation of motion governing the tidal response given by equation (\ref{lequmot})  in the limit  that perturbations are assumed
		to  be adiabatic  ( the RHS of equation (\ref{Nonad}) set to zero ) \textcolor{blue}{ under the Cowling approximation in which perturbation to the 
			gravitational potential is neglected } is
		\begin{align}
			\hspace{0cm}& -{ \omega_f}^2{\mbox{\boldmath$\xi$}}+2{\rm i}{ \omega_f}{\mbox{\boldmath$\Omega_s$}\times}{\mbox{\boldmath$\xi$}}
			=-\frac{F}{\rho}\nabla W - N^2\left(\xi_r+\frac{ U}{g}\right){\bf {\hat {r}}}
			\label{eqmot}
		\end{align}
		Here for ease of notation the subscripts, $m,$ and $\sigma,$ have
		been dropped from the forcing potential ${ U},$  ${\mbox{\boldmath$\xi$}}\equiv (\xi_r,\xi_{\theta},\xi_{\phi})$
		and other perturbations\footnote{We recall that the $\phi$ dependence of these quantities is through a factor $\exp({\rm i}m\phi).$
			But here  we shall allow the $\theta$ dependence of the forcing potential , $U,$ to be through a factor, $Y_{l,m},$ for general $l\ge2.$} and will be taken as read in this and subsequent appendices. The quantity
		$W = (P'+\rho U)/F,$ with $d\ln F/dr= (d\ln P/dr)/\Gamma_1,$\\
		%P^{1/\gamma}, $ 
		%with perturbations to quantities being denoted with a prime. 
	        \textcolor{blue}{The square of the buoyancy frequency is
		$N^2= g(1/(\Gamma_1 P)dP/dr- (1/\rho) d\rho/dr),$  $g$ is the acceleration due
		 to gravity,}
		and the adiabatic
		exponent $\Gamma_1= (\rho/P)d\ln P/d\ln\rho.$ 
		We recall that ${\omega_f}= \sigma +m\Omega_s$
		is the forcing frequency as seen in the rotating frame.
		%The perturbing potential is assumed to be of the form $\Phi' ={\cal C} r^2 Y_2^m(\theta,\phi)\exp({\rm i}{\omega_0}}t)$  where ${\cal C}$ is a constant.
	We remark that the quantity $\xi_{r,eq} =- U/g$ is the standard equilibrium tide.
	
	The components  of the linearised equation of motion in spherical polar coordinates are
	\begin{align}
		\hspace{0cm}& -{ \omega_f}^2\xi_r-2{\rm i}{ \omega_f}\Omega_s\sin\theta\xi_{\phi}                       
		=-\frac{F}{\rho}\frac{\partial W}{\partial r} - N^2\left(\xi_r+\frac{ U}{g}\right) \label{rcomp}\\
		\hspace{0cm}& -{\omega_f}^2\xi_{\theta}-2{\rm i}{ \omega_f}\Omega_s\cos\theta\xi_{\phi}                       
		=-\frac{F}{\rho r}\frac{\partial W}{\partial \theta} \label{thetacomp}\\
		\hspace{0cm}& -{\omega_f}^2\xi_{\phi}+2{\rm i}{ \omega_f}\Omega_s(\sin\theta\xi_{r}+\cos\theta\xi_{\theta})                       
		=-\frac{{\rm i} mWF}{\rho r \sin\theta}
		\label{phicomp}
	\end{align}
	\subsubsection{Decomposition into spheroidal and toroidal components}
      \textcolor{blue}{ We write the radial}  displacement as the 
	sum of equilibrium tide value and a correction $\eta_r,$ \textcolor{blue}{and
	without}  loss of generality  write the horizontal  displacement as the sum of spheroidal and toroidal contributions. Thus
	\begin{align}
		\hspace{0cm}& \xi_r 
		%= \Xi  
		= \xi_{r,eq}+\eta_r\label{rdecomp}\\
		\hspace{0cm}& \xi_{\theta}                      
		=\frac{1}{r\sin\theta}\frac{\partial {\cal T}}{\partial \phi}    + \frac{1}{r} \frac{\partial S}{\partial \theta}\equiv
		\frac{{\rm i}m}{r\sin\theta} {\cal T}    + \frac{1}{r} \frac{\partial S}{\partial \theta}\label{thetadecomp}\\
		\hspace{0cm}& \xi_{\phi}                      
		=-\frac{1}{r}\frac{\partial {\cal T}}{\partial \theta}    + \frac{1}{r\sin\theta} \frac{\partial S}{\partial \phi}\equiv
		-\frac{1}{r}\frac{\partial {\cal T}}{\partial \theta}    + \frac{{\rm i}m}{r\sin\theta}  S
		\label{phidecomp}
	\end{align}

	Making use of  (\ref{thetadecomp}) and (\ref{phidecomp})  we may eliminate $\xi_{\theta}$ and $\xi_{\phi}$ 
	in (\ref{thetacomp}) and (\ref{phicomp}) in favour of $S$ and ${\cal T}$ and then eliminate $W.$
	This leads to the equation 
	\begin{align}
		%\hspace{-0cm}&\frac{{\omega_f}^2}{r^2}\nabla^2_{\perp}S +\frac{2m\Omega_s{\omega_f}}{r^2}S
		%-\frac{2{\rm i}{\omega_f}\Omega_s\cos\theta}{r^2}\nabla^2_{\perp}{\cal T} + \frac{2{\rm i}\Omega_s{\omega_f}\sin\theta}  {r^2}
		%\frac{\partial {\cal T}}{\partial\theta} - \frac{F}{\rho r^2}\nabla^2_{\perp}W= -\frac{2m\Omega_s{\Omega_s}}{r^2}\xi_r\label{ST1}\\
		\hspace{-0cm}&\frac{{\omega_f}^2}{r^2}\nabla^2_{\perp}{\cal T} +\frac{2m\Omega_s{\omega_f}}{r^2}{\cal T}
		+\frac{2{\rm i}{\omega_f}\Omega_s\cos\theta}{r^2}\nabla^2_{\perp}S - \frac{2{\rm i}\Omega_s{\omega_f}\sin\theta}  {r^2}
		\frac{\partial S}{\partial\theta} = - \frac{2{\rm i}{\omega_f}\Omega_s}{ r\sin\theta}\frac{\partial (\sin^2\theta \xi_r)}{\partial \theta}.\label{ST2}
	\end{align}
	which relates ${\cal T}$ to  $S$ and $\xi_r.$ Here $\nabla^2_{\perp}{\cal Q}$ for some quantity ${\cal Q}$ is defined through
	\begin{align}
		\nabla_{\perp}^2 {\cal Q} = \frac{1}{\sin\theta}\frac{\partial}{\partial \theta} \left(\sin\theta\frac{\partial {\cal Q}}{\partial \theta}\right)
		-\frac {m^2 {\cal Q}}{\sin^2\theta}\nonumber
	\end{align}

	\noindent In addition we may make use of  (\ref{thetadecomp}) and (\ref{phidecomp})
	to find an expression for  $F/(\rho r)\nabla^2_{\perp} W$ which is thus found to be given by
%	\begin{align}
%		\frac{F}{\rho r}\nabla^2_{\perp} W=\frac{\omega^2_f}{r}\nabla^2_{\perp} S+2\Omega\omega_f{\rm i}\left(\frac{\mu}{r}\nabla^2_{\perp}{\cal T} 
%		-\frac{(1-\mu^2)}{r}\frac{\partial {\cal T}}{\partial\mu}   +\frac{\partial}{\partial \phi}\left( \xi_r+ \frac{S}{r} \right)\right).\label{RV1}
%	\end{align}
		\begin{align}
		\frac{F}{\rho r}\nabla^2_{\perp} W=\frac{\omega^2_f}{r}\nabla^2_{\perp} S-2\Omega\omega_f{\rm i}\left(\frac{\mu}{r}\nabla^2_{\perp}{\cal T} 
		+\frac{(1-\mu^2)}{r}\frac{\partial {\cal T}}{\partial\mu}   +\frac{\partial}{\partial \phi}\left( \xi_r+ \frac{S}{r} \right)\right).\label{RV1}
	\end{align}
	Using this together with (\ref{rcomp}) enables a second relation between $\xi_r,$ $S,$ and ${\cal T}.$
	To complete the system we need to make use of the continuity equation and adiabatic condition.

	\subsection{Linearised continuity equation and adiabatic  condition}
	The linearised continuity equation is 
	\begin{align}
		\hspace{-0cm}&\rho'  = -\nabla\cdot(\rho{\mbox{\boldmath$\xi$}})\label{contimuity}
	\end{align} 
	and the adiabatic condition is
	\begin{align}
		\hspace{-0cm}&P'  = WF -\rho U
		%P^{1/\gamma} 
		= -\frac{\Gamma_1 P}{F}\nabla\cdot(F{\mbox{\boldmath$\xi$}})
		\label{addy}
	\end{align} 
	Making use of (\ref{thetadecomp}) and (\ref{phidecomp}) equation (\ref{addy}) may be written as
	\begin{align}
		\hspace{-0cm}& WF -\rho U =
		\frac{\Gamma_1 P}{r^2}\left(
		\frac{1}{F}\frac{\partial(r^2 F Ug^{-1})}{\partial r}
		- \frac{1}{F}\frac{\partial(r^2 F\eta_r)}{\partial r} 
		-\nabla_{\perp}^2 S\right).
		\label{addy1}
	\end{align}
	%where
	%\begin{align}
	%	\nabla_{\perp}^2 S = \frac{1}{\sin\theta}\frac{\partial}{\partial \theta} \left(\sin\theta\frac{\partial S}{\partial \theta}\right)
	%	-\frac {n^2 S}{\sin^2\theta}
	%\end{align}

	\subsection{The tidal response in the radiative zone in the limit of small forcing and rotation frequencies}
	%In the limit of vanishing ${ \omega_f},$ with $\Omega_s$ being of the same order,
	%it can be seen from (\ref{rcomp}) - (\ref{phicomp}) that
	%the displacement ${\mbox{\boldmath$\xi$}},$ and thus, $S,$ and, ${\cal T},$ remain finite while $W$ which scales
	%as,  $\omega_f^2,$ vanishes.
	%In a model for which there is a radiative interior and a convective envelope these two regions
	%are treated differently. 
	
	%\subsubsection{The radiative region}
	In this region  in the limit, ${\omega_f}\rightarrow 0,$  given that, $N^2 > 0,$ remains finite,
	equation (\ref{rcomp}) implies that \newline $\eta_r \rightarrow 0,$ and so this can be regarded as small 
	in comparison to the horizontal components of the displacement.
	%\noindent 
%	Inspection of equation (\ref{RV1})  indicates that, assuming global angular scales, the contribution of.  $WF,$  in comparison to the term involving $\nabla_{\perp}^2S$  in equation (\ref{addy1})  is of order
\textcolor{blue} {Inspections of equations(\ref{thetacomp}) and (\ref{phicomp}) indicates that the the contribution of.  $WF,$  in comparison to
the terms involving  horizontal displacements in (\ref{addy}) is of order} 
	$r^2\rho\omega_f^2/(\Gamma_1 P)$ 
	as  ${ \omega_f}\rightarrow 0,$ This suggests we can neglect $W$ in that equation in that  limit   corresponding
	to an anelastic approximation.
	
	We remark that in order to then proceed with (\ref{addy1})  \textcolor{blue}{ while retaining $\eta_r$ there
	  when this quantity} varies on a global length scale,
	we formally require a hierarchical ordering  $\omega_f^2 \ll N^2 \ll \Gamma_1P/(\rho r^2).$
	However, excessive values of $N^2$ can be compensated for by considering
	$\eta_r$ that vary on small scales.   
	
	%For this discussion  to be valid we clearly require $N^2 \gg {\bar \omega_f}^2.$
	\subsection{Determination of $S$ and  ${\cal T}$ in the limit of low forcing and rotation frequencies }\label{DetS}
	After making the anelastic approximation we write
	\begin{align} 
		S= S_1 + S_2, \hspace{3mm} {\rm where}\label{S12}
	\end{align} 
	\begin{align}
		\hspace{-0cm}&  r^2\rho{ U} =
		\Gamma_1 P\left( -\frac{1}{F}\frac{\partial(r^2 F{ U}g^{-1})}{\partial r} +\nabla_{\perp}^2 S_1\right).
		\label{addy2}
	\end{align}
	Equation (\ref{addy1}) then implies that in the anelastic limit
	\begin{align}
		%\hspace{-2cm}&  r^2\rho{\bar U} =
		%\Gamma_1 P\left( 
		\nabla_{\perp}^2 S_2= -\frac{1}{F}\frac{\partial(r^2 F\eta_r)}{\partial r}.
		%\right).
		\label{addy2a}
	\end{align}
	Corresponding to (\ref{S12}) we write ${\cal T}={\cal T}_1+ {\cal T}_2,$ and $ W= W_1 + W_2.$ Equations
	for $S_i, {\cal T}_i, W_i,$ can be obtained from equations
	% (\ref{thetacomp}), (\ref{phicomp}), (\ref{thetadecomp}), and  (\ref{phidecomp}) 
	(\ref{ST2}) and (\ref{RV1}) through the replacements 
	$S\rightarrow S_i, {\cal T} \rightarrow {\cal T}_i, W \rightarrow W_i,$  with $\xi_r \rightarrow \xi_{r,eq}$ for $i=1$ and 
	$\xi_r\rightarrow \eta_r$ for $i=2.$
		The corresponding horizontal displacement components are $(\xi_{\theta,i}, \xi_{\phi,i}).$
	\textcolor{blue}{ Following the traditional approximation in the low forcing frequency and rotation limit,
	we shall subsequently neglect  $\eta_r$ in (\ref{ST2}) and (\ref{RV1})}
	but importantly not in (\ref{addy2a}).  
	We. remark that if $\eta_r$ were to be  neglected in (\ref{addy2a}) as well and thus throughout,  then ${\cal T}_2=S_2=W_2=0.$  
	
	\subsubsection{ Successive approximation}
	The above scheme results in ${\cal T}_1$ and $S_1$ being determined by the equilibrium tide 
	(which we call the lowest order approximation) and $\eta_r$ being subsequently determined
	as a correction in the manner described below. Before doing this we remark that the $\eta_r$ so determined
	can be used to determine an approximation to, ${\cal T}_2$ and $W_2$ from the appropriate forms of  equations
	(\ref{ST2}) and (\ref{RV1}). For tidal forcing $\propto Y_{l,m}(\theta,\phi),$ ${\cal T}_1$ (with $\eta_r$  neglected)
	will be found to involve spherical harmonics of degrees $l\pm1$
	(providing of course that  in the case of the smaller alternative, $l-1 \ge |m|$).  
	We can then see from the appropriate form of equation (\ref{RV1}) that $W_1$ will involve spherical harmonics of degrees $l\pm 2.$
	Equation (\ref{rcomp}) then indicates that these values of $l$ will appear in an expansion of $\eta_r$ in spherical harmonics.
	These  in turn will generate an additional component of degree $l+3$ in the expansion of ${\cal T}_2$ that is potentially resonant.
	Proceeding with successive approximations we see that we can expect toroidal mode resonances with all possible values of $l$
	with the opposite parity to that of the original forcing potential.  Having noted this we now focus on the simplest
	cases with degrees $l\pm1$ as these can be studied at the lowest order of approximation
	
	\subsection{Determination of the toroidal mode response assuming the equilibrium tide for radial motions}
	Here we now determine $S_1$ and ${\cal T}_1$ in the lowest order approximation as indicated above.

	\subsubsection{Calculation of $S_1$}
	
	As $ U$ is known  equation (\ref{addy2}) can  be used to determine $S_1.$
	For the case of interest, for which
	$ U = r^l Y_{l,m}(\theta,\phi)\exp({\rm i}{{\omega_f}}t),$ \footnote{\textcolor{blue}{For convenience and without loss of generality, we replace  amplitude factors such as, $c_{tid},$ by unity. In addition it is a simple matter to repeat the discussion replacing
	 the factor $r^l$ by a general function of $r.$} \\}  we obtain
	\begin{align}
		\hspace{-2mm}&S_1 ={\cal S}_1 Y_{l,m}(\theta,\phi)\exp({\rm i}{{\omega_f}}t), \hspace{3mm} {\rm where}\label{Sdef}
	\end{align} 
	\begin{align}
		\hspace{-0cm}&  r^{l+2}\rho =
		- \Gamma_1 P\left( \frac{1}{F}\frac{\partial(r^{l+2} F g^{-1})}{\partial r} + l(l+1){\cal S}_1\right).
		\label{addy3}
	\end{align}
	Making use of hydrostatic equilibrium this may be reduced to the simple expression
	\begin{align}
		\hspace{-2cm}&  {\cal S}_1 =
		-  \frac{1}{l(l+1)}\frac{d (r^{l+2} g^{-1})}{d r} .
		\label{addy4}
	\end{align}
	\subsubsection{Calculation of ${\cal T}_1.$}\label{T1calc}
	We now proceed to the determination of ${\cal T}_1$ from  the adapted form of equation (\ref{ST2}).
	this may be written in the form
	\begin{align}
		\hspace{-0cm}&\frac{{\omega_f}^2}{r^2}\nabla^2_{\perp}{\cal T}_1 +\frac{2m\Omega_s{\omega}_f}{r^2}{\cal T}_1
		=R_T, \hspace{3mm} {\rm where}\label{ST3}
	\end{align}
	\begin{align}
		\hspace{-0cm}&R_T
		=-\frac{2{\rm i}{\omega}_f\Omega_s\cos\theta}{r^2}\nabla^2_{\perp}S_1 + \frac{2{\rm i}\Omega_s{\omega}_f\sin\theta}  {r^2}
		\frac{\partial S_1}{\partial\theta} -\frac{2{\rm i}{ \omega_f}\Omega_s r^{l-1}}{ g}\frac{\partial ((1-\mu^2) Y_{l,m})}{\partial \mu}.\label{ST4}
	\end{align}
	Making use of (\ref{Sdef}) and (\ref{addy2}) we find
	%  \begin{align}
		%\hspace{-0.1cm}&R_T\exp(-{\rm i}{\bar\sigma}t)
		%=l(l+1){2{\rm i}{\bar\sigma}\Omega\mu}{\cal S }Y_2^m
		%- {2{\rm i}\Omega{\bar\sigma}(1-\mu^2) }{\cal S}\frac{\partial Y_2^m}{\partial\mu} 
		
		%- 2{\rm i}{\bar \sigma}\Omega  {\cal C}\frac{r^3}{g} \frac{\partial ((1-\mu^2) Y_2^m)}{\partial \mu}.\label{ST4}
		%\end{align}
		\begin{align}
			\hspace{-0.1cm}&R_T\exp(-{\rm i}{\omega}t)
			=- {2{\rm i}{\omega_f}\Omega_s}     \frac{d (r^l g^{-1})}{d r} \mu Y_{l,m}
			+ 2{\rm i}\Omega_s{\omega_f}\left(   \frac{1}{l(l+1)r^2}\frac{d (r^{l+2} g^{-1})}{d r}
			- \frac{r^{l-1}}{g} \right) (1-\mu^2)\frac{\partial Y_{l,m}}{\partial \mu}.\label{ST41}
		\end{align}
		
		We solve (\ref{ST3}) by writing ${\cal T}_1$ as a generic sum over spherical harmonics. Thus
		\begin{align}
			{\cal T}_1 = r^2\exp({\rm i}{{\omega_f}}t) \sum_{l'=|m|}^{\infty}{\cal D}_{l'}(r) Y_{l',m}(\theta,\phi), \hspace{3mm} {\rm where}\label{Tser}
		\end{align}
		the coefficients ${\cal D}_{l'}$ are  determined by solving (\ref{ST3}).  \textcolor{blue}{ By inspection of of (\ref{ST41}) %  and (\ref{addy2})
		and making use of} well known properties of spherical harmonics \citep[see][]{AS} \footnote{\textcolor{blue}{Here we refer to the expression of $\mu Y_{l,m}$
		and $(1-\mu^2)dY_{l,m}/d\mu$ as a linear combination of spherical harmonics. These can be found from corresponding relations for Legendre functions.}}
		we find that
				for a given, $l,$ only the terms with $l'=l-1$ and $l'=l+1$ are non zero.  
				In particular the non vanishing expansion coefficients are given by

	%			****WORKING******
				
	%			We use
	%			\begin{align}
	%			&\mu Y_{l,m}=\alpha_{l,m}Y_{l+1,m}+\beta_{l,m}Y_{l-1,m}\nonumber\\
	%		&-(1-\mu^2)\frac{dY_{l,m}}{d\mu}= l\alpha_{l,m}Y_{l+1,m}-(l+1)\beta_{l,m}Y_{l-1,m}
	%		\end{align}
	%		         to show that
	%			\begin{align}
	%			&R_T\exp(-{\rm i}\omega_f t)=- {2{\rm i}\omega_f\Omega_s}\left(\frac{l-1}{l}\left(  r^l d  g^{-1}/d r + 2(l+1)r^{l-1}/g  \right)   \beta_{l,m}Y_{l-1,m}\right.+\nonumber\\
	%			&\left. \frac{(l+2)}{(l+1)}\left(  r^ld  g^{-1}/d r + r^{l-1}/g  \right)   \alpha_{l,m}Y_{l+1,m}\right)
	%			\end{align}
				
	%			To get answer divide $l'$ term by
	%			$$-\omega_f^2l'(l'+1)\left (1- \frac{2m'\Omega_s}{\omega_fl'(l'+1)}\right)$$
				
			%	*******OLD FORMULAE*********
			%	\begin{align}
			%		\hspace{-0cm}& {\cal D}_{l-1} =-\frac{ {2{\rm i}\Omega_s}\left(  (l+1)r^2 d  g^{-1}/d r + (l^2+3l-2)r/g  \right)   \beta_{l,m}}
			%		{{\omega}_fl^2(l-1)\left( 1-2m\Omega_s/({\omega_f}l(l-1) \right)}\hspace{3mm}{\rm and}\label{sing1}
			%	\end{align}
			%	\begin{align}
			%		\hspace{-0cm}&  {\cal D}_{l+1} =-\frac{ {2{\rm i}\Omega_s}\left(  lr^2d  g^{-1}/d r -  (l^2-l-4)r/g  \right)   \alpha_{l,m}}
			%		{{\omega_f}(l+1)^2(l+2)\left( 1-2m\Omega_s/({\omega_f}(l+1)(l+2) )\right)}, \hspace{2mm}{\rm where}\label{sing2}
			%	\end{align}

				%********WORKING******
				
				\begin{align}
					\hspace{-0cm}& {\cal D}_{l-1} =\frac{{2{\rm i}\Omega_s}}{\omega_f l^2
					\left (1- 2m\Omega_s/(\omega_f l(l-1))\right)}\left(  r^l d  g^{-1}/d r + 2(l+1)r^{l-1}/g  \right)  \beta_{l,m}
					%+++++++
					%-\frac{ {2{\rm i}\Omega_s}\left(  (l+1)r^2 d  g^{-1}/d r + (l^2+3l-2)r/g  \right)   \beta_{l,m}}
					%{{\omega}_fl^2(l-1)\left( 1-2m\Omega_s/({\omega_f}l(l-1) \right)}
					\hspace{3mm}{\rm and}\label{sing1}
				\end{align}
				\begin{align}
				 {\cal D}_{l+1} =\frac{ {2{\rm i}\Omega_s}}{\omega_f (l+1)^2
				 \left (1- 2m\Omega_s/(\omega_f(l+1)(l+2))\right)}\left(  r^ld  g^{-1}/d r + r^{l-1}/g  \right)   \alpha_{l,m}
                                   %++++++++++++				
					%\hspace{-0cm}&  {\cal D}_{l+1} =-\frac{ {2{\rm i}\Omega_s}
					%{{\omega_f}(l+1)^2(l+2)\left( 1-2m\Omega_s/({\omega_f}(l+1)(l+2) )\right)}
					%\left(  lr^2d  g^{-1}/d r -  (l^2-l-4)r/g  \right)   \alpha_{l,m}}
					, \hspace{2mm}{\rm where}\label{sing2}
				\end{align}
				\begin{align}
					\alpha_{l,m} =  \sqrt{\frac{(l-m+1)(l+m+1)}{(2l+1)(2l+3)} }    \hspace{2mm} {\rm and} \hspace{2mm}  \beta_{l,m} =  \sqrt{\frac{(l-m)(l+m)}{(2l-1)(2l+1)} }       %\hspace{3mm}{\rm and}
				\end{align}
				Note the potentially vanishing denominators at the $r$ mode resonances
				where ${\omega_f}=2m\Omega/(l'(l'+1).$
				\textcolor{red}{In our tidal problem with, $l=2,$ if  $|m|=1,$  the resonances occur for $l'=l-1=1$ and $l'=l+1=3.$
					For $|m|=2,$ from (\ref{Tser}),   only the case $l'=3$ is present.}
				
				The coefficients ${\cal D}_{l-1} $ and ${\cal D}_{l+1}$ enable ${\cal T}_1$ to be found
				from (\ref{Tser}) and the contribution this makes to the horizontal components of the displacement can then be found from
				(\ref{thetadecomp}) and (\ref{phidecomp}) with the substitution ${\cal T}\rightarrow {\cal T}_1.$
				Significantly ${\cal T}_1$ is affected by  $r$ mode resonances but $S_1$ is not. Accordingly if we are close to resonance
				$S_1$ may be neglected. 
				
				As we shall focus on conditions close to resonance  and accordingly  terms potentially  affected by resonant denominators,
				we remark that an expression for $W_1$
				that contains these is most easily found from (\ref{phicomp}) on setting $W=W_1,$ $\xi_r=0,$
				while including only the contributions involving ${\cal T}_1$ on the left hand side, when that is expressed as a sum 
				of spheroidal and toroidal components, as only this is amplified by resonance.
				This gives     
				\begin{align}
					\frac{W_1F}{\rho } = -\frac{{\rm i}\omega_f^2}{m}\left((1-\mu^2)\frac{\partial {\cal T}_1}{\partial  \mu}
					+\frac{2\Omega_s m\mu}{\omega_f}{\cal T}_1\right).\label{W1res}
					\end {align}

					\subsection{The effect of  radial motions   }\label{Radial motions}
					The above analysis results, at lowest order,  in singularities at single toroidal mode resonances for 
					a forcing potential associated with a  $(l,m)$ pair.
					It is important to recall  that this is exact  only when, $\eta_r,$ is neglected making ${\cal T}_2=S_2=W_2=0.$
					Corrections arising from the inclusion of, $\eta_r,$ will in addition to introducing response components with larger, $l',$
					affect the nature and location of any  individual toroidal mode singularity appearing
					in (\ref{sing1}) add (\ref{sing2}).
					
					These equations can only be used  far enough from the singularity
					that corrections arising from the inclusion of, $\eta_r,$ such as resonant frequency 
					splitting, can be neglected. When $\eta_r$ is not neglected singularities  occur
					at frequencies associated with the normal modes of the system and we could expect the original resonance to split accordingly.
					The frequency width associated with the splitting  should $\rightarrow 0,$ as $\omega_f \rightarrow 0.$

					As indicated above,  the discussion in Section \ref{T1calc}  neglected radial coupling
					between spherical shells occurring through the excitation of radial motion as this was assumed to be small.
					Below we include corrections due to radial motion and show that this causes a single toroidal mode resonance
					to split into  a closely spaced sequence of resonances, each associated with a radial normal mode.
					We derive a second order ordinary differential equation governing each of these.

					\subsubsection{Relating $S_2$ and  $W_2$  to $\eta_r$ through Hough function expansions }
					Following the procedures for obtaining equations relating $S_i, W_i$ and ${\cal T}_i$ outlined at the end of Section \ref{DetS}, 
					we find that these imply that
					\begin{align}
						\frac{1}{r^2} \nabla_{\perp}^2 S_2= \nabla\cdot{\mbox{\boldmath$\xi$}}_2=\frac{F}{r^2\rho\omega_f^2}O(W_2),\hspace{2mm}{\rm where} \hspace{1mm} {\rm  the} \hspace{1mm} 
						{\rm operator} \hspace{1mm} O \hspace{1mm}{\rm is}\hspace{1mm}{\rm defined}\hspace{1mm} {\rm through} \label{S21}
					\end{align}
					
					\begin{align}
						O(W_2)=\frac{\partial}{\partial \mu}\left(D^{-1}\left((1-\mu^2)\frac{\partial W_2}{\partial \mu}-\frac{2m\Omega_s\mu}{\omega_{f}}W_2\right)\right)
						+D^{-1}\left(    \frac{2m\Omega_s\mu}{\omega_{f}}\frac{\partial W_2}{\partial \mu}-\frac{m^2W_2}{1-\mu^2}\right), 
					\end{align}
					where $D=1-(2\Omega_s\mu)^2/\omega_f^2.$
					
					\noindent We remark that. the right hand side of (\ref{S21})  is obtained by using  the appropriate forms of  (\ref{thetacomp}) and  (\ref{phicomp}) to express
					$\nabla\cdot{\mbox{\boldmath$\xi$}}_2$ in terma of $W_2$ after having neglected the radial component of
					${\mbox{\boldmath$\xi$}}_2,$ \textcolor{blue}{or equivalently, $\eta_r,$ on the basis of the smallness of $\omega_f^2/N^2,$} or  the traditional approximation.
					
					The eigenvalues, $\lambda,$ of, $O,$ and the related  eigenfunctions,  ${\cal W}_{\lambda},$ known as Hough functions, satisfy $O({\cal W}_{\lambda})=-\lambda {\cal W}_{\lambda}.$ These define a normalised orthogonal system such that 
					$\int^1_{-1}{\cal W}_{\lambda}{\cal W}_{\lambda'}d\mu~=~\delta_{\lambda,\lambda'}.$
					A particular eigenvalue, $\lambda$ can be regarded as a function of $\Omega_s/\omega_f.$  The one corresponding to a strict toroidal mode resonance
					has $,\lambda =0,$ with, $2\Omega_s/\omega_f = l'(l'+1)/m.$ 
					As would be expected the angular dependence of the eigenfunction is of the same form as that of the dominant
					form of  $W_1$ as resonance is approached, \textcolor{blue}{ with ${\cal T}_1$ being proportional to a spherical harmonic
					of degree $l'$}  (see equation (\ref{W1res})).
					Near to  a strict toroidal mode resonance $\lambda$ is small and we may write
					\begin{align}
						\lambda = C_{l',m}( 2m\Omega_s
						/(l'(l'+1)\omega_f)-1),\label{lamdev}
					\end{align}
					where $C_{l',m} > 0$ is a dimensionless constant of order unity 
					\citep[see][]{PS97}	\footnote{for $l=3$ and  $|m|=2, $ we estimate $C_{3,2}=1/2.$} .
					%\subsubsection{determination of $\eta_r$}======
					
					From  (\ref{addy2a})  and  (\ref{S21})
					%  (\ref{addy2})
					we have
					\begin{align}
						\frac{F}{\rho\omega_f^2}O(W_2)=   -\frac{1}{F}\frac{\partial(r^2 F\eta_r)}{\partial r}.\label{W22}
					\end{align}
					Expanding an arbitrary quantity, ${\cal Q},$ as a series of Hough functions we write ${\cal Q} = \sum_{\lambda}Q_{\lambda}{\cal W}_{\lambda}.$
					In the case of,  $\eta_r,$ and, $W_2,$ equation (\ref{W22}) implies that the expansion coefficients, which depend on, $r,$ are related by 
					\begin{align}
						\frac{F}{\rho\omega_f^2}W_{2,\lambda}=   \frac{1}{\lambda F}\frac{\partial(r^2 F\eta_{r,\lambda})}{\partial r}.\label{W23}
					\end{align}
					\subsubsection{Reduction of the radial component of the linearised equation of motion}
					We now turn to the radial component of the equation of motion (\ref{rcomp}). which we write in the form
					\begin{align}
						\hspace{-2cm}& -{ \omega_f}^2\xi_r-2{\rm i}{ \omega_f}\Omega\sin\theta\xi_{\phi}                       
						=-\frac{F}{\rho}\frac{\partial (W_2 +W_1)}{\partial r} - N^2\eta_r .\label{rcomp1}\\
					\end{align}
					Now in the LHS of the above equation, we neglect,  $\eta_r,$  and,  $\xi_{\phi,2}.$ Thus we set $\xi_r=\xi_{r,eq}$ and $\xi_{\phi}=\xi_{\phi,1}.$
					These approximations are based on a low frequency regime in which the radial displacement correction to the equilibrium tide is small
					and are in line with the traditional approximation. With the help of (\ref{phidecomp}) we thus obtain
					\begin{align}
						\hspace{-2cm}&  
						\frac{F}{\rho}\frac{\partial W_2}{\partial r} + N^2\eta_r=
						-{\bar \omega_f}^2\frac{ U}{g}-\frac{2{\rm i}{\bar \omega_f}\Omega}{r}\left((1-\mu^2)\frac{\partial {\cal T}_1}{\partial \mu} +{\rm i} mS_1\right)
						-\frac{F}{\rho}\frac{\partial W_1}{\partial r} \equiv S_{\eta_{r}}                      
						\label{rcomp2}\\
					\end{align}
					Performing an expansion of (\ref{rcomp2}) in terms of Hough functions, we find with the help of
					(\ref{W23}) that the expansion coefficients are related
					by 
					\begin{align}
						%\frac{F}{\rho}\frac{\partial W_{2,\lambda}}
						%{\partial r} =\
						\frac{\omega^2_f F}{ \rho}\frac{\partial}{\partial r}\left(\frac{\rho}{ F^2}\frac{\partial(r^2 F\eta_{r,\lambda})}{\partial r}\right)
						+\lambda N^2\eta_{r,\lambda}
						=
						\lambda S_{\eta_{r},\lambda}  .  \label{etareq}                  
					\end{align}
					\subsubsection{Eigenvalue problem}\label{EVP}
					Equation (\ref{etareq}) allows the determination of $\eta_{r.\lambda}$ as the calculation  of a forced
					response, the right hand side specifying a known forcing.
					Although $S_{\eta_r}$ diverges at a strict toroidal mode resonance, (see equations (\ref{Tser}) and (\ref{W1res})),
					multiplication by $\lambda$ removes the singularity as can be seen from (\ref{lamdev}).
					Accordingly the resonances will be determined by the solution of the eigenvalue problem obtained
					by setting the left hand side of (\ref{etareq}) to zero. 
					This is the standard equation for $g$ modes under the anelastic approximation
					\citep[see e.g.][]{Chernov2017}.
					These authors find that in a WKBJ approximation  the spectrum  is given by
					\begin {align}
					\omega_f = \frac{\sqrt{\lambda }\int_{0}^{r_c}r^{-1}Ndr}{({n_r}\pi+\psi_{WKBJ})},
				\end{align}
				where ${n_r}$ is an integer and $\psi_{WKBJ}$ is a structure dependent phase factor that for our purposes can be can be evaluated
				for $\lambda=0$ and whose existence and details we discuss below.
				\subsubsection{The outer boundary conditions}
				%do not \textcolor{blue}{  consider further here \footnote{\textcolor{blue}
						The above discussion supposes that the inner solution governing a normal 
						mode is to be matched to one with a similar WKBJ form
						with an appropriate  phase  shift.  Whether this  occurs is determined
						by  matching to a solution in the convective envelope. Following Papaloizou \& Savonije(1997)
						we envisage that $\xi_r$ is. small compared to the horizontal components of the displacement
						and  close to the equilibrium tide and $W$
						should be matched at the lower boundary of the convective envelope.
						It is likely that an imposed  large scale $W$ corresponding to a normal mode produces
						a response on both large and small scales in the convection zone.
						However, we argue that only the \textcolor{blue}{large scale} 
						 response when  expressed in terms of a series of Hough functions
						needs to be retained leading to compatibility with the  discussion of the previous Section.
						This is because, as seen in the numerical results, 
						 \textcolor{blue}{smaller scale  responses  will tend to produce even smaller radial scale}  low frequency
						$g$ modes. These are expected to be damped by thermal and viscous effects close to the boundary. 
						\subsubsection{Toroidal mode resonances}
						The discussion of Section \ref{EVP}  has the consequence that in the low frequency limit each toroidal mode resonance splits into
						a potentially large  set of normal modes, each associated with a value of ${ n_r}.$ This would be limited
						by non adiabatic effects in practice.  Using (\ref{lamdev}) to eliminate ${\lambda}$ the spectrum can be shown to be given by
						
						\begin{align}
				\omega_f - \frac{2m\Omega_s}{l'(l'+1)} =-\frac{({ n_r}\pi+\psi_{WKBJ})^2\omega_f^3}{C_{l',m} (\int_{0}^{r_c}r^{-1}N dr)^2}\sim
							-\frac{({ n_r}\pi+\psi_{WKBJ})^2(2m\Omega_s)^3}{C_{l',m}(l'(l'+1))^3(\int_{0}^{r_c}r^{-1}N dr)^2}.\label{spectrum}
							\end {align}
							We see that the right hand side gives small negative corrections to the basic toroidal mode frequency 
							when,  $C_{l',m} > 0$ and $\Omega_s^2/N^2,$ is small.               
							As indicated in Section \ref{Radial motions} , (\ref{spectrum} ) specifies the frequency 
							departure from strict toroidal mode resonance
							needed in order to make use of (\ref{Tser}) with  (\ref{sing1}) and  (\ref{sing2}).
							 The maximum value of ${ n_r}$ that needs to be considered is limited
							to modest values by consideration of overlap integrals 
							 (eigenfunction mismatched to forcing potential) and non adiabatic effects.

							\section{ The tidal response of the convective envelope in the limit of low forcing
							 frequency as viewed in a frame co-rotating with the star:
								a critical latitude singularity}\label{ConvResp}
							We assume this region is isentropic and thus  $N^2=0.$
							The governing equation of motion  \textcolor{blue}{is obtained from}  (\ref{eqmot}) in 
							which the isentropic condition allows  us to  set, $F=\rho$  
						\textcolor{blue}{and in which} a viscous force per unit mass, ${\bf f}_{\nu}$ \textcolor{blue}{is incorporated.}  Thus we have
							\begin{align}
								\hspace{-2cm}& -{ \omega_f}^2{\mbox{\boldmath$\xi$}}+2{\rm i}{ \omega}{\mbox{\boldmath$\Omega$}\times}{\mbox{\boldmath$\xi$}}
								=-\nabla W + {\bf f}_{\nu}.
								\label{eqmotconv}
							\end{align}
							\textcolor{blue}{In addition we make use of equation (\ref {addy}), which } under adiabatic conditions 
							and the anelastic approximation takes the form
							\begin{align}
								\hspace{-0cm}& \frac{\rho^2U}{\Gamma_1 P}
								%P^{1/\gamma} 
								= \nabla\cdot(\rho{\mbox{\boldmath$\xi$}}).
								\label{addyconv}
							\end{align} 
							In this Section we find it convenient to adopt cylindrical polar coordinates $( \bar r, \phi,z).$
							We anticipate that in the inviscid case $\xi_{\phi}$ will become singular as, $\omega_f\rightarrow 0,$
							requiring the implementation of viscosity. Assuming  this component of the displacement
							is mainly affected,   we set ${\bf f}_{\nu}= (0, f_{\nu,\phi},0)$
							From the components of (\ref{eqmotconv}) we then obtain
							\begin{align}
								&(\omega_f^2-4\Omega_s^2)\xi_{\bar r} =\frac{\partial W}{\partial \bar r} +\frac{2m\Omega_s W}{\bar r\omega_f} 
								+\frac{2{\rm i}\Omega_s f_{\nu,\phi}}{\omega_f}, \label{convcompr}\\
								&(\omega_f^2-4\Omega_s^2)\xi_{\phi} =\frac{2{\rm i}\Omega_s }{\omega_f} \frac{\partial W}{\partial \bar r} 
								+\frac{{\rm i} m W}{\bar r} 
								- f_{\nu,\phi}, \label{convcompphi}\\
								&\omega_f^2\xi_z =\frac{\partial W}{\partial z}. \label{convcompz}
							\end{align}
							Making use of (\ref{convcompr}) - (\ref{convcompz}) with (\ref{addyconv}) we obtain
							\begin{align}
								\omega_f^2(\omega_f^2-4\Omega_s^2)\frac{\rho^2 U}{\Gamma_1 P}=&
								(\omega_f^2-4\Omega_s^2)\frac{\partial}{\partial z}\left(\rho\frac{\partial W}{\partial z}\right)
								+\frac{\omega_f^2}{\bar r}\left(\frac{\partial}{\partial \bar r}
								\left(\rho \bar r\frac{\partial W}{\partial \bar r}\right) -\frac{m^2 \rho W}{\bar r}\right)
								+\frac{2m\Omega_s \omega_f W} {\bar r }\frac{\partial \rho}{\partial \bar r} \nonumber\\
								& +\frac{1}{\bar r} \frac{\partial}{\partial \bar r} \left({2{\rm i}\Omega_s \omega_f \bar  r\rho f_{\nu,\phi} }\right)  -
								\frac{{\rm i} m\omega_f^2 \rho f_{\nu,\phi}}{\bar r}.\label{Weq}
							\end{align}
							In practice we are interested in the case when ${ U}$ corresponds to a forcing potential with $m=2.$ 
							\subsection{Solution in the small $\omega_f$ inviscid limit: a singular response}
							In the limit, $(\omega_f, f_{\nu,\phi})\rightarrow 0,$ (\ref{Weq}) becomes,
							  $\partial( \rho (\partial W/ \partial z))/\partial z=0.$
							This implies that in this limit, $W,$ depends only on $\bar r.$ To find this dependence 
							we write the asymptotic expansion \\
							%$W= \sum_{k=0}^{\infty}\omega_f^{k+1}  W_k,$ with $W_0=W_0(r).$ 
							$W=W_0(\bar r)\omega_f+W_1(\bar r,z)\omega_f^2.$
							In the inviscid limit  (\ref{Weq}) leads to  
							\begin{align}
								\frac{\rho^2 U}{\Gamma_1 P}=\frac{\partial}{\partial z}\left(\rho\frac{\partial W_1}{\partial z}\right)
								 -\frac{m  W_0}{2\Omega_s \bar r }\frac{\partial \rho}{\partial \bar r}
								-\frac{\omega_f}{4\Omega^2_s \bar r}\left(\frac{\partial}{\partial \bar r}
								\left(\rho \bar r\frac{\partial W_0}{\partial \bar r}\right) -\frac{m^2 \rho W_0}{\bar r}\right).
								\label{invlow} 
							\end{align}
							%with neglected terms being of order order, $\omega_f^2,$ 
							
							\noindent  Note that the term $\propto \omega_f$ in (\ref{invlow}) would be absent if a solution 
							consisting of a formal power series in $\omega_f$ was adopted \textcolor{blue}{and all terms of order
							$\omega_f$ and higher were neglected }
							However, this fails to represent the possibility of solutions where $W_0$ varies
							 on a small scale $\propto \omega_f^{1/2}$
							in part of the domain and so with this in mind we retain  the term \textcolor{blue}{$\propto\omega_f$ in (\ref{invlow})}.
							
							\subsubsection{Determination of $W_0$}

							We find $W_0$ by integrating w. r. t. , $z,$ along a line of constant, $\bar r,$ 
							 through the convection zone.  We adopt a simple model where $\rho$ is assumed to  vanish at the upper boundary.
							The limits of integration  are $(z_0, z_*)$\footnote[1]{Note that $z_0$ and $z_*$ can
							 both be positive or both be negative here}, where, $z_0 =\sqrt{r_c^2-\bar r^2},$ is the
							  value of, $z,$ at the inner boundary and, $z_*,$
							 the value at the upper boundary, 
							for, $\bar r< r_c,$ where $r_c$ is the radius of the inner boundary of the convection zone. 
							For, $\bar r > r_c,$ the limits are $(-z_*,z_*).$ 
							
							\noindent  Thus for $\bar r>r_c$ we find
							\begin{align}
								&\hspace{-0cm}\int^{z_*}_{-z_{*}}\frac{\rho^2 U}{\Gamma_1 P}dz
								% =-\frac{n  W_0(r)}{2\Omega_s r }\int^{z_*}_{-z_{*}}\frac{\partial \rho}{\partial r}dz
								=\frac{m  W_0(\bar r)}{2\Omega_s \bar r }
								\int^{z_*}_{-z_{*}}\frac{\rho^2 \bar r}{\Gamma_1 P\sqrt{\bar r^2+z^2} }gdz%\nonumber\\
								%&\hspace{-3cm} 
								-\frac{\omega_f}{4\Omega^2_r \bar r}\left(\frac{\partial}{\partial \bar r}
								\left(\Sigma \bar r\frac{\partial W_0}{\partial \bar r}\right) -\frac{m^2 \Sigma W_0}{\bar r}\right),
								\label{convbdry+}
							\end{align}
							where, $\Sigma=\int^{z_*}_{-z*} \rho dz$ and we have used the fact that
							\begin{align}
								\int^{z_*}_{-z_{*}}\frac{\partial \rho}{\partial \bar r}dz=
								 -\int^{z_*}_{-z_{*}}\frac{\rho^2 \bar r}{\Gamma_1 P\sqrt{\bar r^2+z^2} }gdz
							\end{align}
							\noindent  In addition  for, $\bar r<r_c,$ we obtain
							\begin{align}
								&\hspace{-4cm} \int^{z_*}_{z_{0}}\frac{\rho^2 U}{\Gamma_1 P}dz=
								-\left(\rho\frac{\partial W_1}{\partial z}\right)_{z=z_0} 
								-  \left( \frac{m  W_0(\bar r)}{2\Omega_s \bar r }  +\frac{\omega_f}{4\Omega_s^2}
								\frac{\partial W_0}{\partial \bar r}            \right) 
								\int^{z_*}_{z_{0}}\frac{\partial \rho}{\partial \bar r}dz\nonumber\\
								&\hspace{-1.5cm} -\frac{\omega_f}{4\Omega^2_r \bar r}\left(\frac{\partial}{\partial \bar r}
								\left( \bar r\frac{\partial W_0}{\partial \bar r}\right) -\frac{m^2  W_0}{\bar r}\right)\int^{z_*}_{z_{0}} \rho dz
								\label{convbdry-}
							\end{align}
							The first term on the right hand side can be found by noting that the radial displacement
							at the inner boundary must match the equilibrium tide value
							 \footnote{But note that in the presence of a toroidal mode resonance
								this may be significantly modified }  Thus on the inner boundary we  have for $z \rightarrow  z_0,$
							$z(\partial W_1/\partial z) +\bar r\xi_{\bar r }= -r_c{ U}/g.$

							\subsubsection{Strict low frequency limit}\label{Strict}
							From (\ref{convcompr}) and   (\ref{convcompphi})   in the limit $\omega_f \rightarrow 0$ we have 
							$\xi_{\bar r}= - mW_0/(2 \bar r\Omega_s)$ and $\xi_{\phi}= -{\rm i}/(2\Omega_s)\partial W_0/\partial \bar r.$
							These imply that these displacement components  then depend only on $r.$
							Using the first of these relations in (\ref{convbdry-}) after setting  $ \omega_f\rightarrow 0,$ we obtain
							\begin{align}
								\int^{z_*}_{z_{0}}\frac{\rho^2U}{\Gamma_1 P}dz=
								\frac{(\rho (- m W_0/(2\Omega_s) +r_c{U}/g ))_{z=z_0}}{z_0} 
								-\frac{m W_0(\bar r)}{2\Omega_s \bar r }\int^{z_*}_{z_{0}}\frac{\partial \rho}{\partial \bar r}dz.\label{convbdry--}
							\end{align}
							
							\noindent We see that \textcolor{blue}{in general in the limit
							$ \omega_f \rightarrow 0$} (\ref{convbdry+}) and (\ref{convbdry--}) imply 
							that $W_0(\bar r)$ changes discontinously as $\bar r=r_c$ is passed through. Thus
							\begin{align}
								& W_0|_{\bar r=r_c-} =2\Omega_s r_c{U}/(mg), \hspace{2mm}{\rm and}\nonumber \\
								&  W_0 |_{\bar r=r_c+}=
								2\Omega_s r_c\int^{z_*}_{-z_{*}}\frac{\rho^2 U}{\Gamma_1 P}dz {\bigg /} 
								\left(m\int^{z_*}_{-z_{*}}\frac{\rho^2 r_c}{\Gamma_1 P\sqrt{r_c^2+z^2} }gdz\right).\label{W0eq}
							\end{align}
							This behaviour is reminiscent of that expected to emanate from a 
							critical latitude singularity at the inner convection zone
							boundary.
							Indeed in the limit, $\omega_f  \rightarrow 0, $ this is expected to occur at, $\bar r=r_c.$ 
							Conventionally this discontinuity is resolved by the incorporation of viscosity and we discuss this below.
							However, before doing so we note that it is also potentially resolved through the incorporation of 
							the terms $\propto \omega_f$  in (\ref{convbdry+}) and (\ref{convbdry-}) that involve a second derivative of $W_0.$ 
							The length scale involved can be estimated as, $\propto  \bar r\sqrt{|\omega_f |/(2m\Omega_s)},$ 
							which should be $<< \bar r.$
							For viscous effects to dominate we need $\omega_f$ to be small enough  that  the estimated thickness of the
							viscous transition layer exceeds this. We consider this aspect below.

							\subsection{The effect of viscosity}
							We recall that  to lowest order in, $\omega_f,$  from (\ref{convcompphi}) we found that 
							\begin{align}
								\xi_{\phi}= - {\rm i} \partial W_0/\partial \bar r /(2\Omega_s)\label{xiphiL}.
							\end{align}
							Thus
							we expect $\xi_{\phi}$ to be most affected when $W_0$ changes rapidly producing a strong shear layer.
							Hence in order to investigate the effect of viscosity we consider a very
							simple model for which only the $\phi$ component of the viscous force 
							is included and that this only affects $\xi_{\phi}.$
							In particular  we adopt the simple form $f_{\nu,\phi} = {\rm i}\omega_f \nu\partial ^2 \xi_{\phi} /\partial \bar r^2$
							 in which only 
							variation of, $\xi_{\phi},$ is taken into account
							and then only the highest order derivative 
							with respect to $\bar r$ is retained. In addition for simplicity we shall assume that $\nu$ is constant.
							
							From (\ref{convcompphi}) in the low frequency limit, the viscous force is 
							comparable to inertial forces in magnitude when the length scale
							is $\sim   \sqrt{\nu \omega_f/\Omega_s^2}.$ Anticipating that the actual length 
							scale can be ordered to be significantly larger  than this, as will later be borne out,
							we can find the effect of $f_{\nu,\phi}$ on $\xi_{\phi}$ from (\ref{convcompphi}) 
							iteratively. In this way we find the lowest order viscous correction
							to (\ref{xiphiL}) taking only variation of $W_0$ into account,  which modifies it to read
							\begin {align}
							\xi_{\phi}= - {\rm i} \partial W_0/\partial \bar r /(2\Omega_s) +
							( \omega_f \nu/(8\Omega_s^3) )\partial ^3  W_0 /\partial \bar r^3,\label{xiphiviscapprox}
						\end{align}
						with the lowest order form of $f_{\nu,\phi}$ being
						\begin {align}
						f_{\nu,\phi}=   ( \omega_f \nu/(2\Omega_s) )\partial ^3  W_0 /\partial \bar r^3,
						\label{xiphiviscapprox1}
					\end{align}
					where as above we retain only the variation of $W_0$  and retain only the term with the highest order
					derivative when considering the viscous force.
					This can now be included in  equation (\ref{Weq})   and subsequently by expanding in powers of $\omega_f$
					we find \textcolor{blue}{in the limit $\omega_f\rightarrow 0$}  equation (\ref{invlow})  becomes 
					\begin{align}
						\frac{\rho^2 U}{\Gamma_1 P}-\frac{\partial}{\partial z}\left(\rho\frac{\partial W_1}{\partial z}\right) 
						+\frac{m W_0}{2\Omega_s r }\frac{\partial \rho}{\partial \bar r}=
		-\frac{{\rm i}\rho\nu}{4\Omega_s^2}\frac{\partial ^4  W_0} {\partial \bar r^4}\label{visceq} % \frac{\partial}{\partial r} \left(  f_{\nu,\phi} \right) 
					\end{align}
					
					\subsubsection{Structure of the viscous shear layer}
					\textcolor{blue}{Integrating   (\ref{visceq}) w.r.t. $z$ over $[-z_*,z_*],$ we find that   
					 for $\bar r>r_c$  }
					\begin{align}
						\int^{z_*}_{-z_{*}}\frac{\rho^2 U}{\Gamma_1 P}dz 
				-\frac{m  W_0(r)}{2\Omega_s \bar r }\int^{z_*}_{-z_{*}}\frac{\rho^2 \bar r}{\Gamma_1 P\sqrt{\bar r^2+z^2} }gdz\label{convbdryvisc+}
						=-\frac{{\rm i}\Sigma\nu}{4\Omega_s^2}\frac{\partial ^4  W_0} {\partial \bar r^4}.  
					\end{align}
				\textcolor{blue}{	This gives the form of  equation	  (\ref{convbdry+})  modified by 
					the effect of viscosity in the limit $\omega_f\rightarrow 0.$ }    			
							%where, $\Sigma=\int^{z_*}_{-z*} \rho dz.$ 
					Similarly  for $\bar r < r_c,$ \textcolor{blue}{in the same limit,} 
					 equation (\ref{convbdry-}) is modified by the effect of viscosity to read
					\begin{align}
						z_0\int^{z_*}_{z_{0}}\frac{\rho^2 U}{\Gamma_1 P}dz-
						(\rho (r_c{ U}/g - m W_0/(2\Omega_s) ))_{z=z_0} 
						+\frac{m z_0 W_0(\bar r)}{2\Omega_s \bar r }\int^{z_*}_{z_{0}}\frac{\partial \rho}{\partial \bar r}dz\label{convbdryvisc-}
						= -\frac{{\rm i}\Sigma_{z_0}\nu z_0}{4\Omega_s^2}\frac{\partial ^4  W_0} {\partial \bar r^4},
					\end{align}
					where, $\Sigma_{z_0}=\int^{z_*}_{z_0} \rho dz.$
					\subsubsection{Scaled local coordinate}
					We are interested in a narrow region around, $\bar r=r_c,$ which will be of vanishing width as, $\nu \rightarrow 0.$
					To find its structure
					we define the dimensionless coordinate, $x = |(\bar r-r_c)|/\epsilon,$ where the small quantity, 
					$ \epsilon = [(\nu \bar r^2/(4\Omega_s)]^{1/4}|_{\bar r=r_c}.$
					We replace, $\bar r,$ with, $x,$ in (\ref{convbdryvisc+}) and (\ref{convbdryvisc-})
					 and take the limit, $\epsilon \rightarrow 0,$ which, respectively, leads to
					\begin{align}
						& {\rm i}\frac{\partial^4 W_0}{\partial x^4} 
						= -\alpha_1
						%\int^{z_*}_{-z_{*}}\frac{\rho^2\bar U}{\Gamma_1 P}dz 
						%-\frac{m  W_0(r)}{2\Omega_s r }\int^{z_*}_{-z_{*}}\frac{\rho^2 r}{\Gamma_1 P\sqrt{r^2+z^2} }gdz.
						\label{convbdryvisc1+}
						+\alpha_2W_0, %=\frac{{\rm i}\Sigma\nu}{4\Omega_s^2}\frac{\partial ^4  W_0} {\partial r^4},   
						\hspace{2mm} {\rm with} 
					\end{align}
					\begin{align}
						&\hspace{-0.0cm} \hspace{2mm} \alpha_1 = \frac{\bar r^2\Omega_s}{\Sigma } 
						\left. \int^{z_*}_{-z_{*}}\frac{\rho^2U}{\Gamma_1 P}dz \right |_{\bar r=r_c}\hspace{2mm}{\rm and}
						\hspace{2mm}\alpha_2 = \left. \frac{m \bar r }
						{2 \Sigma }\int^{z_*}_{-z_{*}}\frac{\rho^2 \bar r}{\Gamma_1 P\sqrt{\bar r^2+z^2} }gdz \right |_{\bar r=r_c}
						\hspace{2mm}{\rm for} \hspace{2mm} \bar r > r_c
					\end{align}
					\begin{align}
						%z_0\int^{z_*}_{z_{0}}\frac{\rho^2\bar U}{\Gamma_1 P}dz-
						\hspace{-3.3cm}{\rm and } \hspace{3.7cm}  W_0 = 
						\left. 2\Omega_s \bar r{ U}/(m g) \right |_{\bar r=r_c} \equiv \alpha_3 \hspace{1cm}{\rm for} \hspace{2mm} \bar r <r_c.
						%+\frac{m z_0 W_0(r)}{2\Omega_s r }\int^{z_*}_{z_{0}}\frac{\partial \rho}{\partial r}dz.
						\label{convbdryvisc1-}
						%\frac{{\rm i}\Sigma_{z_0}\nu z_0}{4\Omega_s^2}\frac{\partial ^4  W_0} {\partial r^4},
					\end{align}
					Note that viscosity does not appear in (\ref{convbdryvisc1-}) on account of the
					 vanishing of the  small quantity, $z_0/r_c= \sqrt{r_c^2-\bar r^2}/r_c 
					\sim \sqrt{2x\epsilon/r_c},$ in the limit $\epsilon \rightarrow 0.$ This does not mean 
					that  viscous effects vanish entirely for, $\bar r < r_c,$ \textcolor{blue}{ but to be significant,
					 a smaller scale than inferred for $\bar r > r_c$ would be needed.}
					\subsubsection{ Explicit local solution: width of the shear layer}
					To find the solution we consider (\ref{convbdryvisc1+}) and (\ref{convbdryvisc1-}).
					For $\bar r < r_c$ from (\ref{convbdryvisc1-}) we find very simply that $W_0=\alpha_3$ is constant.
					We then find a solution of (\ref{convbdryvisc1+}) that matches $W_0$ and its first derivative at $x=0$
					and in addition is such that the effect of viscosity $\rightarrow 0$ for $x \rightarrow \infty.$
					This also ensures continuity of $\xi_{\phi}.$ 
					This solution, applying for $\bar r> r_c,$ or $x > 0,$  assuming $m < 0,$ is straightforwardly found to be given by
					\begin{align}
						W_0= \frac{\alpha_1}{\alpha_2}+\frac{1}{2}
						\left( \alpha_3 -\frac{\alpha_1}{\alpha_2}\right) 
						\bigg(( 1+ {\rm i})\exp(-\gamma x)+(1-{\rm i})\exp({\rm i}\gamma x)\bigg),
					\end{align}
					where $\gamma = \exp({\rm i} \pi/8)(|\alpha_2|)^{1/4}.$ Given that the parameters in the above are expected to
					be of order unity, the characteristic scale for the decay of viscous effects is set by considering $x$ to be  of order
					unity. Thus corresponds to a radial width for the shear layer  
					$\sim \epsilon = r_c[(\nu /(4\bar r^2\Omega_s)]^{1/4}|_{\bar r=r_c}.$
					Notably this has a weak power law scaling with $\nu$ that is expected to apply to orbital evolution rates
					associated with the tidal forcing.
					
					\noindent However it is important to note that for \textcolor{blue}{ corrections associated with inertial waves
					 to be formally negligible in } the layer, $\omega_f$ must be small enough that $\epsilon$ exceeds the scale
					$r_c\sqrt{|\omega_f|/ (2m\Omega_s)}$ (see discussion below  equation (\ref{W0eq}). 
					%This implies that to within a factor of order unity
					%we require that 
					%$\nu > r^2\omega_f^2 / \Omega_s.$

				\end{appendix}
				\label{lastpage}
			\end{document}